\documentclass[%
 aip,
 amsmath,amssymb,
 reprint,%
]{revtex4-2}

\usepackage{graphicx}
\usepackage{dcolumn}
\usepackage{bm}

\usepackage[utf8]{inputenc}
\usepackage[T1]{fontenc}
\usepackage{mathptmx}
\usepackage{etoolbox}
\usepackage{xcolor}
\usepackage{comment}
\usepackage{soul}
\newcommand{\Yb}{$\mathrm{^{171}Yb}$~}

\makeatletter
\def\@email#1#2{%
 \endgroup
 \patchcmd{\titleblock@produce}
  {\frontmatter@RRAPformat}
  {\frontmatter@RRAPformat{\produce@RRAP{*#1\href{mailto:#2}{#2}}}\frontmatter@RRAPformat}
  {}{}
}%
\makeatother
\begin{document}

\preprint{AIP/123-QED}

\title{Entanglement-Enhanced Optical Atomic Clocks}
\author{Simone Colombo}
\affiliation{Department of Physics, MIT-Harvard Center for Ultracold Atoms and Research Laboratory of Electronics, Massachusetts Institute of Technology, Cambridge, Massachusetts 02139, USA.}

\author{Edwin Pedrozo-Pe\~{n}afiel}
\affiliation{Department of Physics, MIT-Harvard Center for Ultracold Atoms and Research Laboratory of Electronics, Massachusetts Institute of Technology, Cambridge, Massachusetts 02139, USA.}

\author{Vladan Vuleti\'{c}}
\affiliation{Department of Physics, MIT-Harvard Center for Ultracold Atoms and Research Laboratory of Electronics, Massachusetts Institute of Technology, Cambridge, Massachusetts 02139, USA.}
\email{colombos@mit.edu}\email{epedrozo@mit.edu}\email{vuletic@mit.edu}


\date{\today}

\begin{abstract}
Recent developments in atomic physics have enabled the experimental generation of many-body entangled states to boost the performance of quantum sensors beyond the Standard Quantum Limit (SQL). This limit is imposed by the inherent projection noise of a quantum measurement. In this perspective article, we describe the commonly used experimental methods to create many-body entangled states to operate quantum sensors beyond the SQL. In particular, we focus on the potential of applying quantum entanglement to state-of-the-art optical atomic clocks. In addition, we present recently developed time-reversal protocols that make use of complex states with high quantum Fisher information without requiring sub-SQL measurement resolution. We discuss the prospects for reaching near-Heisenberg limited quantum metrology based on such protocols.

\end{abstract}

\maketitle

\section{\label{sec:level1}introduction}
Optical-transition atomic clocks~\cite{oelker2019demonstration,Schioppo2017Ultrastable,Brewer2019Al,katori2020,bothwell2022resolving} are the most accurate sensors developed by humankind, reaching fractional stabilities below $10^{-18}$.
This mindboggling precision, corresponding to an uncertainty of less than one second over the age of the universe, and the associated technological improvements, enable a broad range of applications in the field of precision metrology, such as the search for dark matter in the low-to intermediate-mass sector \cite{Pospelov2013GNOME,stadnik2015can,delaunay2017probing, berengut2018probing,counts2020evidence,stadnik2020new}, the investigation of nuclear structure and matter \cite{mikami2017probing, flambaum2018isotope, counts2020evidence}, the study of any time variation of fundamental constants~\cite{huntemann2014improvement, safronova2018two, dzuba2018Testing}, testing of the foundations of general relativity\cite{chou2010optical,grotti2018geodesy, katori2020,bothwell2021resolving}~, the detection of low-frequency gravitational waves~\cite{kolkowitz2016gravitational, Sedda_2020}, or geodesy~\cite{lisdat2016clock, bondarescu2015ground}. Moreover, high timekeeping precision enables improvements in navigation systems, both GPS-based and inertial~\cite{major2007quantum, grewal2020global}.

Currently, the main limitations to the clock precision are atomic collisions leading to atomic energy shifts~\cite{le2006accurate, Ludlow2008,campbell2009probingColl,lisdat2009coll,Bloom2014, gao2018systematic}, the Dick noise~\cite{dick1987local} that is associated with the interrupted interrogation of the atomic system, and the standard quantum limit (SQL) resulting from the quantum projection noise of an ensemble of finite atom number.
Schemes for removing the Dick noise have been demonstrated~\cite{Chou2011Quantum,takamoto2011frequency, Nicholson2012,Schioppo2017Ultrastable, Norcia2019,Lodewyck2009Nondestructive, Westergaard_DickeNoise_2010}, while collisional line shifts are usually minimized by deploying a low-density atomic gas or using engineered single-atom traps like 3D optical lattices~\cite{akatsuka2010three, Campbell2017} or arrays of optical tweezers~\cite{madjarov2019atomic, young2020half}. 
Collisional shifts impose a limit on the total number of atoms that are used in optical clocks, which is typically between $10^2$ and $10^4$ atoms~\cite{Ludlow2015}.
At such relatively small atom number, the SQL presents a significant constraint on the clock precision. The SQL can be overcome by engineering quantum correlations (entanglement) between the atoms~\cite{ma2011quantum,Pezze2018}.
%
Appropriate collective entangled states~\cite{wineland1992a,Wineland1994,Kitagawa1993,kuzmich1998atomic, Appel2009,Schleier-Smith2010, Hosten2016, Cox2016a, malia2022distributed} can readily boost the performance of these advanced clocks, particularly for applications that require operation at fixed bandwidth~\cite{Pospelov2013GNOME, stadnik2014axion,smorra2019direct, Sedda_2020}.

The generation of metrologically useful entanglement on the optical clock transition of ytterbium-171 ($\mathrm{^{171}Yb}$) has been recently demonstrated~\cite{Pedrozo2020entanglement}.
%
Yet, full clock operation beyond the SQL in a state-of-the-art optical clock represents one of the most important challenges to be met.
In this article, we will give an overview and perspective on this topic that is central to the development of future optical clocks and other sensors based on quantum interference \cite{Pezze2018}.

\section{Metrological Gain}\label{sec:metrologicalGain}

Optical clocks measure the passing of time in terms of the frequency standard provided by an optical transition to a long-lived atomic excited state with frequency $f_a	\gtrsim 100~$THz, corresponding to more than $10^{14}$ oscillations in a typical interrogation time of $\sim 1$ second.
The fractional stability of an optical clock is expressed as ${\sigma\equiv\delta f/ f_a}$ with ${\delta f}$ being the uncertainty of the frequency estimation.

To measure such high oscillation frequencies, one compares a local oscillator (LO) to the atoms' oscillations.
The LO can be, depending on the type of operation, either an ultrastable laser~\cite{Schioppo2017Ultrastable, takamoto2011frequency, Nicholson2012} (complete clock implementation), or another atomic ensemble~\cite{zheng2022differential, bothwell2022resolving} (differential operation). 
The comparison is performed by measuring the accumulated phase difference between the LO and the atomic reference in a given measurement time $\tau$ (called the spectroscopy or interrogation time).
In a linear protocol, the relation between phase difference $\varphi$ and frequency imbalance ${\Delta f} = f_{a}-f_\mathrm{LO}$ between the LO and the atomic system is given by~\cite{bishof2013optical}
\begin{equation}
    {\varphi(\tau)} = 2\pi \int_0^\tau w(t){\Delta f(t)}\mathrm{d} t
    \label{eq:phi_as_freq}
\end{equation}
where $w(t)$ is the protocol sensitivity response~\cite{bishof2013optical}.
From the accumulated phase, one can extract information about the average frequency difference. %
The clock stability is then limited by the precision in the estimation of the phase $\varphi$.

The goal is then to estimate the phase $\varphi$ with the smallest possible uncertainty $\delta\varphi$ for a fixed set of resources, such as atom number, spectroscopy time, and dark time.
The uncertainty on the phase estimation can have both classical (technical noise) and quantum contributions.
In state-of-the-art optical clocks, the technical noise is of the same order or smaller than the quantum noise.
For atomic sensors that employ a number $N$ of uncorrelated atoms, the quantum noise limits the phase estimation to the SQL, $\delta\varphi_\mathrm{SQL}=1/\sqrt{N}$.
The SQL is not a fundamental limit and can be overcome by proper engineering of quantum correlations (entanglement) between the atoms prior to starting the measurement protocol~\cite{wineland1992a,Wineland1994,Kitagawa1993,kuzmich1998atomic}. 

Metrologically useful entangled states are characterized by their quantum Fisher information~$\mathcal{F}_\mathrm{Q}$~\cite{Pezze2009, hyllus2012fisher, Pezze2018} which quantifies the sensitivity of the state to a change of a parameter, here a phase angle.
Generally, the quantum limit to the phase estimation for a certain quantum state is given by the quantum Cram\'{e}r-Rao bound~\cite{Pezze2018}
\begin{equation}
    \delta \varphi_\mathrm{CR} = \frac{1}{\sqrt{\mathcal{F}_\mathrm{Q}}}.
\end{equation}
Therefore, the quantum Fisher information is the figure of merit for the ultimate sensitivity achievable with a given quantum state. 
The field of quantum metrology investigates the class of quantum states with $\mathcal{F}_\mathrm{Q}>N$, which is a necessary and sufficient condition for a quantum-enhanced sensor (i.e. a sensor operating beyond the SQL)~\cite{Pezze2009}.
The fundamental limit to the phase estimation, the holy grail of quantum metrology, is the Heisenberg limit (HL) for which $\mathcal{F}_\mathrm{Q}=N^2$. 
%

%
%
%
%
%

In this article, we will discuss collective entangled states, i.e. states with $\mathcal{F}_\mathrm{Q}>N$, and how to generate them in optical-transition clocks.
We express then the performances of a quantum enhanced sensors in terms of metrological gain beyond the SQL as
\begin{equation}
    \mathcal{G}\equiv\left(\frac{\mathrm{SNR}}{\mathrm{SNR}_{SQL}}\right)^2,
    \label{eq:def_gain}
\end{equation}
where $\mathrm{SNR}$ and $\mathrm{SNR}_{SQL}$ are the signal-to-noise ratios for the actual sensor and an ideal sensor operating at the SQL, respectively.
Except for relatively simple entangled states with a Gaussian envelope, known as spin squeezed states (SSSs), saturating the Cramér-Rao bound in phase estimation requires probabilistic methods beyond the standard evaluation of expectation values~\cite{Pezze2018,Strobel2014,Krischek2011}.
Hence the metrological gain is bounded by
\begin{equation}
    \mathcal{G}\leq\frac{\mathcal{F}_\mathrm{Q}}{N},
\end{equation}
where the equality applies when the Cramér-Rao bound is saturated.

\section{\label{sec:SqueezinMethods} Squeezing}

%

\begin{figure}[ht!]
\setlength{\unitlength}{1\textwidth}
\includegraphics[width=85mm,scale=1]{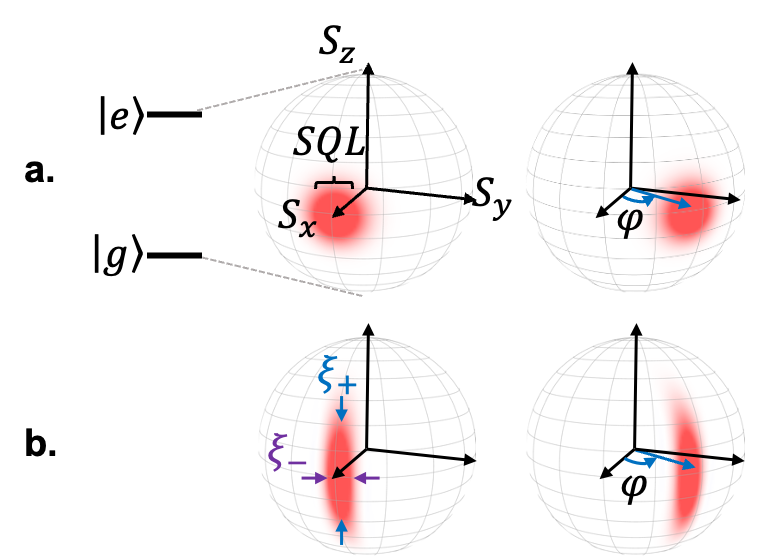}
\caption{\textbf{Generalized Bloch Sphere Representation of an $N$-atom system.} 
\textbf{a.} Ramsey spectroscopy using a (not entangled) coherent spin state. The blurred circle represents the quantum projection noise that gives rise to the Standard
Quantum Limit (SQL), limiting the precision of the determination of the phase $\varphi$. 
\textbf{b.} Ramsey spectroscopy using a spin squeezed state, a special and simple collective entangled state where the quantum noise is redistributed within two orthogonal quadratures. $\xi_{-}$ and $\xi_{+}$ denote the reduced (squeezed) and increased (anti-squeezed) variance along two orthogonal axes, respectively. In this case, the accumulated phase $\varphi$ can be estimated with precision beyond the SQL.} 
\label{fig:BlochSphere_Phase}
\end{figure}

We consider each two-level atom in the clock as a spin-$\frac{1}{2}$ system, and denote by $S=N/2$ the total spin of the ensemble.
Spin squeezed states (SSSs) represent a relatively simple class of collective entangled states.
As shown in Fig.~\ref{fig:BlochSphere_Phase}b, in a SSS the noise is redistributed between two orthogonal quadratures.
The squeezed quadrature, with a variance (normalized to the SQL) of $\xi_-^2=\frac{2}{S}(\Delta S_{min})^2$, is oriented along the phase axis, providing a metrological gain over the SQL by virtue of a smaller quantum noise along the phase direction.
The gain $\mathcal{G}$ of a squeezed state is quantified by the Wineland parameter~\cite{wineland1992a} $ \xi^{-2}_R = \frac{C^2}{\xi_-^2} =\mathcal{G}$, where $C$ is the contrast of the interferometer.
As we will discuss in detail in section \ref{sec:Usefulness}, the maximum metrological gain achievable with squeezed states depends on secondary detrimental effects resulting from the quadrature with increased variance, called antisqueezing, $\xi_+^{2}=\frac{2}{S}(\Delta S_{max})^2$.
%

Establishing any kind of quantum correlation within the atomic ensemble requires an interaction between the atoms, since they need to know about each other’s state. Such an interaction can be direct, driven by state-dependent collisions or spin-spin interactions~\cite{monroe1995demonstration,Riedel2010,Strobel2014,Bohnet2016}, or effective~\cite{Appel2009,Takano2009, Leroux2010,Hosten2016,Cox2016a,Sewell2012}, mediated by an external field, such as a single cavity light mode interacting with all the atoms in the ensemble.
Optical techniques are beneficial for metrological applications since they provide an effective atom-atom interaction that can be turned off after preparation of the spin entangled state\cite{hacker2019deterministic, McConnell-Vuletic2015,Cox2016a,Hosten2016,Leroux2010,braverman2019near,Takano2009}, thus avoiding undesirable collision-induced energy shifts during the operation of the interferometer.

\begin{figure}[ht!]
\setlength{\unitlength}{1\textwidth}
\includegraphics[width=85mm,scale=1]{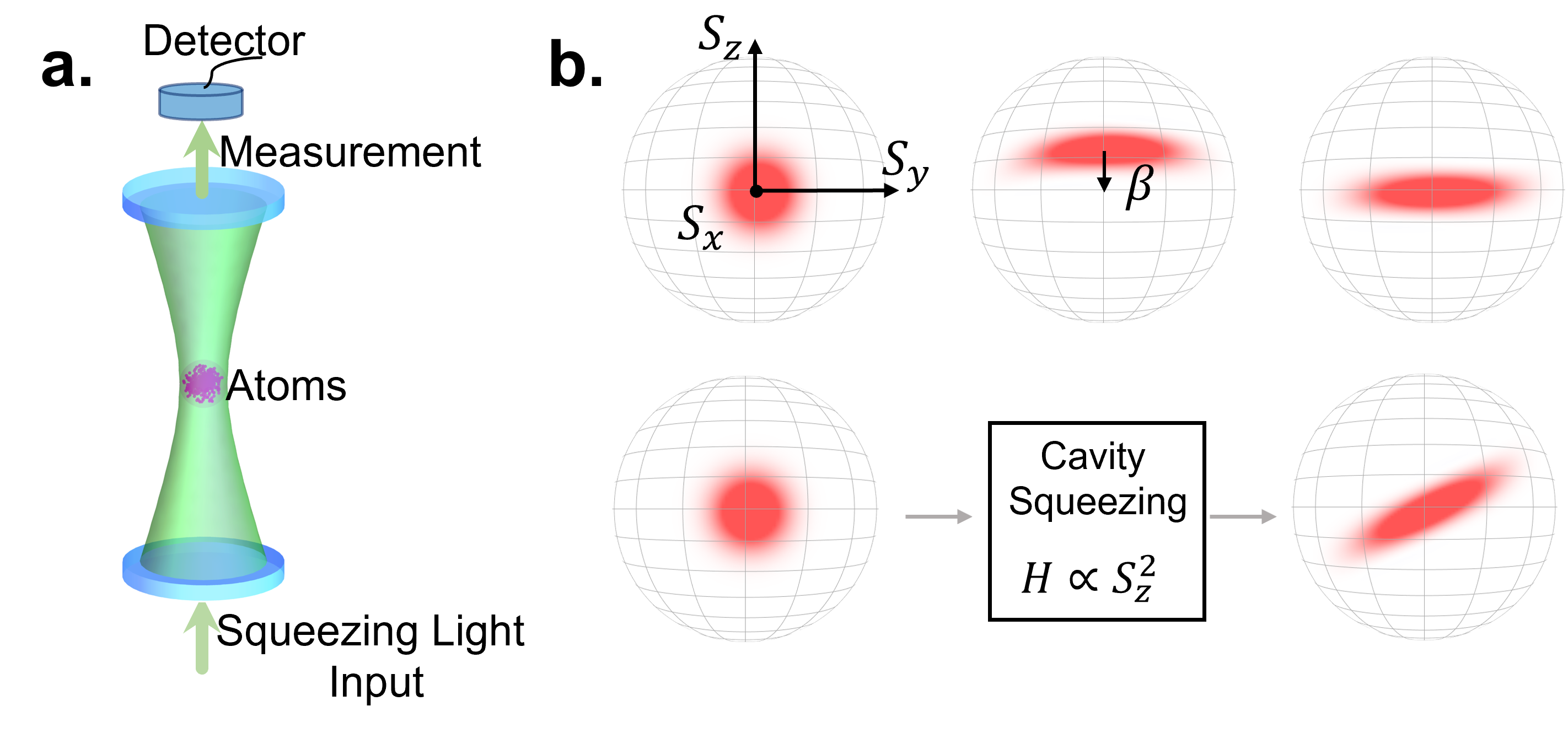}
\caption{\textbf{Atom-cavity system for generation of spin squeezed states.} 
\textbf{a.} An ensemble of $N$ spin-1/2 atoms is coupled to a high-finesse optical cavity to generate the non-linear interaction required to produce spin squeezing. 
\textbf{b.} Top row: SSS generated by measurement-based squeezing. The squeezing is conditioned on the detection of photons that carry information about the collective spin state. After the optical measurement of the spin state via the cavity, a small pulse $\beta$ can be applied to place the quantum state at the desired position on the Bloch sphere. 
Bottom row: deterministic cavity feedback squeezing. A CSS is subject to a one-axis twisting-like Hamiltonian ($H \propto S_z^{2}$) generated by the light-cavity interaction (see main text).}  
\label{fig:cavitySq-MasurementS}
\end{figure}

Incoherent scattering and photon loss set a bound to the maximal achievable entanglement with optical methods.
Hence, the use of high-finesse cavities is preferable since they enhance the interaction relative to the scattering, and have intrinsically smaller optical losses (Fig.~\ref{fig:cavitySq-MasurementS}a).
Moreover, schemes that reduce dissipation~\cite{Saffman2009, trail2010strongly, Leroux2012, Zhang2015, li2021collective} can increase the metrological gain.

\subsection{\label{ssec:measurementBasedSqueezing}Measurement-based squeezing}

A highly robust and conceptually straightforward technique that is able to produce record levels of spin squeezing is based on quantum nondemolition (QND) measurement~\cite{kuzmich1998atomic, Appel2009,Schleier-Smith2010, Hosten2016, Cox2016a, malia2022distributed}, where the uncertainty in one quadrature of the state is reduced through the measurement of photons that have interacted with the atoms (top row of Fig.~\ref{fig:cavitySq-MasurementS}b). However, this method is necessarily imperfect, due to finite total quantum efficiency of light collection and often incomplete use of the collected information. As a result, squeezing via QND measurement in practice has operated far from unitarity, with antisqueezing significantly in excess of the minimum set by the Heisenberg uncertainty principle for a given level of squeezing.


%
In order to avoid incoherent spontaneous emission, so far all cavity QED experiments aimed at the generation of collective entangled atomic states have operated far below the saturation of the atoms, i.e., in the regime of linear atomic response. In this limit, effective photon-photon interactions are weak\cite{li2021collective}.
We can then consider a coherent state of light that interacts with the atom-cavity system and which remains a coherent state after that interaction.
The light state $\left|\alpha(S_z)\right\rangle$ exiting the cavity, where $\alpha$ is the amplitude of the coherent state, depends on the $S_z$ component of the collective atomic spin.
This implies that the light field carries information about the state of the atomic ensemble.
As shown by Li~\textit{et al.}~\cite{li2021collective} this information, or light-atom entanglement, is characterized by the quantum Fisher information $I$ of the light state
\begin{equation}
    \tilde{I} = 4\left|\frac{\mathrm{d} \alpha(S_z)}{\mathrm{d}S_z}\right|^2.
\end{equation}
In particular, the variance $\left(\Delta S_z\right)^2$ associated with estimating the parameter $S_z$, using knowledge of the state $\left|\alpha(S_z)\right\rangle$, is given by the relative Cramér-Rao bound $\left(\Delta S_z\right)^2=1/\tilde{I}$.
%

An external observer can gain knowledge about the $S_z$ observable so that the spin variance is reduced below the SQL
\begin{equation}
    \left(\Delta S_z\right)^2_\mathrm{light}=\frac{2}{S(1+I)},
\end{equation}
where $I=2\tilde{I}/S$ is the normalized Quantum Fisher information of the light field acquired by the observer  ($I=1$ resolves the SQL uncertainty).
Hence, for any $I>1$ a conditionally squeezed state of the collective atomic spin is generated.




\subsection{\label{sec:level2_cavityFeedback}Cavity feedback squeezing}
As discussed above, one can generate strong conditional spin squeezing simply by quantum non-demolition measurement \cite{kuzmich1998atomic, Appel2009,Schleier-Smith2010, Hosten2016, Cox2016a}. The atom-cavity interaction entangles the collective atomic spin with the light, and a high photon detection efficiency is instrumental in obtaining large amounts of spin squeezing. To eliminate the requirement of high detection efficiency, a Hamiltonian method called cavity-feedback squeezing was proposed \cite{Schleier-Smith2010} and experimentally demonstrated \cite{Leroux2010} in 2010.

The method can be understood in terms of a two-fold interaction process\cite{Takano2009}, where quantum spin fluctuations of the atomic ensemble are imprinted onto the light field that then acts back onto the atoms\cite{Schleier-Smith2010}.
By placing the atoms in an optical resonator the atom-light interaction can be significantly enhanced. In general, such strong interaction between an ensemble of $N$ spins and the optical field of an optical resonator can be described through the Hamiltonian \cite{li2021collective}:
%
\begin{align}
    \hat{H}_{dip} = - \hbar \Omega \hat{n}_c \hat{S}_z,
\end{align}
where $ \hat{n}_c$ is the intracavity photon number, and $\hbar \Omega$ is the light shift produced by a single photon inside the cavity.
In a cavity setup, the intracavity photon number $\hat{n}_c=\hat{c}^{\dagger}\hat{c}$ is $\hat{S}_z$-dependent, producing to lowest order the paradigmatic one-axis twisting (OAT) Hamiltonian, initially proposed theoretically by Kitawaga and Ueda in 1993 \cite{Kitagawa1993}:
\begin{align}
    \hat{H}_{OAT} = \hbar \chi \hat{S}_z^2.
    \label{eq:OAT}
\end{align}
This Hamiltonian produces a spin-dependent precession about the $z$-axis that is proportional to $S_z$, causing an elliptical distortion of the initial symmetric noise distribution (bottom row of Fig.~\ref{fig:cavitySq-MasurementS}b). $\chi$ is known as the shearing parameter that quantifies how quickly the initial coherent spin state distribution is ``sheared''. 






\subsection{\label{ssec:directEntanglement} Entanglement via direct atom-atom interaction}

Entanglement can also be generated by direct atom-atom interactions, e.g., collisions in a Bose-Einstein condensate~\cite{Strobel2014, Lucke2016, Hamley2012, fadel2018spatial}, the Coulomb interaction between trapped ions~\cite{Bohnet2016, gilmore2021quantum}, or can be turned on by promoting atoms to their Rydberg level.

Gil \textit{et al.}~\cite{Gil2014SSRydberg} proposed and theoretically investigated an approach to generate an OAT Hamiltonian \eqref{eq:OAT} $\hat{H}\propto \hat{S}^2_z$ directly on the optical clock transition, making use of the strong interaction between atoms in their Rydberg states.
Due to the nature of the van der Waals interaction between Rydberg atoms decaying with the sixth power of the distance, the interaction Hamiltonian corresponds to an effective nearest-neighbour interaction.
%

%

Rydberg dressing methods, where atoms in their ground state are coupled to a Rydberg excited state, are appealing for optical tweezers-array clocks, a new platform where clock atoms are individually trapped in optical tweezers, and arranged in one- or two-dimensional arrays~\cite{madjarov2019atomic, Norcia2019, young2020half}.
In contrast to optical lattice clocks, such platforms offer a high tunability and control of interatomic distance.
This allows to both control the nearest-neighbor interaction strength, and to realize a nearly homogeneous interaction within the whole ensemble.
Recently, Schine \textit{et al.}~\cite{schine2022long} have pairwise entangled Strontium atoms on the optical clock transition via Rydberg dressing.
Generally, this technique has already been demonstrated to be efficient and reliable for entangling atoms~\cite{levine2018high, omran2019CatState} in the radiofrequency domain.

He \textit{et al.}~\cite{he2019engineeringSS} noted that the weak residual spin-orbit interaction in a spin-polarized Fermi gas can generate an effective OAT Hamiltonian that can be used to generate collective entangled states.
In particular, they consider a 3D Fermi-degenerate optical lattice clock~\cite{Campbell2017}, where spin polarized atoms are trapped in a tunable 3D optical lattice with less than one atom per lattice site, and cooled down to the motional ground state (Fermi degeneracy).
The spin-orbit coupling strength and, consequently, the effective nonlinear collective interaction are tuned by varying the trap depth: in a very deep trap the interaction vanishes.
It has been theoretically demonstrated that is possible to generate deterministic entanglement via unitary dynamics in ensembles composed of $N\approx 10^2 - 10^4$ atoms~\cite{he2019engineeringSS}.
The potential gain offered by these states can be as high as $\mathcal{G}\approx 14$~dB after an evolution time $\tau \approx 1$~s.
Although this timescale may look impractically long for many clock applications, this method is particularly interesting because it makes use of usually unwanted interactions to improve the clock performance.

\subsection{\label{sec:level3_CavFeed_Results_RF}
Optically generated Spin Squeezing in Radiofrequency Sensors}

Measurement-based spin squeezing, as well as cavity feedback squeezing, have enabled the generation of entangled states with substantial amount of metrological gain in alkaline atoms \cite{Appel2009, Cox2016a, Hosten2016, Leroux2010, Schleier-Smith2010a}. Using both schemes, spin squeezing has been implemented in proof-of-principles experiments to demonstrate atomic interferometers and atomic clocks operating beyond the SQL in the radiofrequency (RF) and microwave domain \cite{leroux2010orientation, Louchet-Chauvet2010, Hosten2016}. 


Recently, cavity feedback squeezing has been demonstrated in alkaline-earth atoms \cite{Takano2009, braverman2019near}, which are of high interest for optical atomic clocks. Braverman \textit{et al.}~\cite{braverman2019near} have demonstrated a substantial amount of spin squeezing in ytterbium-171 atoms ($\mathrm{^{171}Yb}$), generated via cavity feedback squeezing operating in the near-unitary regime, and offering a metrological gain of $\mathcal{G}=6.5$~dB beyond the SQL~\cite{braverman2019near}, limited by the measurement resolution of the system. 
The squeezing was demonstrated in the ground state of \Yb atoms, which has purely nuclear spin-$1/2$, representing an almost ideal two-level system to coherently manipulate the atomic state. 
Furthermore, using a SSS as an input state in a Ramsey protocol, a reduction in the integration time by a factor of 3.7 over the SQL was achieved~\cite{braverman2019near}.

\begin{figure}[hbtp]
\setlength{\unitlength}{1\textwidth}
\includegraphics[width=85mm,scale=1]{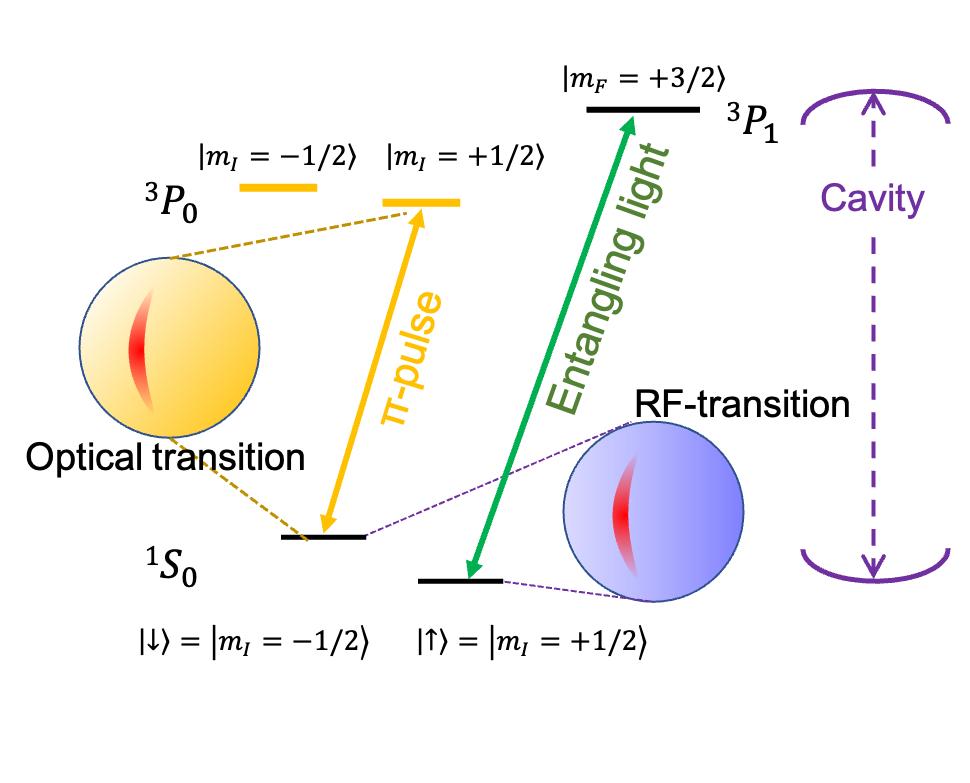}
\caption{\textbf{Scheme for generating entanglement on the optical clock transition.}  The entanglement is first generated via cavity QED techniques within two atomic levels in the RF-domain, here the nuclear sublevels of the ${}^{171}$Yb electronic ground state. Collective quantum states manipulations are easier and have very high fidelity in this domain. The entangled state is then mapped with a coherent optical $\pi$-pulse onto the optical clock transition.
}
\label{fig:strategy_clockEntanglement}
\end{figure}



\section{\label{sec:level3_CavFeed_Results_Optical}Mapping of squeezed states to the Optical Transition}

\begin{figure}[t!]
\setlength{\unitlength}{1\textwidth}
\includegraphics[width=89mm,scale=1]{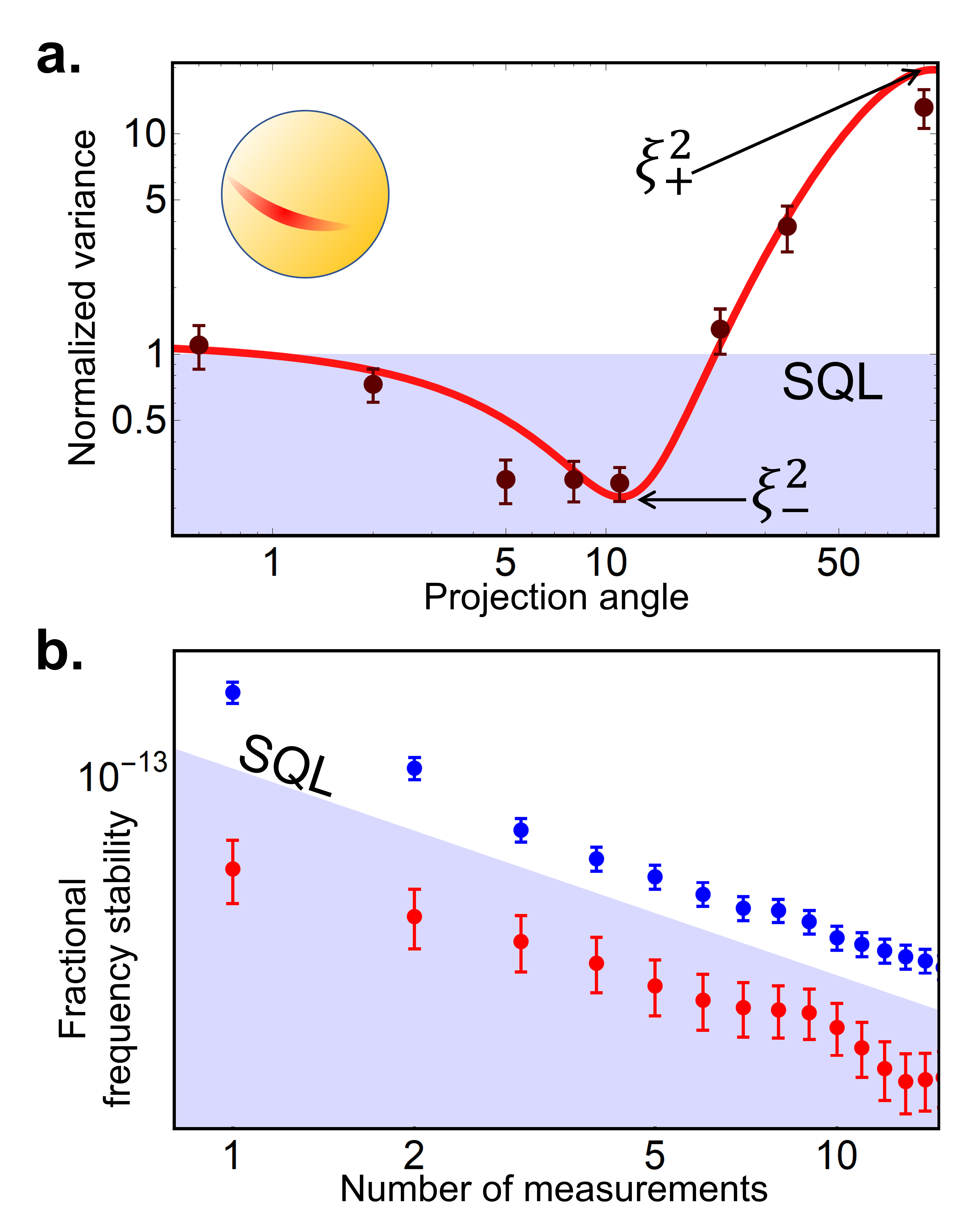}
\caption{\textbf{Entanglement on ytterbium-171 optical clock transition from Pedrozo-Pe\~{n}afiel \textit{et al.}~\cite{Pedrozo2020entanglement}} {\bf a.}Tomography of the generated entangled state, in this case a squeezed state, on the optical clock transition of ytterbium-171. The squeezed $\xi_{-}^2$ and the antisqueezed $\xi_{+}^2$ axes are indicated in the tomography data. {\bf b.} Allan deviation plot (fractional frequency instability) for a Ramsey sequence of the ${}^{171}$Yb clock. Blue symbols represents the data obtained without entanglement, while red symbols correspond to a squeezed input state. The blue area is only accessible with entangled states. 
}
\label{fig:opticalclockResults}
\end{figure}

In 2020, the first demonstration of spin squeezing on an optical clock transition was achieved~\cite{Pedrozo2020entanglement}. %
The strategy used in that work was to generate a SSS in the ground state of \Yb atoms, and transfer it onto the ultra-narrow optical clock transition by applying an optical $\pi$-pulse (see Fig.~\ref{fig:strategy_clockEntanglement}). Then using the SSS on the optical transition as the input state, a Ramsey protocol was demonstrated in the optical domain, where the quality factor of the transition ($Q=\Delta f/f$) was improved by five orders of magnitude compared to microwaves transitions \cite{Ludlow2015}. After the Ramsey protocol was implemented in the optical domain, the squeezed state was mapped back onto the ground state, and read out via the cavity~\cite{braverman2019near, Chen2014, Zhang2012, Hosten2016, Pedrozo2020entanglement, Hobson2019}. Comparing the clock operations using a SSS and a coherent spin state state as input states, a precision of $\mathcal{G}=4.4$~dB below the SQL was demonstrated, corresponding to a 2.8-fold reduction of the averaging time~\cite{Pedrozo2020entanglement} (Fig.~\ref{fig:opticalclockResults} ).

In this particular experiment, the LO stability was the limiting factor and was independently characterized and subtracted from the Allan deviation to quantify  the amount of improvement with respect to the noise imposed by the atomic spins alone. In this respect, improvement of the LO coherence is required to achieve performances at the level of state-of-the-art optical lattice clocks. In particular, optical lattice clocks use highly stable lasers that offer coherence times on the order of $1$ second, corresponding to mHz linewidths~\cite{kessler2012sub}.


\section{\label{sec:EchoProtocols}Beyond Spin Squeezing Via Time-Reversal Protocols}
SSSs are very simple entangled states, and can provide only a limited improvement over the SQL.
To approach the fundamental HL where the metrological gain is $\mathcal{G}=1/N$, it is necessary to generate more complex entangled states~\cite{Pezze2018}. %
For example, a Greenberger-Horne-Zeilinger (GHZ) state, also called "cat-state", is an extreme example of a collective entangled state, and can attain the HL.
Other highly-entangled states, with non-Gaussian envelopes, can carry very large quantum Fisher information $\mathcal{F}_Q$, and hence permit metrology close to the HL~\cite{Davis2016, Kessler2014, PezzeHybridClock2020, Borregaard2013, Pezze2018, schulte2020ramsey}.
Moreover, the metrological gain of some highly entangled states follows the Heisenberg Scaling (HS)~\cite{Saffman2009, Davis2016, Frowis2016, schulte2020ramsey}, i.e. $\mathcal{G}=b/N$ with $b\geq1$; this class of states is a constant factor away from the away from the HL, independent of atom number.

Generating and utilizing highly-entangled states is a difficult task due to their fragility, and remains an important challenge in quantum metrology. 
In practice, the metrological gain is limited by the detector resolution, the curvature of the Bloch sphere, and the impossibility to access all the statistical information carried by entangled states.
In particular, harnessing the resources provided by such states requires advanced statistical methods well beyond the usual evaluation of averages.

To relax those constraints, interaction-based readout methods have been proposed~\cite{Davis2016,Frowis2016, Leibfried2004, Toscano2006, Nolan2017, Macri2016}.
A particularly interesting class of protocols, that allow one to approach the HL in realistic system, is based on Loschmidt-echo-like time-reversal of the many-body Hamiltonian~\cite{Davis2016, Frowis2016, Leibfried2004, Toscano2006, Nolan2017, Macri2016, Volkoff_twist-untwist_2022, colombo2021time, liu2022nonlinear, li2022generalized}.
%
%
%
%
Such protocols are based on the application of a many-body Hamiltonian with both positive and negative sign, corresponding to an effective evolution forward and backward in time.

\begin{figure}[hbtp]
\setlength{\unitlength}{1\textwidth}
\includegraphics[width=89mm,scale=1]{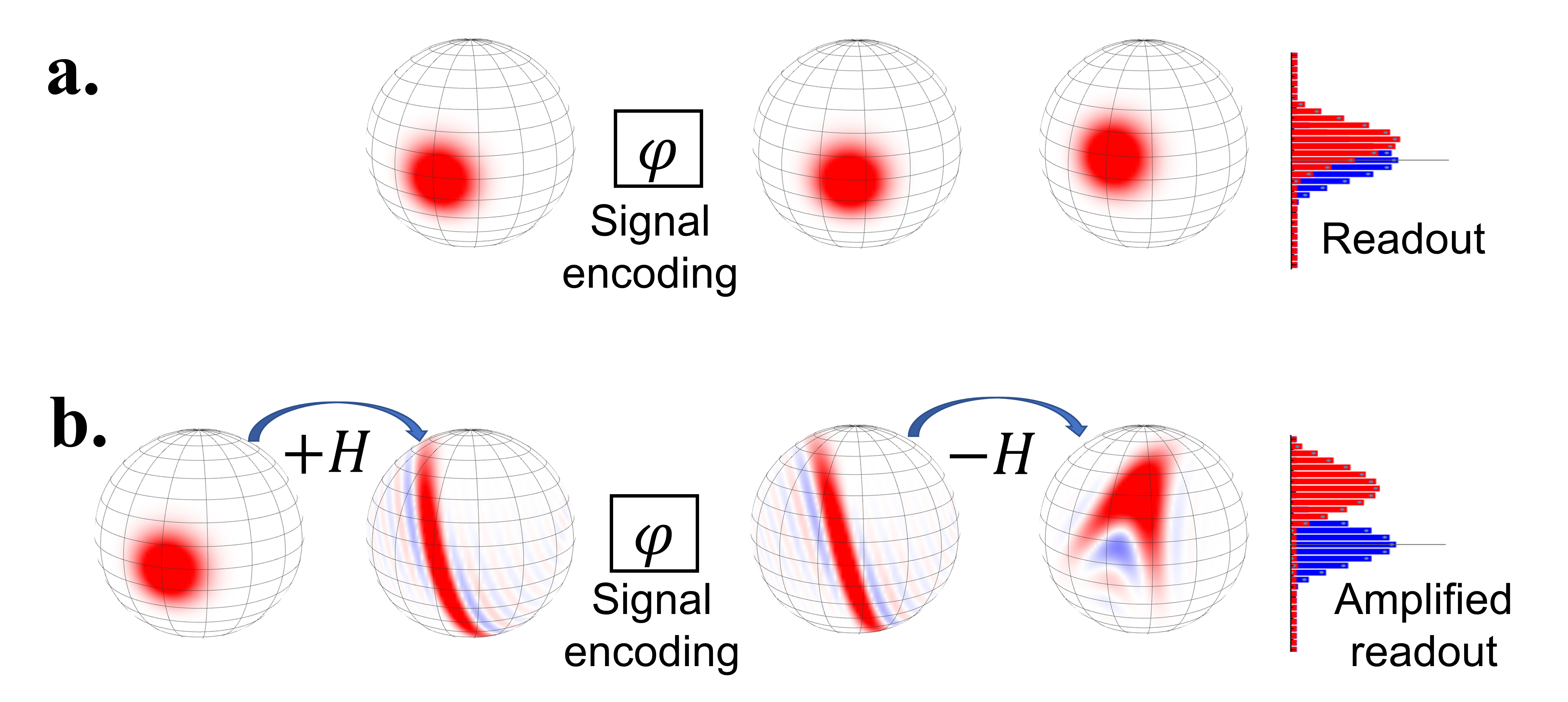}
\caption{\textbf{Signal enhancement by SATIN protocol.} The output of the measurement sequence is represented by the final state projection along the $z$ axis. Blue histograms are the results in the absence of a signal ($\varphi=0$).  {\bf a.} Ramsey protocol with a coherent input state: the imprinted signal $\varphi$ is directly reflected in the readout. {\bf b.} SATIN protocol: the signal $\varphi$ is enhanced in the readout through the time-reversed application of the many-body Hamiltonian. The Fisher information contained in the small features of the highly entangled state is mapped onto a large signal after the time-reversal ($-H$) operation.
}\label{fig:SATIN}
\end{figure}

A Loschmidt echo-like protocol, called Signal Amplification through Time-Reversed Interaction (SATIN)~\cite{colombo2021time}, is described in Fig.~\ref{fig:SATIN}. Here, an initial coherent spin state pointing along $\hat{x}$ evolves under the action of the OAT Hamiltonian, Eq.~\eqref{eq:OAT}, for a prolonged time to generate an over-squeezed state with large quantum Fisher information. 
If the process is sufficiently unitary, time-reversing the Hamiltonian would produce the initial coherent spin state. However, if this entangled state is perturbed, or subjected to a rotation that makes it orthogonal to the unperturbed entangled state, then the subsequent application of the negative Hamiltonian will produce a near-coherent spin state which is now also orthogonal to the initial state. As a major advantage over direct spin squeezing, detecting the distance between these two states requires detection resolution on the order of the SQL only. 

This is the feature that makes SATIN a powerful method for using highly entangled states in atom interferometry, and that enables one to reach near-Heisenberg sensitivity with Heisenberg scaling even for a many-particle system, provided the evolution of the system can be kept nearly unitary, as recently demonstrated by Colombo \textit{et al.} \cite{colombo2021time} .
Specifically, highly non-Gaussian states were generated and used for atom interferometry.
The SATIN protocol allowed to utilize most of the quantum Fisher information carried by these highly entangled states, reaching an improvement of $11.8$~dB below the SQL (a factor of 15 in reduction of averaging time) in phase sensitivity. In addition, that protocol demonstrated HS in metrological gain $\mathcal{G}$, with a linear improvement in $\mathcal{G}$ with respect to the atom number, at a fixed distance to the HL of 12.6 dB. 
In the future, this protocol can be used to perform metrology in the optical domain, by mapping the entanglement onto the optical transition, as was similarly done for a SSS~\cite{Pedrozo2020entanglement}. 


Other time-reversal-like protocols have also been demonstrated in Bose-Einstein condensates through phases shifts in a three-level system for a few neutral atoms~\cite{Linnemann2016}, and in cold trapped ion systems consisting of up to ${\sim}150$ ions, where the coupling with a motional mode plus spin rotations are used for such purpose~\cite{gilmore2021quantum}. 
Other approaches that mimic time-reversal-type protocols have been demonstrated, alternating spin squeezing with state rotations~\cite{Holstein1940}. 
In particular, Hosten \textit{et al.}\cite{Hosten2016a} demonstrated a gain of $\mathcal{G}=8$~dB beyond the SQL without sub-SQL measurement resolution.

The experimental demonstration of the SATIN protocol opens the door for quantum metrology with highly-entangled many-body systems, paving the way to achieve nearly Heisenberg-limited operation of quantum sensors, significantly enhancing the bandwidth at fixed precision, or the precision at fixed bandwidth.
%

\section{\label{sec:Usefulness}Usefulness of Entanglement}
In this article, we discuss mostly clocks operating with Ramsey spectroscopy~\cite{ramsey1950molecular}.
In this case, the spectroscopy time is also called Ramsey time.
For the Ramsey protocol we have $w(t)=1$, and equation (\ref{eq:phi_as_freq}) becomes  ${{\Delta\varphi(\tau)} = 2\pi\,\tau\,\langle {\Delta f}\rangle}$, with $\langle\Delta f\rangle$ being the average frequency difference in the Ramsey time interval $\tau$.
The resulting fractional stability of a clock operated with Ramsey spectroscopy is then   
\begin{equation}
    \sigma(\tau,T, T_c)=\frac{1}{2\pi f_a}\sqrt{\frac{1}{\tau\,D\,T}}\frac{1}{\sqrt{N \mathcal{G}}},
    \label{eq:clockStab}
\end{equation}
with $T$ denoting the total measurement time, and $T_c$ the individual cycle time.
When the clock is operated with a duty cycle $D$ smaller than $100\%$, one should add the Dick noise term to equation \eqref{eq:clockStab}. The latter arises from aliased noise of the LO~\cite{dick1987local}.
In the absence of dark time the Dick noise vanishes\cite{dick1987local,Ludlow2015}, while it is irrelevant in applications where two or more ensembles of atoms are simultaneously probed \cite{Schioppo2017Ultrastable,takamoto2011frequency,Nicholson2012}. 
Schulte \textit{et al.}~\cite{schulte2020prospects} have rigorously theoretically analyzed under which conditions entanglement can provide a precision gain in the presence of Dick noise.
They found that entanglement is useful for atom numbers below a certain threshold that depends on the experimental conditions, such as the LO noise, the dark time, and the Ramsey time $\tau$. 

\subsection{Laser as Local Oscillator}

\begin{figure}[ht!]
\setlength{\unitlength}{1\textwidth}
\includegraphics[width=70mm,scale=1]{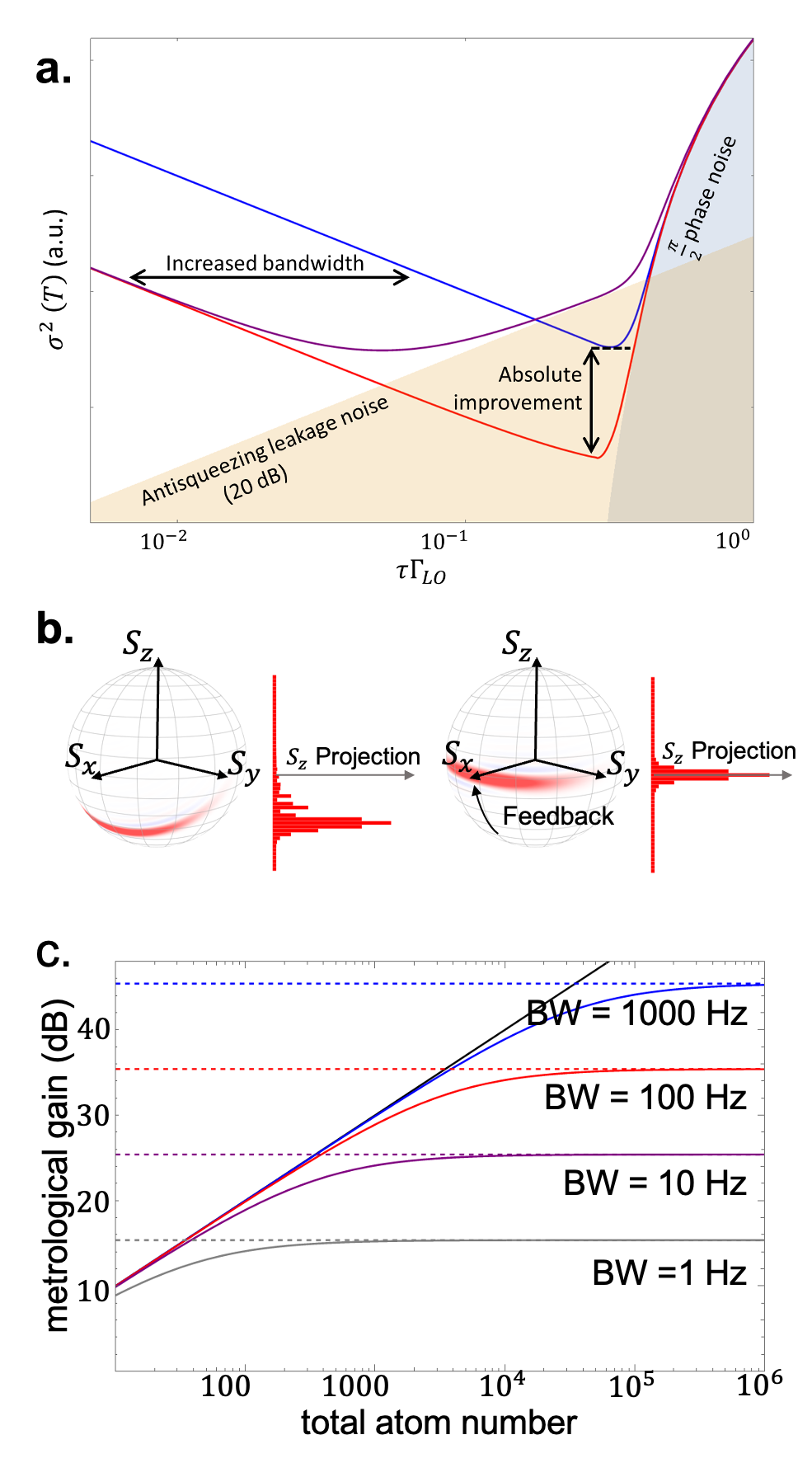} 
\caption{\textbf{Stability of optical clock in a given averaging time $T$ as a function of Ramsey time $\tau$.} 
\textbf{a.} Fractional stability as a function of $\tau$ in the presence of decoherence at rate $\Gamma_{LO}$. The blue, red, and purple lines indicate the operation with an unentangled coherent spin state, a unitary SSS with $11$~dB of metrological gain), and a non-unitary SSS with $11$~dB squeezing and $20$~dB of antisqueezing, respectively. For short Ramsey time, the SSS can always outperform the coherent spin state, corresponding to an effective increase of the sensor bandwidth by $\mathcal{G}$ at fixed precision.
\textbf{b.} A QND measurement can be is used to rotate the SSS closer to the equator, thus avoiding the leakage of the antisqueezed quadrature noise onto the measurement quadrature.
%
\textbf{c.} Maximal metrological gain achievable in a state-of-the-art system at fixed bandwidths as a function of the atom number. We considered here decoherence rates $\Gamma_{nat}=0.01~\mathrm{s}^{-1}$ (natural linewidth), $\Gamma_{deph}=0.025~\mathrm{s}^{-1}$ (dephasing rate), and $\Gamma_{loss}=0.01~\mathrm{s}^{-1}$ (atom loss rate) from the recent clock realization of Young \textit{et al.}~\cite{young2020half}.
} 
\label{fig:gain_Explained}
\end{figure}

In full clock operation, the frequency of the laser LO is locked to the optical atomic transition, and it can be used as a time standard by means of an optical frequency comb.
In a Ramsey sequence, the phase difference $\varphi$ between the atoms and the LO is mapped onto the final population difference between the ground state and the clock state, i.e., $2S_z$, by means of a $\pi/2$ pulse. 
Since $S_z\propto\sin(\varphi)$, the protocol works correctly only if the phase accumulation does not exceed $\pm \pi/2$.
Thus, as shown by Braverman \textit{et al.}~\cite{Braverman2018}, the LO noise limits the clock performance when the standard deviation of the total accumulated phase noise exceeds $\tau\Gamma_{LO}\approx 0.3$ (see Fig.~\ref{fig:gain_Explained}a), where $\Gamma_{LO}$ is the dephasing rate of the LO. 
At this level of noise, the probability of having a total Ramsey phase exceeding $\pm \pi/2$, and thus a wrong reading, is significant, and induces an overall increase of error in the feedback. 

LO noise does not induce spin-spin decoherence; however, in the case of a SSS input state, the approximation that the antisqueezing noise is orthogonal to the measurement fails.
Due to the limited number of atoms utilized in optical clocks ($N\lesssim 10^4$) and the associated non-zero curvature of the generalized Bloch sphere, the anti-squeezing couples into the measurement projection, introducing extra noise and also effectively shortening the average spin vector (i.e., reducing the contrast).
Larger noise and smaller contrast reduce the Wineland~\cite{wineland1992a} parameter and the metrological gain~\cite{Andre2004, leroux2017line, schulte2020prospects}.

To avoid the effect of this leakage, squeezed optical clocks need to operate in a regime with a small accumulated phase $\varphi$.
This implies a Ramsey time $\tau$ limited to a value smaller than that necessary to avoid phase errors exceeding $\pm \pi/2$ in an SQL clock operation.
On the other hand, as can be seen from equation~(\ref{eq:clockStab}), the clock stability $\sigma$ in a time $T$ improves with $\sqrt{\tau}$.
Therefore, to maximize clocks performance, one needs to choose a compromise between a short Ramsey time for greatest phase noise suppression and a long Ramsey time for increased clock frequency stability (see Fig.~\ref{fig:gain_Explained}a).

Schemes that allow to profit from the full metrological gain offered by squeezed states in the presence of local oscillator noise have been proposed~\cite{Borregaard2013,borregaard2013efficient}. 
Borregaard and S\o{}rensen~\cite{Borregaard2013} argue that a large accumulated phase can be strongly reduced by performing a series of QND measurements and feedback before the final strong projective measurement.
Compared to the case of the measurement based squeezing, here the QND measurement does not need to resolve the $S_z$ below the SQL.
It only needs to preserve the atomic coherence, i.e., contrast.
The reading obtained from this weak QND measurement will be used to rotate the squeezed state closer to the equator, where the anitsqueezing noise does not leak into the measurement quadrature (see Fig.~\ref{fig:gain_Explained}b). 
In a similar spirit, the large accumulated phase can be first estimated by using multiple ensembles~\cite{borregaard2013efficient}.
%
%
%
%
%
%


%
%
%


\subsection{Second atomic ensemble as Local Oscillator}\label{ssec:BScurvature}

When two or more ensembles are compared, i.e., for differential operation of the sensor, the LO decoherence is a common-mode noise source and is cancelled.
The coherence time is then ultimately limited by the spontaneous emission rate of the atomic excited state $\Gamma_{nat}$.
However, in state-of-the-art atomic clocks other processes, such as atomic collisions and environmental inhomogeneities, can induce faster atom-atom decoherence.

Escher \textit{et al.}~\cite{escher2011general}  and Demkowicz-Dobrza\'{n}ski \textit{et al.}~\cite{demkowicz2012elusive} have thoroughly investigated the limitations imposed by atomic decoherence on the metrological gain. 
%
They theoretically demonstrated that, for a given uncorrelated noise there is a maximal enhancement that can be achieved, and that such a limit depends on the Ramsey time of the sensor, i.e., on the bandwidth of the measurement.
%
%

In particular, they show that the precision in a total time $T$ as a function of the Ramsey time $\tau$ is limited not only by the LO noise (as in Fig:~\ref{fig:gain_Explained}a), but also by atom-atom dephasing ($\Gamma_{deph}$), and by $\Gamma_{nat}$. 
%
As illustrated in Fig.~\ref{fig:gain_Explained}a, for sufficiently short Ramsey times the full metrological gain is recovered.

\begin{figure}[hb!]
\setlength{\unitlength}{1\textwidth}
\includegraphics[width=58mm,scale=1]{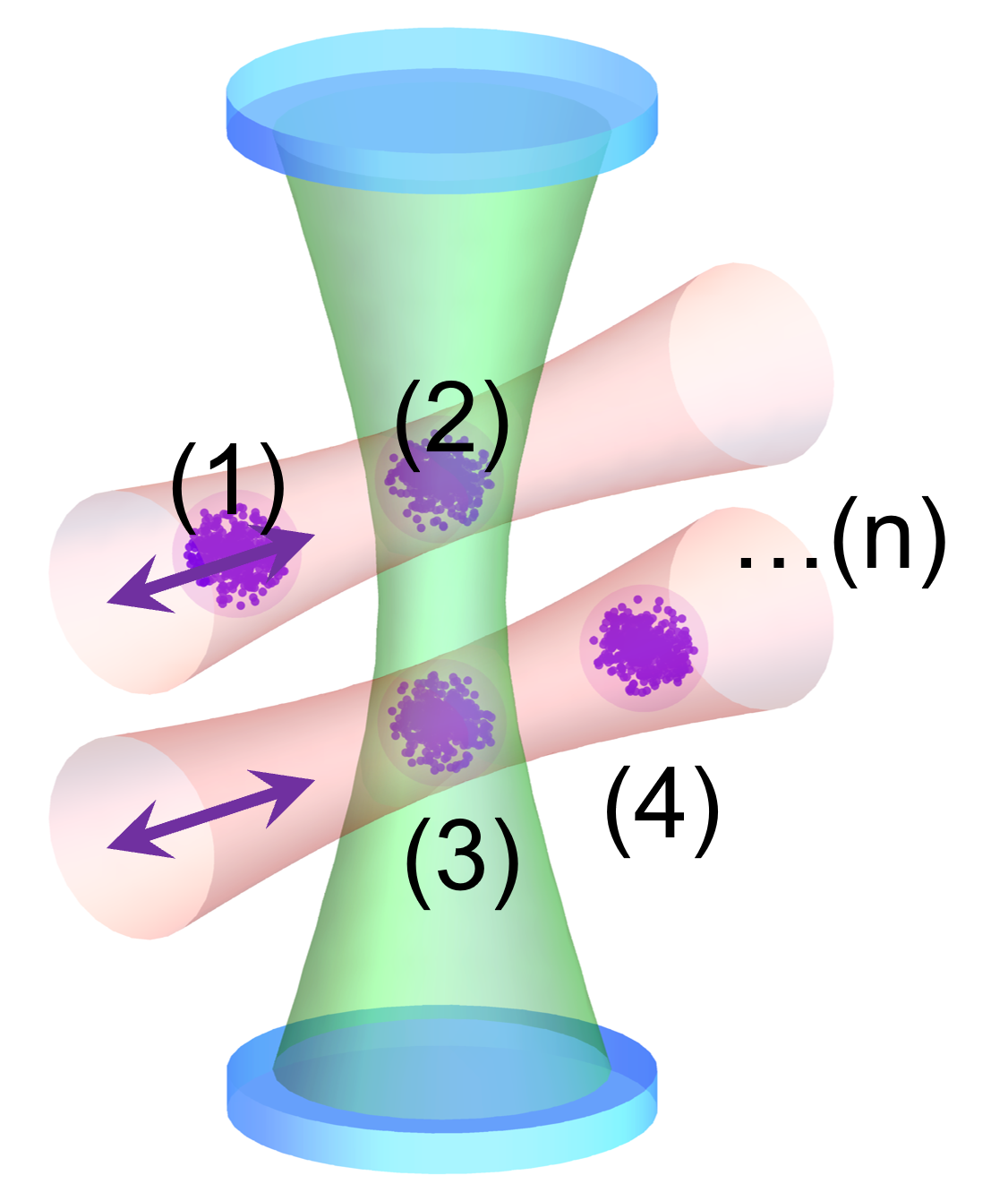}
\caption{\textbf{Network of clocks.} 
Several individually trapped ensembles are prepared in two optical-lattices, and can be shuttled in and out of a high-finesse optical cavity.
%
%
The light-mediated interaction through the cavity field can be used to engineer entanglement between the ensembles.
}
\label{fig:multiEnsemble}
\end{figure}

\subsubsection*{Quantum Network of Clocks}

The comparison of multiple clock ensembles is the core of the idea of a quantum network of clocks, proposed by K\'{o}m\'{a}r~\textit{et al.}\cite{komar2014quantum}. It consists of multiple, coherently interconnected clocks in which the individual devices (nodes) will benefit from the larger resources of the composed system. In particular, this approach may serve both as a way to distribute time at an international scale, as well as a new platform for testing fundamental physics. 
Yet, to achieve this goal, entanglement between the different nodes is necessary to improve the performance beyond the SQL. 
In the best-case scenario, a Greenberger-Horne-Zeilinger (GHZ) state consisting of the product of individual GHZ states of the nodes would allow the network to operate at or near its collective HL. 


%
A possible implementation is illustrated in Fig.~\ref{fig:multiEnsemble}. Two or more spatially separated ensembles can be entangled via their interaction with the same cavity mode.
%
%
Each of these ensembles is effectively an individual clock and, as recently demonstrated by Zheng \textit{et al.}~\cite{zheng2022differential}, they can be easily moved in and out of the cavity.
In this way the atomic ensembles can be coupled to the cavity field on demand, and entangled individually, as well as with each other.

Recently, Nichol \textit{et al.}~\cite{nichol2022quantum} have demonstrated the first quantum network of optical atomic clocks by comparing the frequencies of two entangled $^{88}\mathrm{Sr}^+$ ions separated by 2~m. 
Nichol \textit{et al.} generated heralded entanglement between the two ions (i.e., the two clocks) via a photonic link~\cite{monroe2014large}. 

%
%
%

\subsection{Fixed-bandwidth applications}\label{ssec:Fixed-bandwidth}

Many proposed applications of optical clocks, including gravitational-wave detection~\cite{kolkowitz2016gravitational,Sedda_2020} and the search for axions\cite{Pospelov2013GNOME,budker2014proposal,stadnik2015can, stadnik2020new}, require a large clock bandwidth and short Ramsey time, since the clock needs to track time-varying signals.
Thus, for operations where the Ramsey time $\tau$ is much less than the decoherence time, $\tau\Gamma_{tot}\ll1$, entanglement remains an important resource.
In this situation it is possible to profit from the full metrological gain offered by the entangled input state.
%

%
%

Figure~\ref{fig:gain_Explained}c. shows the maximum gain of an optical clock as a function of the total atom number deployed.
%
%
For bandwidths set in the range 1~Hz to 1~kHz, the transition from Heisenberg-limited performance to decoherence-limited performance occurs at atom numbers ranging from $\sim 50$ to $\sim 5\times 10^4$.
It worth noting that for typical atom numbers used in optical atomic clocks ($10^2$ to $10^4$), a bandwidth of $\approx 10$~Hz or larger implies that the maximum metrological gain is close to the Heisenberg Limit.

\section{\label{sec:Perspectives}Summary and perspective}

In this article, we have described recent advances and offered a perspective on strategies to improve atomic sensors via engineered quantum correlations (entanglement).
The generation and application of spin squeezed states to atom interferometers and atomic clocks have enabled sub-SQL performance of these devices. While in most cases these have been proof-of-principle experiments, some more recent implementations have shown practical applications for sensing beyond the SQL~\cite{gilmore2021quantum} in state-of-the-art sensors.

We have highlighted the generation of entanglement in optical clocks, presenting operating regimes where collective entanglement can readily bring a sensing advantage.
We have also discussed a novel technique, based on effective time-reversal through switching the sign of a many-body Hamiltonian, for generating and harnessing highly entangled states (non-Gaussian states) in atomic sensors~\cite{colombo2021time}. 
Such time-reversal protocols allow one to reach near-Heisenberg-limited quantum metrology in many-atoms sensors.
This sensing paradigm strongly relaxes the most stringent limitations present in standard entangled-enhanced sensing protocols, namely, the detection of states with a large Fisher information, and the curvature of the generalized Bloch sphere.

We expect that the time-reversal technique will become a major paradigm in quantum metrology in the years to come.
Thanks to its applicability to non-Gaussian entangled states~\cite{Davis2016,colombo2021time}, such a paradigm is ideal for the application to optical clocks and other state-of-the-art atomic sensors.
Further investigations to scale up the size of entangled quantum systems, as well as reduce and tame the different decoherence mechanisms, will enable near-Heisenberg limited resolution for large systems with many atoms\cite{escher2011general,demkowicz2012elusive}.
Entangled optical clocks and atomic sensors have the potential to become a leading platform in the search for new physics, ranging from probing different types of physics beyond the Standard Model~\cite{Derevianko2014DarkMatter,Arvanitaki2015DarkMatter,Wcislo2018,stadnik2014axion,Safronova2018RevModPhys} to testing the fundamentals of gravity~\cite{Sedda_2020,Delva2018_gravity,Herrmann2018_gravity}.

\begin{acknowledgments}

We would like to thank Emily Davis, Adam Kaufman, Shimon Kolkowitz, Zeyang Li, Mikhail Lukin, Monika Schleier-Smith, and Jun Ye for inspiring discussions. This work was supported by NSF, NSF CUA, ONR, and DARPA.

\end{acknowledgments}



\bibliography{aipsamp}

\begin{thebibliography}{128}%
\makeatletter
\providecommand \@ifxundefined [1]{%
 \@ifx{#1\undefined}
}%
\providecommand \@ifnum [1]{%
 \ifnum #1\expandafter \@firstoftwo
 \else \expandafter \@secondoftwo
 \fi
}%
\providecommand \@ifx [1]{%
 \ifx #1\expandafter \@firstoftwo
 \else \expandafter \@secondoftwo
 \fi
}%
\providecommand \natexlab [1]{#1}%
\providecommand \enquote  [1]{``#1''}%
\providecommand \bibnamefont  [1]{#1}%
\providecommand \bibfnamefont [1]{#1}%
\providecommand \citenamefont [1]{#1}%
\providecommand \href@noop [0]{\@secondoftwo}%
\providecommand \href [0]{\begingroup \@sanitize@url \@href}%
\providecommand \@href[1]{\@@startlink{#1}\@@href}%
\providecommand \@@href[1]{\endgroup#1\@@endlink}%
\providecommand \@sanitize@url [0]{\catcode `\\12\catcode `\$12\catcode
  `\&12\catcode `\#12\catcode `\^12\catcode `\_12\catcode `\%12\relax}%
\providecommand \@@startlink[1]{}%
\providecommand \@@endlink[0]{}%
\providecommand \url  [0]{\begingroup\@sanitize@url \@url }%
\providecommand \@url [1]{\endgroup\@href {#1}{\urlprefix }}%
\providecommand \urlprefix  [0]{URL }%
\providecommand \Eprint [0]{\href }%
\providecommand \doibase [0]{http://dx.doi.org/}%
\providecommand \selectlanguage [0]{\@gobble}%
\providecommand \bibinfo  [0]{\@secondoftwo}%
\providecommand \bibfield  [0]{\@secondoftwo}%
\providecommand \translation [1]{[#1]}%
\providecommand \BibitemOpen [0]{}%
\providecommand \bibitemStop [0]{}%
\providecommand \bibitemNoStop [0]{.\EOS\space}%
\providecommand \EOS [0]{\spacefactor3000\relax}%
\providecommand \BibitemShut  [1]{\csname bibitem#1\endcsname}%
\let\auto@bib@innerbib\@empty
\bibitem [{\citenamefont {Oelker}\ \emph {et~al.}(2019)\citenamefont {Oelker},
  \citenamefont {Hutson}, \citenamefont {Kennedy}, \citenamefont {Sonderhouse},
  \citenamefont {Bothwell}, \citenamefont {Goban}, \citenamefont {Kedar},
  \citenamefont {Sanner}, \citenamefont {Robinson}, \citenamefont {Marti} \emph
  {et~al.}}]{oelker2019demonstration}%
  \BibitemOpen
  \bibfield  {author} {\bibinfo {author} {\bibfnamefont {E.}~\bibnamefont
  {Oelker}}, \bibinfo {author} {\bibfnamefont {R.}~\bibnamefont {Hutson}},
  \bibinfo {author} {\bibfnamefont {C.}~\bibnamefont {Kennedy}}, \bibinfo
  {author} {\bibfnamefont {L.}~\bibnamefont {Sonderhouse}}, \bibinfo {author}
  {\bibfnamefont {T.}~\bibnamefont {Bothwell}}, \bibinfo {author}
  {\bibfnamefont {A.}~\bibnamefont {Goban}}, \bibinfo {author} {\bibfnamefont
  {D.}~\bibnamefont {Kedar}}, \bibinfo {author} {\bibfnamefont
  {C.}~\bibnamefont {Sanner}}, \bibinfo {author} {\bibfnamefont
  {J.}~\bibnamefont {Robinson}}, \bibinfo {author} {\bibfnamefont
  {G.}~\bibnamefont {Marti}},  \emph {et~al.},\ }\bibfield  {title} {\enquote
  {\bibinfo {title} {Demonstration of $4.8\times10^{-17}$ stability at 1 s for
  two independent optical clocks},}\ }\href@noop {} {\bibfield  {journal}
  {\bibinfo  {journal} {Nature Photonics}\ }\textbf {\bibinfo {volume} {13}},\
  \bibinfo {pages} {714--719} (\bibinfo {year} {2019})}\BibitemShut {NoStop}%
\bibitem [{\citenamefont {Schioppo}\ \emph {et~al.}(2017)\citenamefont
  {Schioppo}, \citenamefont {Brown}, \citenamefont {McGrew}, \citenamefont
  {Hinkley}, \citenamefont {Fasano}, \citenamefont {Beloy}, \citenamefont
  {Yoon}, \citenamefont {Milani}, \citenamefont {Nicolodi}, \citenamefont
  {Sherman}, \citenamefont {Phillips}, \citenamefont {Oates},\ and\
  \citenamefont {Ludlow}}]{Schioppo2017Ultrastable}%
  \BibitemOpen
  \bibfield  {author} {\bibinfo {author} {\bibfnamefont {M.}~\bibnamefont
  {Schioppo}}, \bibinfo {author} {\bibfnamefont {R.~C.}\ \bibnamefont {Brown}},
  \bibinfo {author} {\bibfnamefont {W.~F.}\ \bibnamefont {McGrew}}, \bibinfo
  {author} {\bibfnamefont {N.}~\bibnamefont {Hinkley}}, \bibinfo {author}
  {\bibfnamefont {R.~J.}\ \bibnamefont {Fasano}}, \bibinfo {author}
  {\bibfnamefont {K.}~\bibnamefont {Beloy}}, \bibinfo {author} {\bibfnamefont
  {T.~H.}\ \bibnamefont {Yoon}}, \bibinfo {author} {\bibfnamefont
  {G.}~\bibnamefont {Milani}}, \bibinfo {author} {\bibfnamefont
  {D.}~\bibnamefont {Nicolodi}}, \bibinfo {author} {\bibfnamefont {J.~A.}\
  \bibnamefont {Sherman}}, \bibinfo {author} {\bibfnamefont {N.~B.}\
  \bibnamefont {Phillips}}, \bibinfo {author} {\bibfnamefont {C.~W.}\
  \bibnamefont {Oates}}, \ and\ \bibinfo {author} {\bibfnamefont {A.~D.}\
  \bibnamefont {Ludlow}},\ }\bibfield  {title} {\enquote {\bibinfo {title}
  {Ultrastable optical clock with two cold-atom ensembles},}\ }\href
  {http://dx.doi.org/10.1038/nphoton.2016.231} {\bibfield  {journal} {\bibinfo
  {journal} {Nat. Photonics}\ }\textbf {\bibinfo {volume} {11}},\ \bibinfo
  {pages} {48--52} (\bibinfo {year} {2017})}\BibitemShut {NoStop}%
\bibitem [{\citenamefont {Brewer}\ \emph {et~al.}(2019)\citenamefont {Brewer},
  \citenamefont {Chen}, \citenamefont {Hankin}, \citenamefont {Clements},
  \citenamefont {Chou}, \citenamefont {Wineland}, \citenamefont {Hume},\ and\
  \citenamefont {Leibrandt}}]{Brewer2019Al}%
  \BibitemOpen
  \bibfield  {author} {\bibinfo {author} {\bibfnamefont {S.~M.}\ \bibnamefont
  {Brewer}}, \bibinfo {author} {\bibfnamefont {J.-S.}\ \bibnamefont {Chen}},
  \bibinfo {author} {\bibfnamefont {A.~M.}\ \bibnamefont {Hankin}}, \bibinfo
  {author} {\bibfnamefont {E.~R.}\ \bibnamefont {Clements}}, \bibinfo {author}
  {\bibfnamefont {C.~W.}\ \bibnamefont {Chou}}, \bibinfo {author}
  {\bibfnamefont {D.~J.}\ \bibnamefont {Wineland}}, \bibinfo {author}
  {\bibfnamefont {D.~B.}\ \bibnamefont {Hume}}, \ and\ \bibinfo {author}
  {\bibfnamefont {D.~R.}\ \bibnamefont {Leibrandt}},\ }\bibfield  {title}
  {\enquote {\bibinfo {title} {$^{27}{\mathrm{al}}^{+}$ quantum-logic clock
  with a systematic uncertainty below ${10}^{\ensuremath{-}18}$},}\ }\href
  {\doibase 10.1103/PhysRevLett.123.033201} {\bibfield  {journal} {\bibinfo
  {journal} {Phys. Rev. Lett.}\ }\textbf {\bibinfo {volume} {123}},\ \bibinfo
  {pages} {033201} (\bibinfo {year} {2019})}\BibitemShut {NoStop}%
\bibitem [{\citenamefont {Takamoto}\ \emph {et~al.}(2020)\citenamefont
  {Takamoto}, \citenamefont {Ushijima}, \citenamefont {Ohmae}, \citenamefont
  {Yahagi}, \citenamefont {Kokado}, \citenamefont {Shinkai},\ and\
  \citenamefont {Katori}}]{katori2020}%
  \BibitemOpen
  \bibfield  {author} {\bibinfo {author} {\bibfnamefont {M.}~\bibnamefont
  {Takamoto}}, \bibinfo {author} {\bibfnamefont {I.}~\bibnamefont {Ushijima}},
  \bibinfo {author} {\bibfnamefont {N.}~\bibnamefont {Ohmae}}, \bibinfo
  {author} {\bibfnamefont {T.}~\bibnamefont {Yahagi}}, \bibinfo {author}
  {\bibfnamefont {K.}~\bibnamefont {Kokado}}, \bibinfo {author} {\bibfnamefont
  {H.}~\bibnamefont {Shinkai}}, \ and\ \bibinfo {author} {\bibfnamefont
  {H.}~\bibnamefont {Katori}},\ }\bibfield  {title} {\enquote {\bibinfo {title}
  {Test of general relativity by a pair of transportable optical lattice
  clocks},}\ }\href {https://doi.org/10.1038/s41566-020-0619-8} {\bibfield
  {journal} {\bibinfo  {journal} {Nature Photonics}\ }\textbf {\bibinfo
  {volume} {14}},\ \bibinfo {pages} {411--415} (\bibinfo {year}
  {2020})}\BibitemShut {NoStop}%
\bibitem [{\citenamefont {Bothwell}\ \emph {et~al.}(2022)\citenamefont
  {Bothwell}, \citenamefont {Kennedy}, \citenamefont {Aeppli}, \citenamefont
  {Kedar}, \citenamefont {Robinson}, \citenamefont {Oelker}, \citenamefont
  {Staron},\ and\ \citenamefont {Ye}}]{bothwell2022resolving}%
  \BibitemOpen
  \bibfield  {author} {\bibinfo {author} {\bibfnamefont {T.}~\bibnamefont
  {Bothwell}}, \bibinfo {author} {\bibfnamefont {C.~J.}\ \bibnamefont
  {Kennedy}}, \bibinfo {author} {\bibfnamefont {A.}~\bibnamefont {Aeppli}},
  \bibinfo {author} {\bibfnamefont {D.}~\bibnamefont {Kedar}}, \bibinfo
  {author} {\bibfnamefont {J.~M.}\ \bibnamefont {Robinson}}, \bibinfo {author}
  {\bibfnamefont {E.}~\bibnamefont {Oelker}}, \bibinfo {author} {\bibfnamefont
  {A.}~\bibnamefont {Staron}}, \ and\ \bibinfo {author} {\bibfnamefont
  {J.}~\bibnamefont {Ye}},\ }\bibfield  {title} {\enquote {\bibinfo {title}
  {Resolving the gravitational redshift across a millimetre-scale atomic
  sample},}\ }\href@noop {} {\bibfield  {journal} {\bibinfo  {journal}
  {Nature}\ }\textbf {\bibinfo {volume} {602}},\ \bibinfo {pages} {420--424}
  (\bibinfo {year} {2022})}\BibitemShut {NoStop}%
\bibitem [{\citenamefont {Pospelov}\ \emph {et~al.}(2013)\citenamefont
  {Pospelov}, \citenamefont {Pustelny}, \citenamefont {Ledbetter},
  \citenamefont {Kimball}, \citenamefont {Gawlik},\ and\ \citenamefont
  {Budker}}]{Pospelov2013GNOME}%
  \BibitemOpen
  \bibfield  {author} {\bibinfo {author} {\bibfnamefont {M.}~\bibnamefont
  {Pospelov}}, \bibinfo {author} {\bibfnamefont {S.}~\bibnamefont {Pustelny}},
  \bibinfo {author} {\bibfnamefont {M.~P.}\ \bibnamefont {Ledbetter}}, \bibinfo
  {author} {\bibfnamefont {D.~F.~J.}\ \bibnamefont {Kimball}}, \bibinfo
  {author} {\bibfnamefont {W.}~\bibnamefont {Gawlik}}, \ and\ \bibinfo {author}
  {\bibfnamefont {D.}~\bibnamefont {Budker}},\ }\bibfield  {title} {\enquote
  {\bibinfo {title} {Detecting domain walls of axionlike models using
  terrestrial experiments},}\ }\href {\doibase 10.1103/PhysRevLett.110.021803}
  {\bibfield  {journal} {\bibinfo  {journal} {Phys. Rev. Lett.}\ }\textbf
  {\bibinfo {volume} {110}},\ \bibinfo {pages} {021803} (\bibinfo {year}
  {2013})}\BibitemShut {NoStop}%
\bibitem [{\citenamefont {Stadnik}\ and\ \citenamefont
  {Flambaum}(2015)}]{stadnik2015can}%
  \BibitemOpen
  \bibfield  {author} {\bibinfo {author} {\bibfnamefont {Y.}~\bibnamefont
  {Stadnik}}\ and\ \bibinfo {author} {\bibfnamefont {V.}~\bibnamefont
  {Flambaum}},\ }\bibfield  {title} {\enquote {\bibinfo {title} {Can dark
  matter induce cosmological evolution of the fundamental constants of
  nature?}}\ }\href@noop {} {\bibfield  {journal} {\bibinfo  {journal}
  {Physical Review Letters}\ }\textbf {\bibinfo {volume} {115}},\ \bibinfo
  {pages} {201301} (\bibinfo {year} {2015})}\BibitemShut {NoStop}%
\bibitem [{\citenamefont {Delaunay}\ \emph {et~al.}(2017)\citenamefont
  {Delaunay}, \citenamefont {Ozeri}, \citenamefont {Perez},\ and\ \citenamefont
  {Soreq}}]{delaunay2017probing}%
  \BibitemOpen
  \bibfield  {author} {\bibinfo {author} {\bibfnamefont {C.}~\bibnamefont
  {Delaunay}}, \bibinfo {author} {\bibfnamefont {R.}~\bibnamefont {Ozeri}},
  \bibinfo {author} {\bibfnamefont {G.}~\bibnamefont {Perez}}, \ and\ \bibinfo
  {author} {\bibfnamefont {Y.}~\bibnamefont {Soreq}},\ }\bibfield  {title}
  {\enquote {\bibinfo {title} {Probing atomic higgs-like forces at the
  precision frontier},}\ }\href@noop {} {\bibfield  {journal} {\bibinfo
  {journal} {Physical Review D}\ }\textbf {\bibinfo {volume} {96}},\ \bibinfo
  {pages} {093001} (\bibinfo {year} {2017})}\BibitemShut {NoStop}%
\bibitem [{\citenamefont {Berengut}\ \emph {et~al.}(2018)\citenamefont
  {Berengut}, \citenamefont {Budker}, \citenamefont {Delaunay}, \citenamefont
  {Flambaum}, \citenamefont {Frugiuele}, \citenamefont {Fuchs}, \citenamefont
  {Grojean}, \citenamefont {Harnik}, \citenamefont {Ozeri}, \citenamefont
  {Perez} \emph {et~al.}}]{berengut2018probing}%
  \BibitemOpen
  \bibfield  {author} {\bibinfo {author} {\bibfnamefont {J.~C.}\ \bibnamefont
  {Berengut}}, \bibinfo {author} {\bibfnamefont {D.}~\bibnamefont {Budker}},
  \bibinfo {author} {\bibfnamefont {C.}~\bibnamefont {Delaunay}}, \bibinfo
  {author} {\bibfnamefont {V.~V.}\ \bibnamefont {Flambaum}}, \bibinfo {author}
  {\bibfnamefont {C.}~\bibnamefont {Frugiuele}}, \bibinfo {author}
  {\bibfnamefont {E.}~\bibnamefont {Fuchs}}, \bibinfo {author} {\bibfnamefont
  {C.}~\bibnamefont {Grojean}}, \bibinfo {author} {\bibfnamefont
  {R.}~\bibnamefont {Harnik}}, \bibinfo {author} {\bibfnamefont
  {R.}~\bibnamefont {Ozeri}}, \bibinfo {author} {\bibfnamefont
  {G.}~\bibnamefont {Perez}},  \emph {et~al.},\ }\bibfield  {title} {\enquote
  {\bibinfo {title} {Probing new long-range interactions by isotope shift
  spectroscopy},}\ }\href@noop {} {\bibfield  {journal} {\bibinfo  {journal}
  {Physical review letters}\ }\textbf {\bibinfo {volume} {120}},\ \bibinfo
  {pages} {091801} (\bibinfo {year} {2018})}\BibitemShut {NoStop}%
\bibitem [{\citenamefont {Counts}\ \emph {et~al.}(2020)\citenamefont {Counts},
  \citenamefont {Hur}, \citenamefont {Craik}, \citenamefont {Jeon},
  \citenamefont {Leung}, \citenamefont {Berengut}, \citenamefont {Geddes},
  \citenamefont {Kawasaki}, \citenamefont {Jhe},\ and\ \citenamefont
  {Vuleti{\'c}}}]{counts2020evidence}%
  \BibitemOpen
  \bibfield  {author} {\bibinfo {author} {\bibfnamefont {I.}~\bibnamefont
  {Counts}}, \bibinfo {author} {\bibfnamefont {J.}~\bibnamefont {Hur}},
  \bibinfo {author} {\bibfnamefont {D.~P.~A.}\ \bibnamefont {Craik}}, \bibinfo
  {author} {\bibfnamefont {H.}~\bibnamefont {Jeon}}, \bibinfo {author}
  {\bibfnamefont {C.}~\bibnamefont {Leung}}, \bibinfo {author} {\bibfnamefont
  {J.~C.}\ \bibnamefont {Berengut}}, \bibinfo {author} {\bibfnamefont
  {A.}~\bibnamefont {Geddes}}, \bibinfo {author} {\bibfnamefont
  {A.}~\bibnamefont {Kawasaki}}, \bibinfo {author} {\bibfnamefont
  {W.}~\bibnamefont {Jhe}}, \ and\ \bibinfo {author} {\bibfnamefont
  {V.}~\bibnamefont {Vuleti{\'c}}},\ }\bibfield  {title} {\enquote {\bibinfo
  {title} {Evidence for nonlinear isotope shift in yb+ search for new boson},}\
  }\href@noop {} {\bibfield  {journal} {\bibinfo  {journal} {Physical Review
  Letters}\ }\textbf {\bibinfo {volume} {125}},\ \bibinfo {pages} {123002}
  (\bibinfo {year} {2020})}\BibitemShut {NoStop}%
\bibitem [{\citenamefont {Stadnik}(2020)}]{stadnik2020new}%
  \BibitemOpen
  \bibfield  {author} {\bibinfo {author} {\bibfnamefont {Y.~V.}\ \bibnamefont
  {Stadnik}},\ }\bibfield  {title} {\enquote {\bibinfo {title} {New bounds on
  macroscopic scalar-field topological defects from nontransient signatures due
  to environmental dependence and spatial variations of the fundamental
  constants},}\ }\href@noop {} {\bibfield  {journal} {\bibinfo  {journal}
  {Physical Review D}\ }\textbf {\bibinfo {volume} {102}},\ \bibinfo {pages}
  {115016} (\bibinfo {year} {2020})}\BibitemShut {NoStop}%
\bibitem [{\citenamefont {Mikami}, \citenamefont {Tanaka},\ and\ \citenamefont
  {Yamamoto}(2017)}]{mikami2017probing}%
  \BibitemOpen
  \bibfield  {author} {\bibinfo {author} {\bibfnamefont {K.}~\bibnamefont
  {Mikami}}, \bibinfo {author} {\bibfnamefont {M.}~\bibnamefont {Tanaka}}, \
  and\ \bibinfo {author} {\bibfnamefont {Y.}~\bibnamefont {Yamamoto}},\
  }\bibfield  {title} {\enquote {\bibinfo {title} {Probing new intra-atomic
  force with isotope shifts},}\ }\href@noop {} {\bibfield  {journal} {\bibinfo
  {journal} {The European Physical Journal C}\ }\textbf {\bibinfo {volume}
  {77}},\ \bibinfo {pages} {1--11} (\bibinfo {year} {2017})}\BibitemShut
  {NoStop}%
\bibitem [{\citenamefont {Flambaum}, \citenamefont {Geddes},\ and\
  \citenamefont {Viatkina}(2018)}]{flambaum2018isotope}%
  \BibitemOpen
  \bibfield  {author} {\bibinfo {author} {\bibfnamefont {V.}~\bibnamefont
  {Flambaum}}, \bibinfo {author} {\bibfnamefont {A.}~\bibnamefont {Geddes}}, \
  and\ \bibinfo {author} {\bibfnamefont {A.}~\bibnamefont {Viatkina}},\
  }\bibfield  {title} {\enquote {\bibinfo {title} {Isotope shift, nonlinearity
  of king plots, and the search for new particles},}\ }\href@noop {} {\bibfield
   {journal} {\bibinfo  {journal} {Physical Review A}\ }\textbf {\bibinfo
  {volume} {97}},\ \bibinfo {pages} {032510} (\bibinfo {year}
  {2018})}\BibitemShut {NoStop}%
\bibitem [{\citenamefont {Huntemann}\ \emph {et~al.}(2014)\citenamefont
  {Huntemann}, \citenamefont {Lipphardt}, \citenamefont {Tamm}, \citenamefont
  {Gerginov}, \citenamefont {Weyers},\ and\ \citenamefont
  {Peik}}]{huntemann2014improvement}%
  \BibitemOpen
  \bibfield  {author} {\bibinfo {author} {\bibfnamefont {N.}~\bibnamefont
  {Huntemann}}, \bibinfo {author} {\bibfnamefont {B.}~\bibnamefont
  {Lipphardt}}, \bibinfo {author} {\bibfnamefont {C.}~\bibnamefont {Tamm}},
  \bibinfo {author} {\bibfnamefont {V.}~\bibnamefont {Gerginov}}, \bibinfo
  {author} {\bibfnamefont {S.}~\bibnamefont {Weyers}}, \ and\ \bibinfo {author}
  {\bibfnamefont {E.}~\bibnamefont {Peik}},\ }\bibfield  {title} {\enquote
  {\bibinfo {title} {Improved limit on a temporal variation of
  ${m}_{p}/{m}_{e}$ from comparisons of ${\mathrm{yb}}^{+}$ and cs atomic
  clocks},}\ }\href {\doibase 10.1103/PhysRevLett.113.210802} {\bibfield
  {journal} {\bibinfo  {journal} {Phys. Rev. Lett.}\ }\textbf {\bibinfo
  {volume} {113}},\ \bibinfo {pages} {210802} (\bibinfo {year}
  {2014})}\BibitemShut {NoStop}%
\bibitem [{\citenamefont {Safronova}\ \emph
  {et~al.}(2018{\natexlab{a}})\citenamefont {Safronova}, \citenamefont
  {Porsev}, \citenamefont {Sanner},\ and\ \citenamefont
  {Ye}}]{safronova2018two}%
  \BibitemOpen
  \bibfield  {author} {\bibinfo {author} {\bibfnamefont {M.~S.}\ \bibnamefont
  {Safronova}}, \bibinfo {author} {\bibfnamefont {S.~G.}\ \bibnamefont
  {Porsev}}, \bibinfo {author} {\bibfnamefont {C.}~\bibnamefont {Sanner}}, \
  and\ \bibinfo {author} {\bibfnamefont {J.}~\bibnamefont {Ye}},\ }\bibfield
  {title} {\enquote {\bibinfo {title} {Two clock transitions in neutral yb for
  the highest sensitivity to variations of the fine-structure constant},}\
  }\href@noop {} {\bibfield  {journal} {\bibinfo  {journal} {Physical review
  letters}\ }\textbf {\bibinfo {volume} {120}},\ \bibinfo {pages} {173001}
  (\bibinfo {year} {2018}{\natexlab{a}})}\BibitemShut {NoStop}%
\bibitem [{\citenamefont {Dzuba}, \citenamefont {Flambaum},\ and\ \citenamefont
  {Schiller}(2018)}]{dzuba2018Testing}%
  \BibitemOpen
  \bibfield  {author} {\bibinfo {author} {\bibfnamefont {V.~A.}\ \bibnamefont
  {Dzuba}}, \bibinfo {author} {\bibfnamefont {V.~V.}\ \bibnamefont {Flambaum}},
  \ and\ \bibinfo {author} {\bibfnamefont {S.}~\bibnamefont {Schiller}},\
  }\bibfield  {title} {\enquote {\bibinfo {title} {Testing physics beyond the
  standard model through additional clock transitions in neutral ytterbium},}\
  }\href {\doibase 10.1103/PhysRevA.98.022501} {\bibfield  {journal} {\bibinfo
  {journal} {Phys. Rev. A}\ }\textbf {\bibinfo {volume} {98}},\ \bibinfo
  {pages} {022501} (\bibinfo {year} {2018})}\BibitemShut {NoStop}%
\bibitem [{\citenamefont {Chou}\ \emph {et~al.}(2010)\citenamefont {Chou},
  \citenamefont {Hume}, \citenamefont {Rosenband},\ and\ \citenamefont
  {Wineland}}]{chou2010optical}%
  \BibitemOpen
  \bibfield  {author} {\bibinfo {author} {\bibfnamefont {C.-W.}\ \bibnamefont
  {Chou}}, \bibinfo {author} {\bibfnamefont {D.~B.}\ \bibnamefont {Hume}},
  \bibinfo {author} {\bibfnamefont {T.}~\bibnamefont {Rosenband}}, \ and\
  \bibinfo {author} {\bibfnamefont {D.~J.}\ \bibnamefont {Wineland}},\
  }\bibfield  {title} {\enquote {\bibinfo {title} {Optical clocks and
  relativity},}\ }\href@noop {} {\bibfield  {journal} {\bibinfo  {journal}
  {Science}\ }\textbf {\bibinfo {volume} {329}},\ \bibinfo {pages} {1630--1633}
  (\bibinfo {year} {2010})}\BibitemShut {NoStop}%
\bibitem [{\citenamefont {Grotti}\ \emph {et~al.}(2018)\citenamefont {Grotti},
  \citenamefont {Koller}, \citenamefont {Vogt}, \citenamefont {H{\"a}fner},
  \citenamefont {Sterr}, \citenamefont {Lisdat}, \citenamefont {Denker},
  \citenamefont {Voigt}, \citenamefont {Timmen}, \citenamefont {Rolland} \emph
  {et~al.}}]{grotti2018geodesy}%
  \BibitemOpen
  \bibfield  {author} {\bibinfo {author} {\bibfnamefont {J.}~\bibnamefont
  {Grotti}}, \bibinfo {author} {\bibfnamefont {S.}~\bibnamefont {Koller}},
  \bibinfo {author} {\bibfnamefont {S.}~\bibnamefont {Vogt}}, \bibinfo {author}
  {\bibfnamefont {S.}~\bibnamefont {H{\"a}fner}}, \bibinfo {author}
  {\bibfnamefont {U.}~\bibnamefont {Sterr}}, \bibinfo {author} {\bibfnamefont
  {C.}~\bibnamefont {Lisdat}}, \bibinfo {author} {\bibfnamefont
  {H.}~\bibnamefont {Denker}}, \bibinfo {author} {\bibfnamefont
  {C.}~\bibnamefont {Voigt}}, \bibinfo {author} {\bibfnamefont
  {L.}~\bibnamefont {Timmen}}, \bibinfo {author} {\bibfnamefont
  {A.}~\bibnamefont {Rolland}},  \emph {et~al.},\ }\bibfield  {title} {\enquote
  {\bibinfo {title} {Geodesy and metrology with a transportable optical
  clock},}\ }\href@noop {} {\bibfield  {journal} {\bibinfo  {journal} {Nature
  Physics}\ }\textbf {\bibinfo {volume} {14}},\ \bibinfo {pages} {437--441}
  (\bibinfo {year} {2018})}\BibitemShut {NoStop}%
\bibitem [{\citenamefont {Bothwell}\ \emph {et~al.}(2021)\citenamefont
  {Bothwell}, \citenamefont {Kennedy}, \citenamefont {Aeppli}, \citenamefont
  {Kedar}, \citenamefont {Robinson}, \citenamefont {Oelker}, \citenamefont
  {Staron},\ and\ \citenamefont {Ye}}]{bothwell2021resolving}%
  \BibitemOpen
  \bibfield  {author} {\bibinfo {author} {\bibfnamefont {T.}~\bibnamefont
  {Bothwell}}, \bibinfo {author} {\bibfnamefont {C.~J.}\ \bibnamefont
  {Kennedy}}, \bibinfo {author} {\bibfnamefont {A.}~\bibnamefont {Aeppli}},
  \bibinfo {author} {\bibfnamefont {D.}~\bibnamefont {Kedar}}, \bibinfo
  {author} {\bibfnamefont {J.~M.}\ \bibnamefont {Robinson}}, \bibinfo {author}
  {\bibfnamefont {E.}~\bibnamefont {Oelker}}, \bibinfo {author} {\bibfnamefont
  {A.}~\bibnamefont {Staron}}, \ and\ \bibinfo {author} {\bibfnamefont
  {J.}~\bibnamefont {Ye}},\ }\bibfield  {title} {\enquote {\bibinfo {title}
  {Resolving the gravitational redshift within a millimeter atomic sample},}\
  }\href@noop {} {\bibfield  {journal} {\bibinfo  {journal} {arXiv preprint
  arXiv:2109.12238}\ } (\bibinfo {year} {2021})}\BibitemShut {NoStop}%
\bibitem [{\citenamefont {Kolkowitz}\ \emph {et~al.}(2016)\citenamefont
  {Kolkowitz}, \citenamefont {Pikovski}, \citenamefont {Langellier},
  \citenamefont {Lukin}, \citenamefont {Walsworth},\ and\ \citenamefont
  {Ye}}]{kolkowitz2016gravitational}%
  \BibitemOpen
  \bibfield  {author} {\bibinfo {author} {\bibfnamefont {S.}~\bibnamefont
  {Kolkowitz}}, \bibinfo {author} {\bibfnamefont {I.}~\bibnamefont {Pikovski}},
  \bibinfo {author} {\bibfnamefont {N.}~\bibnamefont {Langellier}}, \bibinfo
  {author} {\bibfnamefont {M.~D.}\ \bibnamefont {Lukin}}, \bibinfo {author}
  {\bibfnamefont {R.~L.}\ \bibnamefont {Walsworth}}, \ and\ \bibinfo {author}
  {\bibfnamefont {J.}~\bibnamefont {Ye}},\ }\bibfield  {title} {\enquote
  {\bibinfo {title} {Gravitational wave detection with optical lattice atomic
  clocks},}\ }\href@noop {} {\bibfield  {journal} {\bibinfo  {journal}
  {Physical Review D}\ }\textbf {\bibinfo {volume} {94}},\ \bibinfo {pages}
  {124043} (\bibinfo {year} {2016})}\BibitemShut {NoStop}%
\bibitem [{\citenamefont {Sedda}\ \emph {et~al.}(2020)\citenamefont {Sedda},
  \citenamefont {Berry}, \citenamefont {Jani}, \citenamefont {Amaro-Seoane},
  \citenamefont {Auclair}, \citenamefont {Baird}, \citenamefont {Baker},
  \citenamefont {Berti}, \citenamefont {Breivik}, \citenamefont {Burrows},
  \citenamefont {Caprini}, \citenamefont {Chen}, \citenamefont {Doneva},
  \citenamefont {Ezquiaga}, \citenamefont {Ford}, \citenamefont {Katz},
  \citenamefont {Kolkowitz}, \citenamefont {McKernan}, \citenamefont {Mueller},
  \citenamefont {Nardini}, \citenamefont {Pikovski}, \citenamefont {Rajendran},
  \citenamefont {Sesana}, \citenamefont {Shao}, \citenamefont {Tamanini},
  \citenamefont {Vartanyan}, \citenamefont {Warburton}, \citenamefont {Witek},
  \citenamefont {Wong},\ and\ \citenamefont {Zevin}}]{Sedda_2020}%
  \BibitemOpen
  \bibfield  {author} {\bibinfo {author} {\bibfnamefont {M.~A.}\ \bibnamefont
  {Sedda}}, \bibinfo {author} {\bibfnamefont {C.~P.~L.}\ \bibnamefont {Berry}},
  \bibinfo {author} {\bibfnamefont {K.}~\bibnamefont {Jani}}, \bibinfo {author}
  {\bibfnamefont {P.}~\bibnamefont {Amaro-Seoane}}, \bibinfo {author}
  {\bibfnamefont {P.}~\bibnamefont {Auclair}}, \bibinfo {author} {\bibfnamefont
  {J.}~\bibnamefont {Baird}}, \bibinfo {author} {\bibfnamefont
  {T.}~\bibnamefont {Baker}}, \bibinfo {author} {\bibfnamefont
  {E.}~\bibnamefont {Berti}}, \bibinfo {author} {\bibfnamefont
  {K.}~\bibnamefont {Breivik}}, \bibinfo {author} {\bibfnamefont
  {A.}~\bibnamefont {Burrows}}, \bibinfo {author} {\bibfnamefont
  {C.}~\bibnamefont {Caprini}}, \bibinfo {author} {\bibfnamefont
  {X.}~\bibnamefont {Chen}}, \bibinfo {author} {\bibfnamefont {D.}~\bibnamefont
  {Doneva}}, \bibinfo {author} {\bibfnamefont {J.~M.}\ \bibnamefont
  {Ezquiaga}}, \bibinfo {author} {\bibfnamefont {K.~E.~S.}\ \bibnamefont
  {Ford}}, \bibinfo {author} {\bibfnamefont {M.~L.}\ \bibnamefont {Katz}},
  \bibinfo {author} {\bibfnamefont {S.}~\bibnamefont {Kolkowitz}}, \bibinfo
  {author} {\bibfnamefont {B.}~\bibnamefont {McKernan}}, \bibinfo {author}
  {\bibfnamefont {G.}~\bibnamefont {Mueller}}, \bibinfo {author} {\bibfnamefont
  {G.}~\bibnamefont {Nardini}}, \bibinfo {author} {\bibfnamefont
  {I.}~\bibnamefont {Pikovski}}, \bibinfo {author} {\bibfnamefont
  {S.}~\bibnamefont {Rajendran}}, \bibinfo {author} {\bibfnamefont
  {A.}~\bibnamefont {Sesana}}, \bibinfo {author} {\bibfnamefont
  {L.}~\bibnamefont {Shao}}, \bibinfo {author} {\bibfnamefont {N.}~\bibnamefont
  {Tamanini}}, \bibinfo {author} {\bibfnamefont {D.}~\bibnamefont {Vartanyan}},
  \bibinfo {author} {\bibfnamefont {N.}~\bibnamefont {Warburton}}, \bibinfo
  {author} {\bibfnamefont {H.}~\bibnamefont {Witek}}, \bibinfo {author}
  {\bibfnamefont {K.}~\bibnamefont {Wong}}, \ and\ \bibinfo {author}
  {\bibfnamefont {M.}~\bibnamefont {Zevin}},\ }\bibfield  {title} {\enquote
  {\bibinfo {title} {The missing link in gravitational-wave astronomy:
  discoveries waiting in the decihertz range},}\ }\href {\doibase
  10.1088/1361-6382/abb5c1} {\bibfield  {journal} {\bibinfo  {journal}
  {Classical and Quantum Gravity}\ }\textbf {\bibinfo {volume} {37}},\ \bibinfo
  {pages} {215011} (\bibinfo {year} {2020})}\BibitemShut {NoStop}%
\bibitem [{\citenamefont {Lisdat}\ \emph {et~al.}(2016)\citenamefont {Lisdat},
  \citenamefont {Grosche}, \citenamefont {Quintin}, \citenamefont {Shi},
  \citenamefont {Raupach}, \citenamefont {Grebing}, \citenamefont {Nicolodi},
  \citenamefont {Stefani}, \citenamefont {Al-Masoudi}, \citenamefont
  {D{\"o}rscher} \emph {et~al.}}]{lisdat2016clock}%
  \BibitemOpen
  \bibfield  {author} {\bibinfo {author} {\bibfnamefont {C.}~\bibnamefont
  {Lisdat}}, \bibinfo {author} {\bibfnamefont {G.}~\bibnamefont {Grosche}},
  \bibinfo {author} {\bibfnamefont {N.}~\bibnamefont {Quintin}}, \bibinfo
  {author} {\bibfnamefont {C.}~\bibnamefont {Shi}}, \bibinfo {author}
  {\bibfnamefont {S.}~\bibnamefont {Raupach}}, \bibinfo {author} {\bibfnamefont
  {C.}~\bibnamefont {Grebing}}, \bibinfo {author} {\bibfnamefont
  {D.}~\bibnamefont {Nicolodi}}, \bibinfo {author} {\bibfnamefont
  {F.}~\bibnamefont {Stefani}}, \bibinfo {author} {\bibfnamefont
  {A.}~\bibnamefont {Al-Masoudi}}, \bibinfo {author} {\bibfnamefont
  {S.}~\bibnamefont {D{\"o}rscher}},  \emph {et~al.},\ }\bibfield  {title}
  {\enquote {\bibinfo {title} {A clock network for geodesy and fundamental
  science},}\ }\href@noop {} {\bibfield  {journal} {\bibinfo  {journal} {Nature
  communications}\ }\textbf {\bibinfo {volume} {7}},\ \bibinfo {pages} {1--7}
  (\bibinfo {year} {2016})}\BibitemShut {NoStop}%
\bibitem [{\citenamefont {Bondarescu}\ \emph {et~al.}(2015)\citenamefont
  {Bondarescu}, \citenamefont {Sch{\"a}rer}, \citenamefont {Lundgren},
  \citenamefont {Het{\'e}nyi}, \citenamefont {Houli{\'e}}, \citenamefont
  {Jetzer},\ and\ \citenamefont {Bondarescu}}]{bondarescu2015ground}%
  \BibitemOpen
  \bibfield  {author} {\bibinfo {author} {\bibfnamefont {R.}~\bibnamefont
  {Bondarescu}}, \bibinfo {author} {\bibfnamefont {A.}~\bibnamefont
  {Sch{\"a}rer}}, \bibinfo {author} {\bibfnamefont {A.}~\bibnamefont
  {Lundgren}}, \bibinfo {author} {\bibfnamefont {G.}~\bibnamefont
  {Het{\'e}nyi}}, \bibinfo {author} {\bibfnamefont {N.}~\bibnamefont
  {Houli{\'e}}}, \bibinfo {author} {\bibfnamefont {P.}~\bibnamefont {Jetzer}},
  \ and\ \bibinfo {author} {\bibfnamefont {M.}~\bibnamefont {Bondarescu}},\
  }\bibfield  {title} {\enquote {\bibinfo {title} {Ground-based optical atomic
  clocks as a tool to monitor vertical surface motion},}\ }\href@noop {}
  {\bibfield  {journal} {\bibinfo  {journal} {Geophysical Journal
  International}\ }\textbf {\bibinfo {volume} {202}},\ \bibinfo {pages}
  {1770--1774} (\bibinfo {year} {2015})}\BibitemShut {NoStop}%
\bibitem [{\citenamefont {Major}(2007)}]{major2007quantum}%
  \BibitemOpen
  \bibfield  {author} {\bibinfo {author} {\bibfnamefont {F.~G.}\ \bibnamefont
  {Major}},\ }\href@noop {} {\emph {\bibinfo {title} {The quantum beat:
  principles and applications of atomic clocks}}},\ Vol.~\bibinfo {volume} {2}\
  (\bibinfo  {publisher} {Springer},\ \bibinfo {year} {2007})\BibitemShut
  {NoStop}%
\bibitem [{\citenamefont {Grewal}, \citenamefont {Andrews},\ and\ \citenamefont
  {Bartone}(2020)}]{grewal2020global}%
  \BibitemOpen
  \bibfield  {author} {\bibinfo {author} {\bibfnamefont {M.~S.}\ \bibnamefont
  {Grewal}}, \bibinfo {author} {\bibfnamefont {A.~P.}\ \bibnamefont {Andrews}},
  \ and\ \bibinfo {author} {\bibfnamefont {C.~G.}\ \bibnamefont {Bartone}},\
  }\href@noop {} {\emph {\bibinfo {title} {Global navigation satellite systems,
  inertial navigation, and integration}}}\ (\bibinfo  {publisher} {John Wiley
  \& Sons},\ \bibinfo {year} {2020})\BibitemShut {NoStop}%
\bibitem [{\citenamefont {Le~Targat}\ \emph {et~al.}(2006)\citenamefont
  {Le~Targat}, \citenamefont {Baillard}, \citenamefont {Fouch{\'e}},
  \citenamefont {Brusch}, \citenamefont {Tcherbakoff}, \citenamefont {Rovera},\
  and\ \citenamefont {Lemonde}}]{le2006accurate}%
  \BibitemOpen
  \bibfield  {author} {\bibinfo {author} {\bibfnamefont {R.}~\bibnamefont
  {Le~Targat}}, \bibinfo {author} {\bibfnamefont {X.}~\bibnamefont {Baillard}},
  \bibinfo {author} {\bibfnamefont {M.}~\bibnamefont {Fouch{\'e}}}, \bibinfo
  {author} {\bibfnamefont {A.}~\bibnamefont {Brusch}}, \bibinfo {author}
  {\bibfnamefont {O.}~\bibnamefont {Tcherbakoff}}, \bibinfo {author}
  {\bibfnamefont {G.~D.}\ \bibnamefont {Rovera}}, \ and\ \bibinfo {author}
  {\bibfnamefont {P.}~\bibnamefont {Lemonde}},\ }\bibfield  {title} {\enquote
  {\bibinfo {title} {Accurate optical lattice clock with sr 87 atoms},}\
  }\href@noop {} {\bibfield  {journal} {\bibinfo  {journal} {Physical Review
  Letters}\ }\textbf {\bibinfo {volume} {97}},\ \bibinfo {pages} {130801}
  (\bibinfo {year} {2006})}\BibitemShut {NoStop}%
\bibitem [{\citenamefont {Ludlow}\ \emph {et~al.}(2008)\citenamefont {Ludlow},
  \citenamefont {Zelevinsky}, \citenamefont {Campbell}, \citenamefont {Blatt},
  \citenamefont {Boyd}, \citenamefont {de~Miranda}, \citenamefont {Martin},
  \citenamefont {Thomsen}, \citenamefont {Foreman}, \citenamefont {Ye},
  \citenamefont {Fortier}, \citenamefont {Stalnaker}, \citenamefont {Diddams},
  \citenamefont {Le~Coq}, \citenamefont {Barber}, \citenamefont {Poli},
  \citenamefont {Lemke}, \citenamefont {Beck},\ and\ \citenamefont
  {Oates}}]{Ludlow2008}%
  \BibitemOpen
  \bibfield  {author} {\bibinfo {author} {\bibfnamefont {A.~D.}\ \bibnamefont
  {Ludlow}}, \bibinfo {author} {\bibfnamefont {T.}~\bibnamefont {Zelevinsky}},
  \bibinfo {author} {\bibfnamefont {G.~K.}\ \bibnamefont {Campbell}}, \bibinfo
  {author} {\bibfnamefont {S.}~\bibnamefont {Blatt}}, \bibinfo {author}
  {\bibfnamefont {M.~M.}\ \bibnamefont {Boyd}}, \bibinfo {author}
  {\bibfnamefont {M.~H.~G.}\ \bibnamefont {de~Miranda}}, \bibinfo {author}
  {\bibfnamefont {M.~J.}\ \bibnamefont {Martin}}, \bibinfo {author}
  {\bibfnamefont {J.~W.}\ \bibnamefont {Thomsen}}, \bibinfo {author}
  {\bibfnamefont {S.~M.}\ \bibnamefont {Foreman}}, \bibinfo {author}
  {\bibfnamefont {J.}~\bibnamefont {Ye}}, \bibinfo {author} {\bibfnamefont
  {T.~M.}\ \bibnamefont {Fortier}}, \bibinfo {author} {\bibfnamefont {J.~E.}\
  \bibnamefont {Stalnaker}}, \bibinfo {author} {\bibfnamefont {S.~A.}\
  \bibnamefont {Diddams}}, \bibinfo {author} {\bibfnamefont {Y.}~\bibnamefont
  {Le~Coq}}, \bibinfo {author} {\bibfnamefont {Z.~W.}\ \bibnamefont {Barber}},
  \bibinfo {author} {\bibfnamefont {N.}~\bibnamefont {Poli}}, \bibinfo {author}
  {\bibfnamefont {N.~D.}\ \bibnamefont {Lemke}}, \bibinfo {author}
  {\bibfnamefont {K.~M.}\ \bibnamefont {Beck}}, \ and\ \bibinfo {author}
  {\bibfnamefont {C.~W.}\ \bibnamefont {Oates}},\ }\bibfield  {title} {\enquote
  {\bibinfo {title} {{Sr} lattice clock at $1 \times 10^{-16}$ fractional
  uncertainty by remote optical evaluation with a {Ca} clock},}\ }\href
  {\doibase 10.1126/science.1153341} {\bibfield  {journal} {\bibinfo  {journal}
  {Science}\ }\textbf {\bibinfo {volume} {319}},\ \bibinfo {pages} {1805--1808}
  (\bibinfo {year} {2008})}\BibitemShut {NoStop}%
\bibitem [{\citenamefont {Campbell}\ \emph {et~al.}(2009)\citenamefont
  {Campbell}, \citenamefont {Boyd}, \citenamefont {Thomsen}, \citenamefont
  {Martin}, \citenamefont {Blatt}, \citenamefont {Swallows}, \citenamefont
  {Nicholson}, \citenamefont {Fortier}, \citenamefont {Oates}, \citenamefont
  {Diddams} \emph {et~al.}}]{campbell2009probingColl}%
  \BibitemOpen
  \bibfield  {author} {\bibinfo {author} {\bibfnamefont {G.}~\bibnamefont
  {Campbell}}, \bibinfo {author} {\bibfnamefont {M.}~\bibnamefont {Boyd}},
  \bibinfo {author} {\bibfnamefont {J.}~\bibnamefont {Thomsen}}, \bibinfo
  {author} {\bibfnamefont {M.}~\bibnamefont {Martin}}, \bibinfo {author}
  {\bibfnamefont {S.}~\bibnamefont {Blatt}}, \bibinfo {author} {\bibfnamefont
  {M.}~\bibnamefont {Swallows}}, \bibinfo {author} {\bibfnamefont {T.~L.}\
  \bibnamefont {Nicholson}}, \bibinfo {author} {\bibfnamefont {T.}~\bibnamefont
  {Fortier}}, \bibinfo {author} {\bibfnamefont {C.~W.}\ \bibnamefont {Oates}},
  \bibinfo {author} {\bibfnamefont {S.~A.}\ \bibnamefont {Diddams}},  \emph
  {et~al.},\ }\bibfield  {title} {\enquote {\bibinfo {title} {Probing
  interactions between ultracold fermions},}\ }\href@noop {} {\bibfield
  {journal} {\bibinfo  {journal} {science}\ }\textbf {\bibinfo {volume}
  {324}},\ \bibinfo {pages} {360--363} (\bibinfo {year} {2009})}\BibitemShut
  {NoStop}%
\bibitem [{\citenamefont {Lisdat}\ \emph {et~al.}(2009)\citenamefont {Lisdat},
  \citenamefont {Winfred}, \citenamefont {Middelmann}, \citenamefont {Riehle},\
  and\ \citenamefont {Sterr}}]{lisdat2009coll}%
  \BibitemOpen
  \bibfield  {author} {\bibinfo {author} {\bibfnamefont {C.}~\bibnamefont
  {Lisdat}}, \bibinfo {author} {\bibfnamefont {J.~S. R.~V.}\ \bibnamefont
  {Winfred}}, \bibinfo {author} {\bibfnamefont {T.}~\bibnamefont {Middelmann}},
  \bibinfo {author} {\bibfnamefont {F.}~\bibnamefont {Riehle}}, \ and\ \bibinfo
  {author} {\bibfnamefont {U.}~\bibnamefont {Sterr}},\ }\bibfield  {title}
  {\enquote {\bibinfo {title} {Collisional losses, decoherence, and frequency
  shifts in optical lattice clocks with bosons},}\ }\href {\doibase
  10.1103/PhysRevLett.103.090801} {\bibfield  {journal} {\bibinfo  {journal}
  {Phys. Rev. Lett.}\ }\textbf {\bibinfo {volume} {103}},\ \bibinfo {pages}
  {090801} (\bibinfo {year} {2009})}\BibitemShut {NoStop}%
\bibitem [{\citenamefont {Bloom}\ \emph {et~al.}(2014)\citenamefont {Bloom},
  \citenamefont {Nicholson}, \citenamefont {Williams}, \citenamefont
  {Campbell}, \citenamefont {Bishof}, \citenamefont {Zhang}, \citenamefont
  {Zhang}, \citenamefont {Bromley},\ and\ \citenamefont {Ye}}]{Bloom2014}%
  \BibitemOpen
  \bibfield  {author} {\bibinfo {author} {\bibfnamefont {B.~J.}\ \bibnamefont
  {Bloom}}, \bibinfo {author} {\bibfnamefont {T.~L.}\ \bibnamefont
  {Nicholson}}, \bibinfo {author} {\bibfnamefont {J.~R.}\ \bibnamefont
  {Williams}}, \bibinfo {author} {\bibfnamefont {S.~L.}\ \bibnamefont
  {Campbell}}, \bibinfo {author} {\bibfnamefont {M.}~\bibnamefont {Bishof}},
  \bibinfo {author} {\bibfnamefont {X.}~\bibnamefont {Zhang}}, \bibinfo
  {author} {\bibfnamefont {W.}~\bibnamefont {Zhang}}, \bibinfo {author}
  {\bibfnamefont {S.~L.}\ \bibnamefont {Bromley}}, \ and\ \bibinfo {author}
  {\bibfnamefont {J.}~\bibnamefont {Ye}},\ }\bibfield  {title} {\enquote
  {\bibinfo {title} {An optical lattice clock with accuracy and stability at
  the $10^{-18}$ level},}\ }\href {http://dx.doi.org/10.1038/nature12941}
  {\bibfield  {journal} {\bibinfo  {journal} {Nature (London)}\ }\textbf
  {\bibinfo {volume} {506}},\ \bibinfo {pages} {71--75} (\bibinfo {year}
  {2014})}\BibitemShut {NoStop}%
\bibitem [{\citenamefont {Gao}\ \emph {et~al.}(2018)\citenamefont {Gao},
  \citenamefont {Zhou}, \citenamefont {Han}, \citenamefont {Li}, \citenamefont
  {Zhang}, \citenamefont {Yao}, \citenamefont {Li}, \citenamefont {Qiao},
  \citenamefont {Ai}, \citenamefont {Lou} \emph {et~al.}}]{gao2018systematic}%
  \BibitemOpen
  \bibfield  {author} {\bibinfo {author} {\bibfnamefont {Q.}~\bibnamefont
  {Gao}}, \bibinfo {author} {\bibfnamefont {M.}~\bibnamefont {Zhou}}, \bibinfo
  {author} {\bibfnamefont {C.}~\bibnamefont {Han}}, \bibinfo {author}
  {\bibfnamefont {S.}~\bibnamefont {Li}}, \bibinfo {author} {\bibfnamefont
  {S.}~\bibnamefont {Zhang}}, \bibinfo {author} {\bibfnamefont
  {Y.}~\bibnamefont {Yao}}, \bibinfo {author} {\bibfnamefont {B.}~\bibnamefont
  {Li}}, \bibinfo {author} {\bibfnamefont {H.}~\bibnamefont {Qiao}}, \bibinfo
  {author} {\bibfnamefont {D.}~\bibnamefont {Ai}}, \bibinfo {author}
  {\bibfnamefont {G.}~\bibnamefont {Lou}},  \emph {et~al.},\ }\bibfield
  {title} {\enquote {\bibinfo {title} {Systematic evaluation of a 171yb optical
  clock by synchronous comparison between two lattice systems},}\ }\href@noop
  {} {\bibfield  {journal} {\bibinfo  {journal} {Scientific Reports}\ }\textbf
  {\bibinfo {volume} {8}},\ \bibinfo {pages} {1--8} (\bibinfo {year}
  {2018})}\BibitemShut {NoStop}%
\bibitem [{\citenamefont {Dick}(1987)}]{dick1987local}%
  \BibitemOpen
  \bibfield  {author} {\bibinfo {author} {\bibfnamefont {G.~J.}\ \bibnamefont
  {Dick}},\ }\href@noop {} {\enquote {\bibinfo {title} {Local oscillator
  induced instabilities in trapped ion frequency standards},}\ }\bibinfo {type}
  {Tech. Rep.}\ (\bibinfo  {institution} {CALIFORNIA INST OF TECH PASADENA JET
  PROPULSION LAB},\ \bibinfo {year} {1987})\BibitemShut {NoStop}%
\bibitem [{\citenamefont {Chou}\ \emph {et~al.}(2011)\citenamefont {Chou},
  \citenamefont {Hume}, \citenamefont {Thorpe}, \citenamefont {Wineland},\ and\
  \citenamefont {Rosenband}}]{Chou2011Quantum}%
  \BibitemOpen
  \bibfield  {author} {\bibinfo {author} {\bibfnamefont {C.~W.}\ \bibnamefont
  {Chou}}, \bibinfo {author} {\bibfnamefont {D.~B.}\ \bibnamefont {Hume}},
  \bibinfo {author} {\bibfnamefont {M.~J.}\ \bibnamefont {Thorpe}}, \bibinfo
  {author} {\bibfnamefont {D.~J.}\ \bibnamefont {Wineland}}, \ and\ \bibinfo
  {author} {\bibfnamefont {T.}~\bibnamefont {Rosenband}},\ }\bibfield  {title}
  {\enquote {\bibinfo {title} {Quantum coherence between two atoms beyond
  $q={10}^{15}$},}\ }\href {\doibase 10.1103/PhysRevLett.106.160801} {\bibfield
   {journal} {\bibinfo  {journal} {Phys. Rev. Lett.}\ }\textbf {\bibinfo
  {volume} {106}},\ \bibinfo {pages} {160801} (\bibinfo {year}
  {2011})}\BibitemShut {NoStop}%
\bibitem [{\citenamefont {Takamoto}, \citenamefont {Takano},\ and\
  \citenamefont {Katori}(2011)}]{takamoto2011frequency}%
  \BibitemOpen
  \bibfield  {author} {\bibinfo {author} {\bibfnamefont {M.}~\bibnamefont
  {Takamoto}}, \bibinfo {author} {\bibfnamefont {T.}~\bibnamefont {Takano}}, \
  and\ \bibinfo {author} {\bibfnamefont {H.}~\bibnamefont {Katori}},\
  }\bibfield  {title} {\enquote {\bibinfo {title} {Frequency comparison of
  optical lattice clocks beyond the dick limit},}\ }\href@noop {} {\bibfield
  {journal} {\bibinfo  {journal} {Nature Photonics}\ }\textbf {\bibinfo
  {volume} {5}},\ \bibinfo {pages} {288} (\bibinfo {year} {2011})}\BibitemShut
  {NoStop}%
\bibitem [{\citenamefont {Nicholson}\ \emph {et~al.}(2012)\citenamefont
  {Nicholson}, \citenamefont {Martin}, \citenamefont {Williams}, \citenamefont
  {Bloom}, \citenamefont {Bishof}, \citenamefont {Swallows}, \citenamefont
  {Campbell},\ and\ \citenamefont {Ye}}]{Nicholson2012}%
  \BibitemOpen
  \bibfield  {author} {\bibinfo {author} {\bibfnamefont {T.~L.}\ \bibnamefont
  {Nicholson}}, \bibinfo {author} {\bibfnamefont {M.~J.}\ \bibnamefont
  {Martin}}, \bibinfo {author} {\bibfnamefont {J.~R.}\ \bibnamefont
  {Williams}}, \bibinfo {author} {\bibfnamefont {B.~J.}\ \bibnamefont {Bloom}},
  \bibinfo {author} {\bibfnamefont {M.}~\bibnamefont {Bishof}}, \bibinfo
  {author} {\bibfnamefont {M.~D.}\ \bibnamefont {Swallows}}, \bibinfo {author}
  {\bibfnamefont {S.~L.}\ \bibnamefont {Campbell}}, \ and\ \bibinfo {author}
  {\bibfnamefont {J.}~\bibnamefont {Ye}},\ }\bibfield  {title} {\enquote
  {\bibinfo {title} {Comparison of two independent sr optical clocks with $1
  \times {10}^{-17}$ stability at ${10}^{3}$ s},}\ }\href {\doibase
  10.1103/PhysRevLett.109.230801} {\bibfield  {journal} {\bibinfo  {journal}
  {Phys. Rev. Lett.}\ }\textbf {\bibinfo {volume} {109}},\ \bibinfo {pages}
  {230801} (\bibinfo {year} {2012})}\BibitemShut {NoStop}%
\bibitem [{\citenamefont {Norcia}\ \emph {et~al.}(2019)\citenamefont {Norcia},
  \citenamefont {Young}, \citenamefont {Eckner}, \citenamefont {Oelker},
  \citenamefont {Ye},\ and\ \citenamefont {Kaufman}}]{Norcia2019}%
  \BibitemOpen
  \bibfield  {author} {\bibinfo {author} {\bibfnamefont {M.~A.}\ \bibnamefont
  {Norcia}}, \bibinfo {author} {\bibfnamefont {A.~W.}\ \bibnamefont {Young}},
  \bibinfo {author} {\bibfnamefont {W.~J.}\ \bibnamefont {Eckner}}, \bibinfo
  {author} {\bibfnamefont {E.}~\bibnamefont {Oelker}}, \bibinfo {author}
  {\bibfnamefont {J.}~\bibnamefont {Ye}}, \ and\ \bibinfo {author}
  {\bibfnamefont {A.~M.}\ \bibnamefont {Kaufman}},\ }\bibfield  {title}
  {\enquote {\bibinfo {title} {Seconds-scale coherence on an optical clock
  transition in a tweezer array},}\ }\href {\doibase 10.1126/science.aay0644}
  {\bibfield  {journal} {\bibinfo  {journal} {Science}\ }\textbf {\bibinfo
  {volume} {366}},\ \bibinfo {pages} {93--97} (\bibinfo {year} {2019})},\
  \Eprint
  {http://arxiv.org/abs/https://science.sciencemag.org/content/366/6461/93.full.pdf}
  {https://science.sciencemag.org/content/366/6461/93.full.pdf} \BibitemShut
  {NoStop}%
\bibitem [{\citenamefont {Lodewyck}, \citenamefont {Westergaard},\ and\
  \citenamefont {Lemonde}(2009)}]{Lodewyck2009Nondestructive}%
  \BibitemOpen
  \bibfield  {author} {\bibinfo {author} {\bibfnamefont {J.}~\bibnamefont
  {Lodewyck}}, \bibinfo {author} {\bibfnamefont {P.~G.}\ \bibnamefont
  {Westergaard}}, \ and\ \bibinfo {author} {\bibfnamefont {P.}~\bibnamefont
  {Lemonde}},\ }\bibfield  {title} {\enquote {\bibinfo {title} {Nondestructive
  measurement of the transition probability in a sr optical lattice clock},}\
  }\href {\doibase 10.1103/PhysRevA.79.061401} {\bibfield  {journal} {\bibinfo
  {journal} {Phys. Rev. A}\ }\textbf {\bibinfo {volume} {79}},\ \bibinfo
  {pages} {061401} (\bibinfo {year} {2009})}\BibitemShut {NoStop}%
\bibitem [{\citenamefont {Westergaard}, \citenamefont {Lodewyck},\ and\
  \citenamefont {Lemonde}(2010)}]{Westergaard_DickeNoise_2010}%
  \BibitemOpen
  \bibfield  {author} {\bibinfo {author} {\bibfnamefont {P.~G.}\ \bibnamefont
  {Westergaard}}, \bibinfo {author} {\bibfnamefont {J.}~\bibnamefont
  {Lodewyck}}, \ and\ \bibinfo {author} {\bibfnamefont {P.}~\bibnamefont
  {Lemonde}},\ }\bibfield  {title} {\enquote {\bibinfo {title} {Minimizing the
  dick effect in an optical lattice clock},}\ }\href {\doibase
  10.1109/TUFFC.2010.1457} {\bibfield  {journal} {\bibinfo  {journal} {IEEE
  Transactions on Ultrasonics, Ferroelectrics, and Frequency Control}\ }\textbf
  {\bibinfo {volume} {57}},\ \bibinfo {pages} {623--628} (\bibinfo {year}
  {2010})}\BibitemShut {NoStop}%
\bibitem [{\citenamefont {Akatsuka}, \citenamefont {Takamoto},\ and\
  \citenamefont {Katori}(2010)}]{akatsuka2010three}%
  \BibitemOpen
  \bibfield  {author} {\bibinfo {author} {\bibfnamefont {T.}~\bibnamefont
  {Akatsuka}}, \bibinfo {author} {\bibfnamefont {M.}~\bibnamefont {Takamoto}},
  \ and\ \bibinfo {author} {\bibfnamefont {H.}~\bibnamefont {Katori}},\
  }\bibfield  {title} {\enquote {\bibinfo {title} {Three-dimensional optical
  lattice clock with bosonic sr 88 atoms},}\ }\href@noop {} {\bibfield
  {journal} {\bibinfo  {journal} {Physical Review A}\ }\textbf {\bibinfo
  {volume} {81}},\ \bibinfo {pages} {023402} (\bibinfo {year}
  {2010})}\BibitemShut {NoStop}%
\bibitem [{\citenamefont {Campbell}\ \emph {et~al.}(2017)\citenamefont
  {Campbell}, \citenamefont {Hutson}, \citenamefont {Marti}, \citenamefont
  {Goban}, \citenamefont {Darkwah~Oppong}, \citenamefont {McNally},
  \citenamefont {Sonderhouse}, \citenamefont {Robinson}, \citenamefont {Zhang},
  \citenamefont {Bloom},\ and\ \citenamefont {Ye}}]{Campbell2017}%
  \BibitemOpen
  \bibfield  {author} {\bibinfo {author} {\bibfnamefont {S.~L.}\ \bibnamefont
  {Campbell}}, \bibinfo {author} {\bibfnamefont {R.~B.}\ \bibnamefont
  {Hutson}}, \bibinfo {author} {\bibfnamefont {G.~E.}\ \bibnamefont {Marti}},
  \bibinfo {author} {\bibfnamefont {A.}~\bibnamefont {Goban}}, \bibinfo
  {author} {\bibfnamefont {N.}~\bibnamefont {Darkwah~Oppong}}, \bibinfo
  {author} {\bibfnamefont {R.~L.}\ \bibnamefont {McNally}}, \bibinfo {author}
  {\bibfnamefont {L.}~\bibnamefont {Sonderhouse}}, \bibinfo {author}
  {\bibfnamefont {J.~M.}\ \bibnamefont {Robinson}}, \bibinfo {author}
  {\bibfnamefont {W.}~\bibnamefont {Zhang}}, \bibinfo {author} {\bibfnamefont
  {B.~J.}\ \bibnamefont {Bloom}}, \ and\ \bibinfo {author} {\bibfnamefont
  {J.}~\bibnamefont {Ye}},\ }\bibfield  {title} {\enquote {\bibinfo {title} {A
  fermi-degenerate three-dimensional optical lattice clock},}\ }\href {\doibase
  10.1126/science.aam5538} {\bibfield  {journal} {\bibinfo  {journal}
  {Science}\ }\textbf {\bibinfo {volume} {358}},\ \bibinfo {pages} {90--94}
  (\bibinfo {year} {2017})}\BibitemShut {NoStop}%
\bibitem [{\citenamefont {Madjarov}\ \emph {et~al.}(2019)\citenamefont
  {Madjarov}, \citenamefont {Cooper}, \citenamefont {Shaw}, \citenamefont
  {Covey}, \citenamefont {Schkolnik}, \citenamefont {Yoon}, \citenamefont
  {Williams},\ and\ \citenamefont {Endres}}]{madjarov2019atomic}%
  \BibitemOpen
  \bibfield  {author} {\bibinfo {author} {\bibfnamefont {I.~S.}\ \bibnamefont
  {Madjarov}}, \bibinfo {author} {\bibfnamefont {A.}~\bibnamefont {Cooper}},
  \bibinfo {author} {\bibfnamefont {A.~L.}\ \bibnamefont {Shaw}}, \bibinfo
  {author} {\bibfnamefont {J.~P.}\ \bibnamefont {Covey}}, \bibinfo {author}
  {\bibfnamefont {V.}~\bibnamefont {Schkolnik}}, \bibinfo {author}
  {\bibfnamefont {T.~H.}\ \bibnamefont {Yoon}}, \bibinfo {author}
  {\bibfnamefont {J.~R.}\ \bibnamefont {Williams}}, \ and\ \bibinfo {author}
  {\bibfnamefont {M.}~\bibnamefont {Endres}},\ }\bibfield  {title} {\enquote
  {\bibinfo {title} {An atomic-array optical clock with single-atom readout},}\
  }\href@noop {} {\bibfield  {journal} {\bibinfo  {journal} {Physical Review
  X}\ }\textbf {\bibinfo {volume} {9}},\ \bibinfo {pages} {041052} (\bibinfo
  {year} {2019})}\BibitemShut {NoStop}%
\bibitem [{\citenamefont {Young}\ \emph {et~al.}(2020)\citenamefont {Young},
  \citenamefont {Eckner}, \citenamefont {Milner}, \citenamefont {Kedar},
  \citenamefont {Norcia}, \citenamefont {Oelker}, \citenamefont {Schine},
  \citenamefont {Ye},\ and\ \citenamefont {Kaufman}}]{young2020half}%
  \BibitemOpen
  \bibfield  {author} {\bibinfo {author} {\bibfnamefont {A.~W.}\ \bibnamefont
  {Young}}, \bibinfo {author} {\bibfnamefont {W.~J.}\ \bibnamefont {Eckner}},
  \bibinfo {author} {\bibfnamefont {W.~R.}\ \bibnamefont {Milner}}, \bibinfo
  {author} {\bibfnamefont {D.}~\bibnamefont {Kedar}}, \bibinfo {author}
  {\bibfnamefont {M.~A.}\ \bibnamefont {Norcia}}, \bibinfo {author}
  {\bibfnamefont {E.}~\bibnamefont {Oelker}}, \bibinfo {author} {\bibfnamefont
  {N.}~\bibnamefont {Schine}}, \bibinfo {author} {\bibfnamefont
  {J.}~\bibnamefont {Ye}}, \ and\ \bibinfo {author} {\bibfnamefont {A.~M.}\
  \bibnamefont {Kaufman}},\ }\bibfield  {title} {\enquote {\bibinfo {title}
  {Half-minute-scale atomic coherence and high relative stability in a tweezer
  clock},}\ }\href@noop {} {\bibfield  {journal} {\bibinfo  {journal} {Nature}\
  }\textbf {\bibinfo {volume} {588}},\ \bibinfo {pages} {408--413} (\bibinfo
  {year} {2020})}\BibitemShut {NoStop}%
\bibitem [{\citenamefont {Ludlow}\ \emph {et~al.}(2015)\citenamefont {Ludlow},
  \citenamefont {Boyd}, \citenamefont {Ye}, \citenamefont {Peik},\ and\
  \citenamefont {Schmidt}}]{Ludlow2015}%
  \BibitemOpen
  \bibfield  {author} {\bibinfo {author} {\bibfnamefont {A.~D.}\ \bibnamefont
  {Ludlow}}, \bibinfo {author} {\bibfnamefont {M.~M.}\ \bibnamefont {Boyd}},
  \bibinfo {author} {\bibfnamefont {J.}~\bibnamefont {Ye}}, \bibinfo {author}
  {\bibfnamefont {E.}~\bibnamefont {Peik}}, \ and\ \bibinfo {author}
  {\bibfnamefont {P.~O.}\ \bibnamefont {Schmidt}},\ }\bibfield  {title}
  {\enquote {\bibinfo {title} {Optical atomic clocks},}\ }\href {\doibase
  10.1103/RevModPhys.87.637} {\bibfield  {journal} {\bibinfo  {journal} {Rev.
  Mod. Phys.}\ }\textbf {\bibinfo {volume} {87}},\ \bibinfo {pages} {637--701}
  (\bibinfo {year} {2015})}\BibitemShut {NoStop}%
\bibitem [{\citenamefont {Ma}\ \emph {et~al.}(2011)\citenamefont {Ma},
  \citenamefont {Wang}, \citenamefont {Sun},\ and\ \citenamefont
  {Nori}}]{ma2011quantum}%
  \BibitemOpen
  \bibfield  {author} {\bibinfo {author} {\bibfnamefont {J.}~\bibnamefont
  {Ma}}, \bibinfo {author} {\bibfnamefont {X.}~\bibnamefont {Wang}}, \bibinfo
  {author} {\bibfnamefont {C.-P.}\ \bibnamefont {Sun}}, \ and\ \bibinfo
  {author} {\bibfnamefont {F.}~\bibnamefont {Nori}},\ }\bibfield  {title}
  {\enquote {\bibinfo {title} {Quantum spin squeezing},}\ }\href@noop {}
  {\bibfield  {journal} {\bibinfo  {journal} {Phys. Rep.}\ }\textbf {\bibinfo
  {volume} {509}},\ \bibinfo {pages} {89--165} (\bibinfo {year}
  {2011})}\BibitemShut {NoStop}%
\bibitem [{\citenamefont {Pezz{\`e}}\ \emph {et~al.}(2018)\citenamefont
  {Pezz{\`e}}, \citenamefont {Smerzi}, \citenamefont {Oberthaler},
  \citenamefont {Schmied},\ and\ \citenamefont {Treutlein}}]{Pezze2018}%
  \BibitemOpen
  \bibfield  {author} {\bibinfo {author} {\bibfnamefont {L.}~\bibnamefont
  {Pezz{\`e}}}, \bibinfo {author} {\bibfnamefont {A.}~\bibnamefont {Smerzi}},
  \bibinfo {author} {\bibfnamefont {M.~K.}\ \bibnamefont {Oberthaler}},
  \bibinfo {author} {\bibfnamefont {R.}~\bibnamefont {Schmied}}, \ and\
  \bibinfo {author} {\bibfnamefont {P.}~\bibnamefont {Treutlein}},\ }\bibfield
  {title} {\enquote {\bibinfo {title} {Quantum metrology with nonclassical
  states of atomic ensembles},}\ }\href@noop {} {\bibfield  {journal} {\bibinfo
   {journal} {Rev. Mod. Phys.}\ }\textbf {\bibinfo {volume} {90}},\ \bibinfo
  {pages} {035005} (\bibinfo {year} {2018})}\BibitemShut {NoStop}%
\bibitem [{\citenamefont {Wineland}\ \emph {et~al.}(1992)\citenamefont
  {Wineland}, \citenamefont {Bollinger}, \citenamefont {Itano}, \citenamefont
  {Moore},\ and\ \citenamefont {Heinzen}}]{wineland1992a}%
  \BibitemOpen
  \bibfield  {author} {\bibinfo {author} {\bibfnamefont {D.~J.}\ \bibnamefont
  {Wineland}}, \bibinfo {author} {\bibfnamefont {J.~J.}\ \bibnamefont
  {Bollinger}}, \bibinfo {author} {\bibfnamefont {W.~M.}\ \bibnamefont
  {Itano}}, \bibinfo {author} {\bibfnamefont {F.~L.}\ \bibnamefont {Moore}}, \
  and\ \bibinfo {author} {\bibfnamefont {D.~J.}\ \bibnamefont {Heinzen}},\
  }\bibfield  {title} {\enquote {\bibinfo {title} {Spin squeezing and reduced
  quantum noise in spectroscopy},}\ }\href {\doibase 10.1103/PhysRevA.46.R6797}
  {\bibfield  {journal} {\bibinfo  {journal} {Phys. Rev. A}\ }\textbf {\bibinfo
  {volume} {46}},\ \bibinfo {pages} {R6797--R6800} (\bibinfo {year}
  {1992})}\BibitemShut {NoStop}%
\bibitem [{\citenamefont {Wineland}\ \emph {et~al.}(1994)\citenamefont
  {Wineland}, \citenamefont {Bollinger}, \citenamefont {Itano},\ and\
  \citenamefont {Heinzen}}]{Wineland1994}%
  \BibitemOpen
  \bibfield  {author} {\bibinfo {author} {\bibfnamefont {D.~J.}\ \bibnamefont
  {Wineland}}, \bibinfo {author} {\bibfnamefont {J.~J.}\ \bibnamefont
  {Bollinger}}, \bibinfo {author} {\bibfnamefont {W.~M.}\ \bibnamefont
  {Itano}}, \ and\ \bibinfo {author} {\bibfnamefont {D.~J.}\ \bibnamefont
  {Heinzen}},\ }\bibfield  {title} {\enquote {\bibinfo {title} {Squeezed atomic
  states and projection noise in spectroscopy},}\ }\href {\doibase
  10.1103/PhysRevA.50.67} {\bibfield  {journal} {\bibinfo  {journal} {Phys.
  Rev. A}\ }\textbf {\bibinfo {volume} {50}},\ \bibinfo {pages} {67--88}
  (\bibinfo {year} {1994})}\BibitemShut {NoStop}%
\bibitem [{\citenamefont {Kitagawa}\ and\ \citenamefont
  {Ueda}(1993)}]{Kitagawa1993}%
  \BibitemOpen
  \bibfield  {author} {\bibinfo {author} {\bibfnamefont {M.}~\bibnamefont
  {Kitagawa}}\ and\ \bibinfo {author} {\bibfnamefont {M.}~\bibnamefont
  {Ueda}},\ }\bibfield  {title} {\enquote {\bibinfo {title} {Squeezed spin
  states},}\ }\href {\doibase 10.1103/PhysRevA.47.5138} {\bibfield  {journal}
  {\bibinfo  {journal} {Phys. Rev. A}\ }\textbf {\bibinfo {volume} {47}},\
  \bibinfo {pages} {5138--5143} (\bibinfo {year} {1993})}\BibitemShut {NoStop}%
\bibitem [{\citenamefont {Kuzmich}, \citenamefont {Bigelow},\ and\
  \citenamefont {Mandel}(1998)}]{kuzmich1998atomic}%
  \BibitemOpen
  \bibfield  {author} {\bibinfo {author} {\bibfnamefont {A.}~\bibnamefont
  {Kuzmich}}, \bibinfo {author} {\bibfnamefont {N.}~\bibnamefont {Bigelow}}, \
  and\ \bibinfo {author} {\bibfnamefont {L.}~\bibnamefont {Mandel}},\
  }\bibfield  {title} {\enquote {\bibinfo {title} {Atomic quantum
  non-demolition measurements and squeezing},}\ }\href@noop {} {\bibfield
  {journal} {\bibinfo  {journal} {EPL (Europhysics Letters)}\ }\textbf
  {\bibinfo {volume} {42}},\ \bibinfo {pages} {481} (\bibinfo {year}
  {1998})}\BibitemShut {NoStop}%
\bibitem [{\citenamefont {Appel}\ \emph {et~al.}(2009)\citenamefont {Appel},
  \citenamefont {Windpassinger}, \citenamefont {Oblak}, \citenamefont {Hoff},
  \citenamefont {Kj{\ae}rgaard},\ and\ \citenamefont {Polzik}}]{Appel2009}%
  \BibitemOpen
  \bibfield  {author} {\bibinfo {author} {\bibfnamefont {J.}~\bibnamefont
  {Appel}}, \bibinfo {author} {\bibfnamefont {P.~J.}\ \bibnamefont
  {Windpassinger}}, \bibinfo {author} {\bibfnamefont {D.}~\bibnamefont
  {Oblak}}, \bibinfo {author} {\bibfnamefont {U.~B.}\ \bibnamefont {Hoff}},
  \bibinfo {author} {\bibfnamefont {N.}~\bibnamefont {Kj{\ae}rgaard}}, \ and\
  \bibinfo {author} {\bibfnamefont {E.~S.}\ \bibnamefont {Polzik}},\ }\bibfield
   {title} {\enquote {\bibinfo {title} {Mesoscopic atomic entanglement for
  precision measurements beyond the standard quantum limit},}\ }\href
  {http://www.pnas.org/content/106/27/10960.abstract} {\bibfield  {journal}
  {\bibinfo  {journal} {Proc. Natl. Acad. Sci. U.S.A.}\ }\textbf {\bibinfo
  {volume} {106}},\ \bibinfo {pages} {10960--10965} (\bibinfo {year}
  {2009})}\BibitemShut {NoStop}%
\bibitem [{\citenamefont {Schleier-Smith}, \citenamefont {Leroux},\ and\
  \citenamefont {Vuleti\'{c}}(2010{\natexlab{a}})}]{Schleier-Smith2010}%
  \BibitemOpen
  \bibfield  {author} {\bibinfo {author} {\bibfnamefont {M.~H.}\ \bibnamefont
  {Schleier-Smith}}, \bibinfo {author} {\bibfnamefont {I.~D.}\ \bibnamefont
  {Leroux}}, \ and\ \bibinfo {author} {\bibfnamefont {V.}~\bibnamefont
  {Vuleti\'{c}}},\ }\bibfield  {title} {\enquote {\bibinfo {title} {Squeezing
  the collective spin of a dilute atomic ensemble by cavity feedback},}\ }\href
  {\doibase 10.1103/PhysRevA.81.021804} {\bibfield  {journal} {\bibinfo
  {journal} {Phys. Rev. A}\ }\textbf {\bibinfo {volume} {81}},\ \bibinfo
  {pages} {021804(R)} (\bibinfo {year} {2010}{\natexlab{a}})}\BibitemShut
  {NoStop}%
\bibitem [{\citenamefont {Hosten}\ \emph
  {et~al.}(2016{\natexlab{a}})\citenamefont {Hosten}, \citenamefont {Engelsen},
  \citenamefont {Krishnakumar},\ and\ \citenamefont {Kasevich}}]{Hosten2016}%
  \BibitemOpen
  \bibfield  {author} {\bibinfo {author} {\bibfnamefont {O.}~\bibnamefont
  {Hosten}}, \bibinfo {author} {\bibfnamefont {N.~J.}\ \bibnamefont
  {Engelsen}}, \bibinfo {author} {\bibfnamefont {R.}~\bibnamefont
  {Krishnakumar}}, \ and\ \bibinfo {author} {\bibfnamefont {M.~A.}\
  \bibnamefont {Kasevich}},\ }\bibfield  {title} {\enquote {\bibinfo {title}
  {Measurement noise 100 times lower than the quantum-projection limit using
  entangled atoms},}\ }\href {https://doi.org/10.1038/nature16176} {\bibfield
  {journal} {\bibinfo  {journal} {Nature (London)}\ }\textbf {\bibinfo {volume}
  {529}},\ \bibinfo {pages} {505--508} (\bibinfo {year}
  {2016}{\natexlab{a}})}\BibitemShut {NoStop}%
\bibitem [{\citenamefont {Cox}\ \emph {et~al.}(2016)\citenamefont {Cox},
  \citenamefont {Greve}, \citenamefont {Weiner},\ and\ \citenamefont
  {Thompson}}]{Cox2016a}%
  \BibitemOpen
  \bibfield  {author} {\bibinfo {author} {\bibfnamefont {K.~C.}\ \bibnamefont
  {Cox}}, \bibinfo {author} {\bibfnamefont {G.~P.}\ \bibnamefont {Greve}},
  \bibinfo {author} {\bibfnamefont {J.~M.}\ \bibnamefont {Weiner}}, \ and\
  \bibinfo {author} {\bibfnamefont {J.~K.}\ \bibnamefont {Thompson}},\
  }\bibfield  {title} {\enquote {\bibinfo {title} {Deterministic squeezed
  states with collective measurements and feedback},}\ }\href {\doibase
  10.1103/PhysRevLett.116.093602} {\bibfield  {journal} {\bibinfo  {journal}
  {Phys. Rev. Lett.}\ }\textbf {\bibinfo {volume} {116}},\ \bibinfo {pages}
  {093602} (\bibinfo {year} {2016})}\BibitemShut {NoStop}%
\bibitem [{\citenamefont {Malia}\ \emph {et~al.}(2022)\citenamefont {Malia},
  \citenamefont {Wu}, \citenamefont {Mart{\'\i}nez-Rinc{\'o}n},\ and\
  \citenamefont {Kasevich}}]{malia2022distributed}%
  \BibitemOpen
  \bibfield  {author} {\bibinfo {author} {\bibfnamefont {B.~K.}\ \bibnamefont
  {Malia}}, \bibinfo {author} {\bibfnamefont {Y.}~\bibnamefont {Wu}}, \bibinfo
  {author} {\bibfnamefont {J.}~\bibnamefont {Mart{\'\i}nez-Rinc{\'o}n}}, \ and\
  \bibinfo {author} {\bibfnamefont {M.~A.}\ \bibnamefont {Kasevich}},\
  }\bibfield  {title} {\enquote {\bibinfo {title} {Distributed quantum sensing
  with a mode-entangled network of spin-squeezed atomic states},}\ }\href@noop
  {} {\bibfield  {journal} {\bibinfo  {journal} {arXiv preprint
  arXiv:2205.06382}\ } (\bibinfo {year} {2022})}\BibitemShut {NoStop}%
\bibitem [{\citenamefont {Stadnik}\ and\ \citenamefont
  {Flambaum}(2014)}]{stadnik2014axion}%
  \BibitemOpen
  \bibfield  {author} {\bibinfo {author} {\bibfnamefont {Y.}~\bibnamefont
  {Stadnik}}\ and\ \bibinfo {author} {\bibfnamefont {V.}~\bibnamefont
  {Flambaum}},\ }\bibfield  {title} {\enquote {\bibinfo {title} {Axion-induced
  effects in atoms, molecules, and nuclei: Parity nonconservation, anapole
  moments, electric dipole moments, and spin-gravity and spin-axion momentum
  couplings},}\ }\href@noop {} {\bibfield  {journal} {\bibinfo  {journal}
  {Physical Review D}\ }\textbf {\bibinfo {volume} {89}},\ \bibinfo {pages}
  {043522} (\bibinfo {year} {2014})}\BibitemShut {NoStop}%
\bibitem [{\citenamefont {Smorra}\ \emph {et~al.}(2019)\citenamefont {Smorra},
  \citenamefont {Stadnik}, \citenamefont {Blessing}, \citenamefont {Bohman},
  \citenamefont {Borchert}, \citenamefont {Devlin}, \citenamefont {Erlewein},
  \citenamefont {Harrington}, \citenamefont {Higuchi}, \citenamefont {Mooser}
  \emph {et~al.}}]{smorra2019direct}%
  \BibitemOpen
  \bibfield  {author} {\bibinfo {author} {\bibfnamefont {C.}~\bibnamefont
  {Smorra}}, \bibinfo {author} {\bibfnamefont {Y.}~\bibnamefont {Stadnik}},
  \bibinfo {author} {\bibfnamefont {P.}~\bibnamefont {Blessing}}, \bibinfo
  {author} {\bibfnamefont {M.}~\bibnamefont {Bohman}}, \bibinfo {author}
  {\bibfnamefont {M.}~\bibnamefont {Borchert}}, \bibinfo {author}
  {\bibfnamefont {J.}~\bibnamefont {Devlin}}, \bibinfo {author} {\bibfnamefont
  {S.}~\bibnamefont {Erlewein}}, \bibinfo {author} {\bibfnamefont
  {J.}~\bibnamefont {Harrington}}, \bibinfo {author} {\bibfnamefont
  {T.}~\bibnamefont {Higuchi}}, \bibinfo {author} {\bibfnamefont
  {A.}~\bibnamefont {Mooser}},  \emph {et~al.},\ }\bibfield  {title} {\enquote
  {\bibinfo {title} {Direct limits on the interaction of antiprotons with
  axion-like dark matter},}\ }\href@noop {} {\bibfield  {journal} {\bibinfo
  {journal} {Nature}\ }\textbf {\bibinfo {volume} {575}},\ \bibinfo {pages}
  {310--314} (\bibinfo {year} {2019})}\BibitemShut {NoStop}%
\bibitem [{\citenamefont {Pedrozo-Pe\~{n}afiel}\ \emph
  {et~al.}(2020)\citenamefont {Pedrozo-Pe\~{n}afiel}, \citenamefont {Colombo},
  \citenamefont {Shu}, \citenamefont {Adiyatullin}, \citenamefont {Li},
  \citenamefont {Mendez}, \citenamefont {Braverman}, \citenamefont {Kawasaki},
  \citenamefont {Akamatsu}, \citenamefont {Xiao},\ and\ \citenamefont
  {Vuleti\'{c}}}]{Pedrozo2020entanglement}%
  \BibitemOpen
  \bibfield  {author} {\bibinfo {author} {\bibfnamefont {E.}~\bibnamefont
  {Pedrozo-Pe\~{n}afiel}}, \bibinfo {author} {\bibfnamefont {S.}~\bibnamefont
  {Colombo}}, \bibinfo {author} {\bibfnamefont {C.}~\bibnamefont {Shu}},
  \bibinfo {author} {\bibfnamefont {A.~F.}\ \bibnamefont {Adiyatullin}},
  \bibinfo {author} {\bibfnamefont {Z.}~\bibnamefont {Li}}, \bibinfo {author}
  {\bibfnamefont {E.}~\bibnamefont {Mendez}}, \bibinfo {author} {\bibfnamefont
  {B.}~\bibnamefont {Braverman}}, \bibinfo {author} {\bibfnamefont
  {A.}~\bibnamefont {Kawasaki}}, \bibinfo {author} {\bibfnamefont
  {D.}~\bibnamefont {Akamatsu}}, \bibinfo {author} {\bibfnamefont
  {Y.}~\bibnamefont {Xiao}}, \ and\ \bibinfo {author} {\bibfnamefont
  {V.}~\bibnamefont {Vuleti\'{c}}},\ }\bibfield  {title} {\enquote {\bibinfo
  {title} {Entanglement on an optical atomic-clock transition},}\ }\href
  {https://doi.org/10.1038/s41586-020-3006-1} {\bibfield  {journal} {\bibinfo
  {journal} {Nature (London)}\ }\textbf {\bibinfo {volume} {588}},\ \bibinfo
  {pages} {414--418} (\bibinfo {year} {2020})}\BibitemShut {NoStop}%
\bibitem [{\citenamefont {Zheng}\ \emph {et~al.}(2022)\citenamefont {Zheng},
  \citenamefont {Dolde}, \citenamefont {Lochab}, \citenamefont {Merriman},
  \citenamefont {Li},\ and\ \citenamefont {Kolkowitz}}]{zheng2022differential}%
  \BibitemOpen
  \bibfield  {author} {\bibinfo {author} {\bibfnamefont {X.}~\bibnamefont
  {Zheng}}, \bibinfo {author} {\bibfnamefont {J.}~\bibnamefont {Dolde}},
  \bibinfo {author} {\bibfnamefont {V.}~\bibnamefont {Lochab}}, \bibinfo
  {author} {\bibfnamefont {B.~N.}\ \bibnamefont {Merriman}}, \bibinfo {author}
  {\bibfnamefont {H.}~\bibnamefont {Li}}, \ and\ \bibinfo {author}
  {\bibfnamefont {S.}~\bibnamefont {Kolkowitz}},\ }\bibfield  {title} {\enquote
  {\bibinfo {title} {Differential clock comparisons with a multiplexed optical
  lattice clock},}\ }\href@noop {} {\bibfield  {journal} {\bibinfo  {journal}
  {Nature}\ }\textbf {\bibinfo {volume} {602}},\ \bibinfo {pages} {425--430}
  (\bibinfo {year} {2022})}\BibitemShut {NoStop}%
\bibitem [{\citenamefont {Bishof}\ \emph {et~al.}(2013)\citenamefont {Bishof},
  \citenamefont {Zhang}, \citenamefont {Martin},\ and\ \citenamefont
  {Ye}}]{bishof2013optical}%
  \BibitemOpen
  \bibfield  {author} {\bibinfo {author} {\bibfnamefont {M.}~\bibnamefont
  {Bishof}}, \bibinfo {author} {\bibfnamefont {X.}~\bibnamefont {Zhang}},
  \bibinfo {author} {\bibfnamefont {M.~J.}\ \bibnamefont {Martin}}, \ and\
  \bibinfo {author} {\bibfnamefont {J.}~\bibnamefont {Ye}},\ }\bibfield
  {title} {\enquote {\bibinfo {title} {Optical spectrum analyzer with
  quantum-limited noise floor},}\ }\href@noop {} {\bibfield  {journal}
  {\bibinfo  {journal} {Physical review letters}\ }\textbf {\bibinfo {volume}
  {111}},\ \bibinfo {pages} {093604} (\bibinfo {year} {2013})}\BibitemShut
  {NoStop}%
\bibitem [{\citenamefont {Pezzé}\ and\ \citenamefont
  {Smerzi}(2009)}]{Pezze2009}%
  \BibitemOpen
  \bibfield  {author} {\bibinfo {author} {\bibfnamefont {L.}~\bibnamefont
  {Pezzé}}\ and\ \bibinfo {author} {\bibfnamefont {A.}~\bibnamefont
  {Smerzi}},\ }\bibfield  {title} {\enquote {\bibinfo {title} {{Entanglement,
  Nonlinear Dynamics, and the Heisenberg Limit}},}\ }\href {\doibase
  10.1103/PhysRevLett.102.100401} {\bibfield  {journal} {\bibinfo  {journal}
  {Phys. Rev. Lett.}\ }\textbf {\bibinfo {volume} {102}},\ \bibinfo {pages}
  {100401} (\bibinfo {year} {2009})}\BibitemShut {NoStop}%
\bibitem [{\citenamefont {Hyllus}\ \emph {et~al.}(2012)\citenamefont {Hyllus},
  \citenamefont {Laskowski}, \citenamefont {Krischek}, \citenamefont
  {Schwemmer}, \citenamefont {Wieczorek}, \citenamefont {Weinfurter},
  \citenamefont {Pezz{\'e}},\ and\ \citenamefont {Smerzi}}]{hyllus2012fisher}%
  \BibitemOpen
  \bibfield  {author} {\bibinfo {author} {\bibfnamefont {P.}~\bibnamefont
  {Hyllus}}, \bibinfo {author} {\bibfnamefont {W.}~\bibnamefont {Laskowski}},
  \bibinfo {author} {\bibfnamefont {R.}~\bibnamefont {Krischek}}, \bibinfo
  {author} {\bibfnamefont {C.}~\bibnamefont {Schwemmer}}, \bibinfo {author}
  {\bibfnamefont {W.}~\bibnamefont {Wieczorek}}, \bibinfo {author}
  {\bibfnamefont {H.}~\bibnamefont {Weinfurter}}, \bibinfo {author}
  {\bibfnamefont {L.}~\bibnamefont {Pezz{\'e}}}, \ and\ \bibinfo {author}
  {\bibfnamefont {A.}~\bibnamefont {Smerzi}},\ }\bibfield  {title} {\enquote
  {\bibinfo {title} {Fisher information and multiparticle entanglement},}\
  }\href@noop {} {\bibfield  {journal} {\bibinfo  {journal} {Physical Review
  A}\ }\textbf {\bibinfo {volume} {85}},\ \bibinfo {pages} {022321} (\bibinfo
  {year} {2012})}\BibitemShut {NoStop}%
\bibitem [{\citenamefont {Strobel}\ \emph {et~al.}(2014)\citenamefont
  {Strobel}, \citenamefont {Muessel}, \citenamefont {Linnemann}, \citenamefont
  {Zibold}, \citenamefont {Hume}, \citenamefont {Pezz\'{e}}, \citenamefont
  {Smerzi},\ and\ \citenamefont {Oberthaler}}]{Strobel2014}%
  \BibitemOpen
  \bibfield  {author} {\bibinfo {author} {\bibfnamefont {H.}~\bibnamefont
  {Strobel}}, \bibinfo {author} {\bibfnamefont {W.}~\bibnamefont {Muessel}},
  \bibinfo {author} {\bibfnamefont {D.}~\bibnamefont {Linnemann}}, \bibinfo
  {author} {\bibfnamefont {T.}~\bibnamefont {Zibold}}, \bibinfo {author}
  {\bibfnamefont {D.~B.}\ \bibnamefont {Hume}}, \bibinfo {author}
  {\bibfnamefont {L.}~\bibnamefont {Pezz\'{e}}}, \bibinfo {author}
  {\bibfnamefont {A.}~\bibnamefont {Smerzi}}, \ and\ \bibinfo {author}
  {\bibfnamefont {M.~K.}\ \bibnamefont {Oberthaler}},\ }\bibfield  {title}
  {\enquote {\bibinfo {title} {Fisher information and entanglement of
  {non-Gaussian} spin states},}\ }\href {\doibase 10.1126/science.1250147}
  {\bibfield  {journal} {\bibinfo  {journal} {Science}\ }\textbf {\bibinfo
  {volume} {345}},\ \bibinfo {pages} {424--427} (\bibinfo {year} {2014})},\
  \Eprint
  {http://arxiv.org/abs/https://science.sciencemag.org/content/345/6195/424.full.pdf}
  {https://science.sciencemag.org/content/345/6195/424.full.pdf} \BibitemShut
  {NoStop}%
\bibitem [{\citenamefont {Krischek}\ \emph {et~al.}(2011)\citenamefont
  {Krischek}, \citenamefont {Schwemmer}, \citenamefont {Wieczorek},
  \citenamefont {Weinfurter}, \citenamefont {Hyllus}, \citenamefont {Pezz\'e},\
  and\ \citenamefont {Smerzi}}]{Krischek2011}%
  \BibitemOpen
  \bibfield  {author} {\bibinfo {author} {\bibfnamefont {R.}~\bibnamefont
  {Krischek}}, \bibinfo {author} {\bibfnamefont {C.}~\bibnamefont {Schwemmer}},
  \bibinfo {author} {\bibfnamefont {W.}~\bibnamefont {Wieczorek}}, \bibinfo
  {author} {\bibfnamefont {H.}~\bibnamefont {Weinfurter}}, \bibinfo {author}
  {\bibfnamefont {P.}~\bibnamefont {Hyllus}}, \bibinfo {author} {\bibfnamefont
  {L.}~\bibnamefont {Pezz\'e}}, \ and\ \bibinfo {author} {\bibfnamefont
  {A.}~\bibnamefont {Smerzi}},\ }\bibfield  {title} {\enquote {\bibinfo {title}
  {Useful multiparticle entanglement and sub-shot-noise sensitivity in
  experimental phase estimation},}\ }\href {\doibase
  10.1103/PhysRevLett.107.080504} {\bibfield  {journal} {\bibinfo  {journal}
  {Phys. Rev. Lett.}\ }\textbf {\bibinfo {volume} {107}},\ \bibinfo {pages}
  {080504} (\bibinfo {year} {2011})}\BibitemShut {NoStop}%
\bibitem [{\citenamefont {Monroe}\ \emph {et~al.}(1995)\citenamefont {Monroe},
  \citenamefont {Meekhof}, \citenamefont {King}, \citenamefont {Itano},\ and\
  \citenamefont {Wineland}}]{monroe1995demonstration}%
  \BibitemOpen
  \bibfield  {author} {\bibinfo {author} {\bibfnamefont {C.}~\bibnamefont
  {Monroe}}, \bibinfo {author} {\bibfnamefont {D.~M.}\ \bibnamefont {Meekhof}},
  \bibinfo {author} {\bibfnamefont {B.~E.}\ \bibnamefont {King}}, \bibinfo
  {author} {\bibfnamefont {W.~M.}\ \bibnamefont {Itano}}, \ and\ \bibinfo
  {author} {\bibfnamefont {D.~J.}\ \bibnamefont {Wineland}},\ }\bibfield
  {title} {\enquote {\bibinfo {title} {Demonstration of a fundamental quantum
  logic gate},}\ }\href@noop {} {\bibfield  {journal} {\bibinfo  {journal}
  {Physical review letters}\ }\textbf {\bibinfo {volume} {75}},\ \bibinfo
  {pages} {4714} (\bibinfo {year} {1995})}\BibitemShut {NoStop}%
\bibitem [{\citenamefont {Riedel}\ \emph {et~al.}(2010)\citenamefont {Riedel},
  \citenamefont {B{\"o}hi}, \citenamefont {Li}, \citenamefont {H{\"a}nsch},
  \citenamefont {Sinatra},\ and\ \citenamefont {Treutlein}}]{Riedel2010}%
  \BibitemOpen
  \bibfield  {author} {\bibinfo {author} {\bibfnamefont {M.~F.}\ \bibnamefont
  {Riedel}}, \bibinfo {author} {\bibfnamefont {P.}~\bibnamefont {B{\"o}hi}},
  \bibinfo {author} {\bibfnamefont {Y.}~\bibnamefont {Li}}, \bibinfo {author}
  {\bibfnamefont {T.~W.}\ \bibnamefont {H{\"a}nsch}}, \bibinfo {author}
  {\bibfnamefont {A.}~\bibnamefont {Sinatra}}, \ and\ \bibinfo {author}
  {\bibfnamefont {P.}~\bibnamefont {Treutlein}},\ }\bibfield  {title} {\enquote
  {\bibinfo {title} {Atom-chip-based generation of entanglement for quantum
  metrology},}\ }\href {https://doi.org/10.1038/nature08988} {\bibfield
  {journal} {\bibinfo  {journal} {Nature (London)}\ }\textbf {\bibinfo {volume}
  {464}},\ \bibinfo {pages} {1170} (\bibinfo {year} {2010})}\BibitemShut
  {NoStop}%
\bibitem [{\citenamefont {Bohnet}\ \emph {et~al.}(2016)\citenamefont {Bohnet},
  \citenamefont {Sawyer}, \citenamefont {Britton}, \citenamefont {Wall},
  \citenamefont {Rey}, \citenamefont {Foss-Feig},\ and\ \citenamefont
  {Bollinger}}]{Bohnet2016}%
  \BibitemOpen
  \bibfield  {author} {\bibinfo {author} {\bibfnamefont {J.~G.}\ \bibnamefont
  {Bohnet}}, \bibinfo {author} {\bibfnamefont {B.~C.}\ \bibnamefont {Sawyer}},
  \bibinfo {author} {\bibfnamefont {J.~W.}\ \bibnamefont {Britton}}, \bibinfo
  {author} {\bibfnamefont {M.~L.}\ \bibnamefont {Wall}}, \bibinfo {author}
  {\bibfnamefont {A.~M.}\ \bibnamefont {Rey}}, \bibinfo {author} {\bibfnamefont
  {M.}~\bibnamefont {Foss-Feig}}, \ and\ \bibinfo {author} {\bibfnamefont
  {J.~J.}\ \bibnamefont {Bollinger}},\ }\bibfield  {title} {\enquote {\bibinfo
  {title} {Quantum spin dynamics and entanglement generation with hundreds of
  trapped ions},}\ }\href
  {http://science.sciencemag.org/content/352/6291/1297.abstract} {\bibfield
  {journal} {\bibinfo  {journal} {Science}\ }\textbf {\bibinfo {volume}
  {352}},\ \bibinfo {pages} {1297} (\bibinfo {year} {2016})}\BibitemShut
  {NoStop}%
\bibitem [{\citenamefont {Takano}\ \emph {et~al.}(2009)\citenamefont {Takano},
  \citenamefont {Fuyama}, \citenamefont {Namiki},\ and\ \citenamefont
  {Takahashi}}]{Takano2009}%
  \BibitemOpen
  \bibfield  {author} {\bibinfo {author} {\bibfnamefont {T.}~\bibnamefont
  {Takano}}, \bibinfo {author} {\bibfnamefont {M.}~\bibnamefont {Fuyama}},
  \bibinfo {author} {\bibfnamefont {R.}~\bibnamefont {Namiki}}, \ and\ \bibinfo
  {author} {\bibfnamefont {Y.}~\bibnamefont {Takahashi}},\ }\bibfield  {title}
  {\enquote {\bibinfo {title} {Spin squeezing of a cold atomic ensemble with
  the nuclear spin of one-half},}\ }\href {\doibase
  10.1103/PhysRevLett.102.033601} {\bibfield  {journal} {\bibinfo  {journal}
  {Phys. Rev. Lett.}\ }\textbf {\bibinfo {volume} {102}},\ \bibinfo {pages}
  {033601} (\bibinfo {year} {2009})}\BibitemShut {NoStop}%
\bibitem [{\citenamefont {Leroux}, \citenamefont {Schleier-Smith},\ and\
  \citenamefont {Vuleti{\'c}}(2010)}]{Leroux2010}%
  \BibitemOpen
  \bibfield  {author} {\bibinfo {author} {\bibfnamefont {I.~D.}\ \bibnamefont
  {Leroux}}, \bibinfo {author} {\bibfnamefont {M.~H.}\ \bibnamefont
  {Schleier-Smith}}, \ and\ \bibinfo {author} {\bibfnamefont {V.}~\bibnamefont
  {Vuleti{\'c}}},\ }\bibfield  {title} {\enquote {\bibinfo {title}
  {Implementation of cavity squeezing of a collective atomic spin},}\ }\href
  {\doibase 10.1103/PhysRevLett.104.073602} {\bibfield  {journal} {\bibinfo
  {journal} {Phys. Rev. Lett.}\ }\textbf {\bibinfo {volume} {104}},\ \bibinfo
  {pages} {073602} (\bibinfo {year} {2010})}\BibitemShut {NoStop}%
\bibitem [{\citenamefont {Sewell}\ \emph {et~al.}(2012)\citenamefont {Sewell},
  \citenamefont {Koschorreck}, \citenamefont {Napolitano}, \citenamefont
  {Dubost}, \citenamefont {Behbood},\ and\ \citenamefont
  {Mitchell}}]{Sewell2012}%
  \BibitemOpen
  \bibfield  {author} {\bibinfo {author} {\bibfnamefont {R.~J.}\ \bibnamefont
  {Sewell}}, \bibinfo {author} {\bibfnamefont {M.}~\bibnamefont {Koschorreck}},
  \bibinfo {author} {\bibfnamefont {M.}~\bibnamefont {Napolitano}}, \bibinfo
  {author} {\bibfnamefont {B.}~\bibnamefont {Dubost}}, \bibinfo {author}
  {\bibfnamefont {N.}~\bibnamefont {Behbood}}, \ and\ \bibinfo {author}
  {\bibfnamefont {M.~W.}\ \bibnamefont {Mitchell}},\ }\bibfield  {title}
  {\enquote {\bibinfo {title} {Magnetic sensitivity beyond the projection noise
  limit by spin squeezing},}\ }\href {\doibase 10.1103/PhysRevLett.109.253605}
  {\bibfield  {journal} {\bibinfo  {journal} {Phys. Rev. Lett.}\ }\textbf
  {\bibinfo {volume} {109}},\ \bibinfo {pages} {253605} (\bibinfo {year}
  {2012})}\BibitemShut {NoStop}%
\bibitem [{\citenamefont {Hacker}\ \emph {et~al.}(2019)\citenamefont {Hacker},
  \citenamefont {Welte}, \citenamefont {Daiss}, \citenamefont {Shaukat},
  \citenamefont {Ritter}, \citenamefont {Li},\ and\ \citenamefont
  {Rempe}}]{hacker2019deterministic}%
  \BibitemOpen
  \bibfield  {author} {\bibinfo {author} {\bibfnamefont {B.}~\bibnamefont
  {Hacker}}, \bibinfo {author} {\bibfnamefont {S.}~\bibnamefont {Welte}},
  \bibinfo {author} {\bibfnamefont {S.}~\bibnamefont {Daiss}}, \bibinfo
  {author} {\bibfnamefont {A.}~\bibnamefont {Shaukat}}, \bibinfo {author}
  {\bibfnamefont {S.}~\bibnamefont {Ritter}}, \bibinfo {author} {\bibfnamefont
  {L.}~\bibnamefont {Li}}, \ and\ \bibinfo {author} {\bibfnamefont
  {G.}~\bibnamefont {Rempe}},\ }\bibfield  {title} {\enquote {\bibinfo {title}
  {Deterministic creation of entangled atom--light schr{\"o}dinger-cat
  states},}\ }\href@noop {} {\bibfield  {journal} {\bibinfo  {journal} {Nature
  Photonics}\ }\textbf {\bibinfo {volume} {13}},\ \bibinfo {pages} {110--115}
  (\bibinfo {year} {2019})}\BibitemShut {NoStop}%
\bibitem [{\citenamefont {McConnell}\ \emph {et~al.}(2015)\citenamefont
  {McConnell}, \citenamefont {Zhang}, \citenamefont {Hu}, \citenamefont
  {Ćuk},\ and\ \citenamefont {Vuleti\'{c}}}]{McConnell-Vuletic2015}%
  \BibitemOpen
  \bibfield  {author} {\bibinfo {author} {\bibfnamefont {R.}~\bibnamefont
  {McConnell}}, \bibinfo {author} {\bibfnamefont {H.}~\bibnamefont {Zhang}},
  \bibinfo {author} {\bibfnamefont {J.}~\bibnamefont {Hu}}, \bibinfo {author}
  {\bibfnamefont {S.}~\bibnamefont {Ćuk}}, \ and\ \bibinfo {author}
  {\bibfnamefont {V.}~\bibnamefont {Vuleti\'{c}}},\ }\bibfield  {title}
  {\enquote {\bibinfo {title} {Entanglement with negative {Wigner} function of
  almost 3,000 atoms heralded by one photon},}\ }\href
  {https://doi.org/10.1038/nature14293} {\bibfield  {journal} {\bibinfo
  {journal} {Nature (London)}\ }\textbf {\bibinfo {volume} {519}},\ \bibinfo
  {pages} {439--442} (\bibinfo {year} {2015})}\BibitemShut {NoStop}%
\bibitem [{\citenamefont {Braverman}\ \emph {et~al.}(2019)\citenamefont
  {Braverman}, \citenamefont {Kawasaki}, \citenamefont {Pedrozo-Pe{\~n}afiel},
  \citenamefont {Colombo}, \citenamefont {Shu}, \citenamefont {Li},
  \citenamefont {Mendez}, \citenamefont {Yamoah}, \citenamefont {Salvi},
  \citenamefont {Akamatsu} \emph {et~al.}}]{braverman2019near}%
  \BibitemOpen
  \bibfield  {author} {\bibinfo {author} {\bibfnamefont {B.}~\bibnamefont
  {Braverman}}, \bibinfo {author} {\bibfnamefont {A.}~\bibnamefont {Kawasaki}},
  \bibinfo {author} {\bibfnamefont {E.}~\bibnamefont {Pedrozo-Pe{\~n}afiel}},
  \bibinfo {author} {\bibfnamefont {S.}~\bibnamefont {Colombo}}, \bibinfo
  {author} {\bibfnamefont {C.}~\bibnamefont {Shu}}, \bibinfo {author}
  {\bibfnamefont {Z.}~\bibnamefont {Li}}, \bibinfo {author} {\bibfnamefont
  {E.}~\bibnamefont {Mendez}}, \bibinfo {author} {\bibfnamefont
  {M.}~\bibnamefont {Yamoah}}, \bibinfo {author} {\bibfnamefont
  {L.}~\bibnamefont {Salvi}}, \bibinfo {author} {\bibfnamefont
  {D.}~\bibnamefont {Akamatsu}},  \emph {et~al.},\ }\bibfield  {title}
  {\enquote {\bibinfo {title} {Near-unitary spin squeezing in {Yb}~171},}\
  }\href@noop {} {\bibfield  {journal} {\bibinfo  {journal} {Phys. Rev. Lett.}\
  }\textbf {\bibinfo {volume} {122}},\ \bibinfo {pages} {223203} (\bibinfo
  {year} {2019})}\BibitemShut {NoStop}%
\bibitem [{\citenamefont {Saffman}\ \emph {et~al.}(2009)\citenamefont
  {Saffman}, \citenamefont {Oblak}, \citenamefont {Appel},\ and\ \citenamefont
  {Polzik}}]{Saffman2009}%
  \BibitemOpen
  \bibfield  {author} {\bibinfo {author} {\bibfnamefont {M.}~\bibnamefont
  {Saffman}}, \bibinfo {author} {\bibfnamefont {D.}~\bibnamefont {Oblak}},
  \bibinfo {author} {\bibfnamefont {J.}~\bibnamefont {Appel}}, \ and\ \bibinfo
  {author} {\bibfnamefont {E.~S.}\ \bibnamefont {Polzik}},\ }\bibfield  {title}
  {\enquote {\bibinfo {title} {Spin squeezing of atomic ensembles by multicolor
  quantum nondemolition measurements},}\ }\href {\doibase
  10.1103/PhysRevA.79.023831} {\bibfield  {journal} {\bibinfo  {journal} {Phys.
  Rev. A}\ }\textbf {\bibinfo {volume} {79}},\ \bibinfo {pages} {023831}
  (\bibinfo {year} {2009})}\BibitemShut {NoStop}%
\bibitem [{\citenamefont {Trail}, \citenamefont {Jessen},\ and\ \citenamefont
  {Deutsch}(2010)}]{trail2010strongly}%
  \BibitemOpen
  \bibfield  {author} {\bibinfo {author} {\bibfnamefont {C.~M.}\ \bibnamefont
  {Trail}}, \bibinfo {author} {\bibfnamefont {P.~S.}\ \bibnamefont {Jessen}}, \
  and\ \bibinfo {author} {\bibfnamefont {I.~H.}\ \bibnamefont {Deutsch}},\
  }\bibfield  {title} {\enquote {\bibinfo {title} {Strongly enhanced spin
  squeezing via quantum control},}\ }\href@noop {} {\bibfield  {journal}
  {\bibinfo  {journal} {Physical review letters}\ }\textbf {\bibinfo {volume}
  {105}},\ \bibinfo {pages} {193602} (\bibinfo {year} {2010})}\BibitemShut
  {NoStop}%
\bibitem [{\citenamefont {Leroux}\ \emph {et~al.}(2012)\citenamefont {Leroux},
  \citenamefont {Schleier-Smith}, \citenamefont {Zhang},\ and\ \citenamefont
  {Vuleti\'{c}}}]{Leroux2012}%
  \BibitemOpen
  \bibfield  {author} {\bibinfo {author} {\bibfnamefont {I.~D.}\ \bibnamefont
  {Leroux}}, \bibinfo {author} {\bibfnamefont {M.~H.}\ \bibnamefont
  {Schleier-Smith}}, \bibinfo {author} {\bibfnamefont {H.}~\bibnamefont
  {Zhang}}, \ and\ \bibinfo {author} {\bibfnamefont {V.}~\bibnamefont
  {Vuleti\'{c}}},\ }\bibfield  {title} {\enquote {\bibinfo {title} {Unitary
  cavity spin squeezing by quantum erasure},}\ }\href@noop {} {\bibfield
  {journal} {\bibinfo  {journal} {Physical Review A}\ }\textbf {\bibinfo
  {volume} {85}},\ \bibinfo {pages} {013803} (\bibinfo {year}
  {2012})}\BibitemShut {NoStop}%
\bibitem [{\citenamefont {Zhang}\ \emph {et~al.}(2015)\citenamefont {Zhang},
  \citenamefont {Zou}, \citenamefont {Zou}, \citenamefont {Jiang},\ and\
  \citenamefont {Guo}}]{Zhang2015}%
  \BibitemOpen
  \bibfield  {author} {\bibinfo {author} {\bibfnamefont {Y.-L.}\ \bibnamefont
  {Zhang}}, \bibinfo {author} {\bibfnamefont {C.-L.}\ \bibnamefont {Zou}},
  \bibinfo {author} {\bibfnamefont {X.-B.}\ \bibnamefont {Zou}}, \bibinfo
  {author} {\bibfnamefont {L.}~\bibnamefont {Jiang}}, \ and\ \bibinfo {author}
  {\bibfnamefont {G.-C.}\ \bibnamefont {Guo}},\ }\bibfield  {title} {\enquote
  {\bibinfo {title} {Detuning-enhanced cavity spin squeezing},}\ }\href
  {\doibase 10.1103/PhysRevA.91.033625} {\bibfield  {journal} {\bibinfo
  {journal} {Phys. Rev. A}\ }\textbf {\bibinfo {volume} {91}},\ \bibinfo
  {pages} {033625} (\bibinfo {year} {2015})}\BibitemShut {NoStop}%
\bibitem [{\citenamefont {Li}\ \emph {et~al.}(2022{\natexlab{a}})\citenamefont
  {Li}, \citenamefont {Braverman}, \citenamefont {Colombo}, \citenamefont
  {Shu}, \citenamefont {Kawasaki}, \citenamefont {Adiyatullin}, \citenamefont
  {Pedrozo-Pe\~nafiel}, \citenamefont {Mendez},\ and\ \citenamefont
  {Vuleti\ifmmode~\acute{c}\else \'{c}\fi{}}}]{li2021collective}%
  \BibitemOpen
  \bibfield  {author} {\bibinfo {author} {\bibfnamefont {Z.}~\bibnamefont
  {Li}}, \bibinfo {author} {\bibfnamefont {B.}~\bibnamefont {Braverman}},
  \bibinfo {author} {\bibfnamefont {S.}~\bibnamefont {Colombo}}, \bibinfo
  {author} {\bibfnamefont {C.}~\bibnamefont {Shu}}, \bibinfo {author}
  {\bibfnamefont {A.}~\bibnamefont {Kawasaki}}, \bibinfo {author}
  {\bibfnamefont {A.~F.}\ \bibnamefont {Adiyatullin}}, \bibinfo {author}
  {\bibfnamefont {E.}~\bibnamefont {Pedrozo-Pe\~nafiel}}, \bibinfo {author}
  {\bibfnamefont {E.}~\bibnamefont {Mendez}}, \ and\ \bibinfo {author}
  {\bibfnamefont {V.}~\bibnamefont {Vuleti\ifmmode~\acute{c}\else
  \'{c}\fi{}}},\ }\bibfield  {title} {\enquote {\bibinfo {title} {Collective
  spin-light and light-mediated spin-spin interactions in an optical cavity},}\
  }\href {\doibase 10.1103/PRXQuantum.3.020308} {\bibfield  {journal} {\bibinfo
   {journal} {PRX Quantum}\ }\textbf {\bibinfo {volume} {3}},\ \bibinfo {pages}
  {020308} (\bibinfo {year} {2022}{\natexlab{a}})}\BibitemShut {NoStop}%
\bibitem [{\citenamefont {L\"{u}cke}\ \emph {et~al.}(2011)\citenamefont
  {L\"{u}cke}, \citenamefont {Scherer}, \citenamefont {Kruse}, \citenamefont
  {Pezz\'{e}}, \citenamefont {Deuretzbacher}, \citenamefont {Hyllus},
  \citenamefont {Topic}, \citenamefont {Peise}, \citenamefont {Ertmer},
  \citenamefont {Arlt}, \citenamefont {Santos}, \citenamefont {Smerzi},\ and\
  \citenamefont {Klempt}}]{Lucke2016}%
  \BibitemOpen
  \bibfield  {author} {\bibinfo {author} {\bibfnamefont {B.}~\bibnamefont
  {L\"{u}cke}}, \bibinfo {author} {\bibfnamefont {M.}~\bibnamefont {Scherer}},
  \bibinfo {author} {\bibfnamefont {J.}~\bibnamefont {Kruse}}, \bibinfo
  {author} {\bibfnamefont {L.}~\bibnamefont {Pezz\'{e}}}, \bibinfo {author}
  {\bibfnamefont {F.}~\bibnamefont {Deuretzbacher}}, \bibinfo {author}
  {\bibfnamefont {P.}~\bibnamefont {Hyllus}}, \bibinfo {author} {\bibfnamefont
  {O.}~\bibnamefont {Topic}}, \bibinfo {author} {\bibfnamefont
  {J.}~\bibnamefont {Peise}}, \bibinfo {author} {\bibfnamefont
  {W.}~\bibnamefont {Ertmer}}, \bibinfo {author} {\bibfnamefont
  {J.}~\bibnamefont {Arlt}}, \bibinfo {author} {\bibfnamefont {L.}~\bibnamefont
  {Santos}}, \bibinfo {author} {\bibfnamefont {A.}~\bibnamefont {Smerzi}}, \
  and\ \bibinfo {author} {\bibfnamefont {C.}~\bibnamefont {Klempt}},\
  }\bibfield  {title} {\enquote {\bibinfo {title} {Twin matter waves for
  interferometry beyond the classical limit},}\ }\href {\doibase
  10.1126/science.1208798} {\bibfield  {journal} {\bibinfo  {journal}
  {Science}\ }\textbf {\bibinfo {volume} {334}},\ \bibinfo {pages} {773--776}
  (\bibinfo {year} {2011})},\ \Eprint
  {http://arxiv.org/abs/https://science.sciencemag.org/content/sci/334/6057/773.full.pdf}
  {https://science.sciencemag.org/content/sci/334/6057/773.full.pdf}
  \BibitemShut {NoStop}%
\bibitem [{\citenamefont {Hamley}\ \emph {et~al.}(2012)\citenamefont {Hamley},
  \citenamefont {Gerving}, \citenamefont {Hoang}, \citenamefont {Bookjans},\
  and\ \citenamefont {Chapman}}]{Hamley2012}%
  \BibitemOpen
  \bibfield  {author} {\bibinfo {author} {\bibfnamefont {C.~D.}\ \bibnamefont
  {Hamley}}, \bibinfo {author} {\bibfnamefont {C.}~\bibnamefont {Gerving}},
  \bibinfo {author} {\bibfnamefont {T.}~\bibnamefont {Hoang}}, \bibinfo
  {author} {\bibfnamefont {E.}~\bibnamefont {Bookjans}}, \ and\ \bibinfo
  {author} {\bibfnamefont {M.~S.}\ \bibnamefont {Chapman}},\ }\bibfield
  {title} {\enquote {\bibinfo {title} {Spin-nematic squeezed vacuum in a
  quantum gas},}\ }\href {https://doi.org/10.1038/nphys2245} {\bibfield
  {journal} {\bibinfo  {journal} {Nat. Phys.}\ }\textbf {\bibinfo {volume}
  {8}},\ \bibinfo {pages} {305} (\bibinfo {year} {2012})}\BibitemShut {NoStop}%
\bibitem [{\citenamefont {Fadel}\ \emph {et~al.}(2018)\citenamefont {Fadel},
  \citenamefont {Zibold}, \citenamefont {D{\'e}camps},\ and\ \citenamefont
  {Treutlein}}]{fadel2018spatial}%
  \BibitemOpen
  \bibfield  {author} {\bibinfo {author} {\bibfnamefont {M.}~\bibnamefont
  {Fadel}}, \bibinfo {author} {\bibfnamefont {T.}~\bibnamefont {Zibold}},
  \bibinfo {author} {\bibfnamefont {B.}~\bibnamefont {D{\'e}camps}}, \ and\
  \bibinfo {author} {\bibfnamefont {P.}~\bibnamefont {Treutlein}},\ }\bibfield
  {title} {\enquote {\bibinfo {title} {Spatial entanglement patterns and
  einstein-podolsky-rosen steering in bose-einstein condensates},}\ }\href@noop
  {} {\bibfield  {journal} {\bibinfo  {journal} {Science}\ }\textbf {\bibinfo
  {volume} {360}},\ \bibinfo {pages} {409--413} (\bibinfo {year}
  {2018})}\BibitemShut {NoStop}%
\bibitem [{\citenamefont {Gilmore}\ \emph {et~al.}(2021)\citenamefont
  {Gilmore}, \citenamefont {Affolter}, \citenamefont {Lewis-Swan},
  \citenamefont {Barberena}, \citenamefont {Jordan}, \citenamefont {Rey},\ and\
  \citenamefont {Bollinger}}]{gilmore2021quantum}%
  \BibitemOpen
  \bibfield  {author} {\bibinfo {author} {\bibfnamefont {K.~A.}\ \bibnamefont
  {Gilmore}}, \bibinfo {author} {\bibfnamefont {M.}~\bibnamefont {Affolter}},
  \bibinfo {author} {\bibfnamefont {R.~J.}\ \bibnamefont {Lewis-Swan}},
  \bibinfo {author} {\bibfnamefont {D.}~\bibnamefont {Barberena}}, \bibinfo
  {author} {\bibfnamefont {E.}~\bibnamefont {Jordan}}, \bibinfo {author}
  {\bibfnamefont {A.~M.}\ \bibnamefont {Rey}}, \ and\ \bibinfo {author}
  {\bibfnamefont {J.~J.}\ \bibnamefont {Bollinger}},\ }\bibfield  {title}
  {\enquote {\bibinfo {title} {Quantum-enhanced sensing of displacements and
  electric fields with two-dimensional trapped-ion crystals},}\ }\href@noop {}
  {\bibfield  {journal} {\bibinfo  {journal} {Science}\ }\textbf {\bibinfo
  {volume} {373}},\ \bibinfo {pages} {673--678} (\bibinfo {year}
  {2021})}\BibitemShut {NoStop}%
\bibitem [{\citenamefont {Gil}\ \emph {et~al.}(2014)\citenamefont {Gil},
  \citenamefont {Mukherjee}, \citenamefont {Bridge}, \citenamefont {Jones},\
  and\ \citenamefont {Pohl}}]{Gil2014SSRydberg}%
  \BibitemOpen
  \bibfield  {author} {\bibinfo {author} {\bibfnamefont {L.~I.~R.}\
  \bibnamefont {Gil}}, \bibinfo {author} {\bibfnamefont {R.}~\bibnamefont
  {Mukherjee}}, \bibinfo {author} {\bibfnamefont {E.~M.}\ \bibnamefont
  {Bridge}}, \bibinfo {author} {\bibfnamefont {M.~P.~A.}\ \bibnamefont
  {Jones}}, \ and\ \bibinfo {author} {\bibfnamefont {T.}~\bibnamefont {Pohl}},\
  }\bibfield  {title} {\enquote {\bibinfo {title} {Spin squeezing in a rydberg
  lattice clock},}\ }\href {\doibase 10.1103/PhysRevLett.112.103601} {\bibfield
   {journal} {\bibinfo  {journal} {Phys. Rev. Lett.}\ }\textbf {\bibinfo
  {volume} {112}},\ \bibinfo {pages} {103601} (\bibinfo {year}
  {2014})}\BibitemShut {NoStop}%
\bibitem [{\citenamefont {Schine}\ \emph {et~al.}(2022)\citenamefont {Schine},
  \citenamefont {Young}, \citenamefont {Eckner}, \citenamefont {Martin},\ and\
  \citenamefont {Kaufman}}]{schine2022long}%
  \BibitemOpen
  \bibfield  {author} {\bibinfo {author} {\bibfnamefont {N.}~\bibnamefont
  {Schine}}, \bibinfo {author} {\bibfnamefont {A.~W.}\ \bibnamefont {Young}},
  \bibinfo {author} {\bibfnamefont {W.~J.}\ \bibnamefont {Eckner}}, \bibinfo
  {author} {\bibfnamefont {M.~J.}\ \bibnamefont {Martin}}, \ and\ \bibinfo
  {author} {\bibfnamefont {A.~M.}\ \bibnamefont {Kaufman}},\ }\bibfield
  {title} {\enquote {\bibinfo {title} {Long-lived bell states in an array of
  optical clock qubits},}\ }\href@noop {} {\bibfield  {journal} {\bibinfo
  {journal} {Nature Physics}\ ,\ \bibinfo {pages} {1--7}} (\bibinfo {year}
  {2022})}\BibitemShut {NoStop}%
\bibitem [{\citenamefont {Levine}\ \emph {et~al.}(2018)\citenamefont {Levine},
  \citenamefont {Keesling}, \citenamefont {Omran}, \citenamefont {Bernien},
  \citenamefont {Schwartz}, \citenamefont {Zibrov}, \citenamefont {Endres},
  \citenamefont {Greiner}, \citenamefont {Vuleti{\'c}},\ and\ \citenamefont
  {Lukin}}]{levine2018high}%
  \BibitemOpen
  \bibfield  {author} {\bibinfo {author} {\bibfnamefont {H.}~\bibnamefont
  {Levine}}, \bibinfo {author} {\bibfnamefont {A.}~\bibnamefont {Keesling}},
  \bibinfo {author} {\bibfnamefont {A.}~\bibnamefont {Omran}}, \bibinfo
  {author} {\bibfnamefont {H.}~\bibnamefont {Bernien}}, \bibinfo {author}
  {\bibfnamefont {S.}~\bibnamefont {Schwartz}}, \bibinfo {author}
  {\bibfnamefont {A.~S.}\ \bibnamefont {Zibrov}}, \bibinfo {author}
  {\bibfnamefont {M.}~\bibnamefont {Endres}}, \bibinfo {author} {\bibfnamefont
  {M.}~\bibnamefont {Greiner}}, \bibinfo {author} {\bibfnamefont
  {V.}~\bibnamefont {Vuleti{\'c}}}, \ and\ \bibinfo {author} {\bibfnamefont
  {M.~D.}\ \bibnamefont {Lukin}},\ }\bibfield  {title} {\enquote {\bibinfo
  {title} {High-fidelity control and entanglement of rydberg-atom qubits},}\
  }\href@noop {} {\bibfield  {journal} {\bibinfo  {journal} {Physical review
  letters}\ }\textbf {\bibinfo {volume} {121}},\ \bibinfo {pages} {123603}
  (\bibinfo {year} {2018})}\BibitemShut {NoStop}%
\bibitem [{\citenamefont {Omran}\ \emph {et~al.}(2019)\citenamefont {Omran},
  \citenamefont {Levine}, \citenamefont {Keesling}, \citenamefont {Semeghini},
  \citenamefont {Wang}, \citenamefont {Ebadi}, \citenamefont {Bernien},
  \citenamefont {Zibrov}, \citenamefont {Pichler}, \citenamefont {Choi} \emph
  {et~al.}}]{omran2019CatState}%
  \BibitemOpen
  \bibfield  {author} {\bibinfo {author} {\bibfnamefont {A.}~\bibnamefont
  {Omran}}, \bibinfo {author} {\bibfnamefont {H.}~\bibnamefont {Levine}},
  \bibinfo {author} {\bibfnamefont {A.}~\bibnamefont {Keesling}}, \bibinfo
  {author} {\bibfnamefont {G.}~\bibnamefont {Semeghini}}, \bibinfo {author}
  {\bibfnamefont {T.~T.}\ \bibnamefont {Wang}}, \bibinfo {author}
  {\bibfnamefont {S.}~\bibnamefont {Ebadi}}, \bibinfo {author} {\bibfnamefont
  {H.}~\bibnamefont {Bernien}}, \bibinfo {author} {\bibfnamefont {A.~S.}\
  \bibnamefont {Zibrov}}, \bibinfo {author} {\bibfnamefont {H.}~\bibnamefont
  {Pichler}}, \bibinfo {author} {\bibfnamefont {S.}~\bibnamefont {Choi}},
  \emph {et~al.},\ }\bibfield  {title} {\enquote {\bibinfo {title} {Generation
  and manipulation of {Schr{\"o}dinger} cat states in {Rydberg} atom arrays},}\
  }\href@noop {} {\bibfield  {journal} {\bibinfo  {journal} {Science}\ }\textbf
  {\bibinfo {volume} {365}},\ \bibinfo {pages} {570--574} (\bibinfo {year}
  {2019})}\BibitemShut {NoStop}%
\bibitem [{\citenamefont {He}\ \emph {et~al.}(2019)\citenamefont {He},
  \citenamefont {Perlin}, \citenamefont {Muleady}, \citenamefont {Lewis-Swan},
  \citenamefont {Hutson}, \citenamefont {Ye},\ and\ \citenamefont
  {Rey}}]{he2019engineeringSS}%
  \BibitemOpen
  \bibfield  {author} {\bibinfo {author} {\bibfnamefont {P.}~\bibnamefont
  {He}}, \bibinfo {author} {\bibfnamefont {M.~A.}\ \bibnamefont {Perlin}},
  \bibinfo {author} {\bibfnamefont {S.~R.}\ \bibnamefont {Muleady}}, \bibinfo
  {author} {\bibfnamefont {R.~J.}\ \bibnamefont {Lewis-Swan}}, \bibinfo
  {author} {\bibfnamefont {R.~B.}\ \bibnamefont {Hutson}}, \bibinfo {author}
  {\bibfnamefont {J.}~\bibnamefont {Ye}}, \ and\ \bibinfo {author}
  {\bibfnamefont {A.~M.}\ \bibnamefont {Rey}},\ }\bibfield  {title} {\enquote
  {\bibinfo {title} {Engineering spin squeezing in a 3d optical lattice with
  interacting spin-orbit-coupled fermions},}\ }\href {\doibase
  10.1103/PhysRevResearch.1.033075} {\bibfield  {journal} {\bibinfo  {journal}
  {Phys. Rev. Research}\ }\textbf {\bibinfo {volume} {1}},\ \bibinfo {pages}
  {033075} (\bibinfo {year} {2019})}\BibitemShut {NoStop}%
\bibitem [{\citenamefont {Schleier-Smith}, \citenamefont {Leroux},\ and\
  \citenamefont {Vuleti\'{c}}(2010{\natexlab{b}})}]{Schleier-Smith2010a}%
  \BibitemOpen
  \bibfield  {author} {\bibinfo {author} {\bibfnamefont {M.~H.}\ \bibnamefont
  {Schleier-Smith}}, \bibinfo {author} {\bibfnamefont {I.~D.}\ \bibnamefont
  {Leroux}}, \ and\ \bibinfo {author} {\bibfnamefont {V.}~\bibnamefont
  {Vuleti\'{c}}},\ }\bibfield  {title} {\enquote {\bibinfo {title} {States of
  an ensemble of two-level atoms with reduced quantum uncertainty},}\ }\href
  {\doibase 10.1103/PhysRevLett.104.073604} {\bibfield  {journal} {\bibinfo
  {journal} {Phys. Rev. Lett.}\ }\textbf {\bibinfo {volume} {104}},\ \bibinfo
  {pages} {073604} (\bibinfo {year} {2010}{\natexlab{b}})}\BibitemShut
  {NoStop}%
\bibitem [{\citenamefont {Leroux}, \citenamefont {Schleier-Smith},\ and\
  \citenamefont {Vuleti\'{c}}(2010)}]{leroux2010orientation}%
  \BibitemOpen
  \bibfield  {author} {\bibinfo {author} {\bibfnamefont {I.~D.}\ \bibnamefont
  {Leroux}}, \bibinfo {author} {\bibfnamefont {M.~H.}\ \bibnamefont
  {Schleier-Smith}}, \ and\ \bibinfo {author} {\bibfnamefont {V.}~\bibnamefont
  {Vuleti\'{c}}},\ }\bibfield  {title} {\enquote {\bibinfo {title}
  {Orientation-dependent entanglement lifetime in a squeezed atomic clock},}\
  }\href@noop {} {\bibfield  {journal} {\bibinfo  {journal} {Phys. Rev. Lett.}\
  }\textbf {\bibinfo {volume} {104}},\ \bibinfo {pages} {250801} (\bibinfo
  {year} {2010})}\BibitemShut {NoStop}%
\bibitem [{\citenamefont {Louchet-Chauvet}\ \emph {et~al.}(2010)\citenamefont
  {Louchet-Chauvet}, \citenamefont {Appel}, \citenamefont {Renema},
  \citenamefont {Oblak}, \citenamefont {Kjaergaard},\ and\ \citenamefont
  {Polzik}}]{Louchet-Chauvet2010}%
  \BibitemOpen
  \bibfield  {author} {\bibinfo {author} {\bibfnamefont {A.}~\bibnamefont
  {Louchet-Chauvet}}, \bibinfo {author} {\bibfnamefont {J.}~\bibnamefont
  {Appel}}, \bibinfo {author} {\bibfnamefont {J.~J.}\ \bibnamefont {Renema}},
  \bibinfo {author} {\bibfnamefont {D.}~\bibnamefont {Oblak}}, \bibinfo
  {author} {\bibfnamefont {N.}~\bibnamefont {Kjaergaard}}, \ and\ \bibinfo
  {author} {\bibfnamefont {E.~S.}\ \bibnamefont {Polzik}},\ }\bibfield  {title}
  {\enquote {\bibinfo {title} {Entanglement-assisted atomic clock beyond the
  projection noise limit},}\ }\href
  {http://stacks.iop.org/1367-2630/12/i=6/a=065032} {\bibfield  {journal}
  {\bibinfo  {journal} {New J. Phys.}\ }\textbf {\bibinfo {volume} {12}},\
  \bibinfo {pages} {065032} (\bibinfo {year} {2010})}\BibitemShut {NoStop}%
\bibitem [{\citenamefont {Chen}\ \emph {et~al.}(2014)\citenamefont {Chen},
  \citenamefont {Bohnet}, \citenamefont {Weiner}, \citenamefont {Cox},\ and\
  \citenamefont {Thompson}}]{Chen2014}%
  \BibitemOpen
  \bibfield  {author} {\bibinfo {author} {\bibfnamefont {Z.}~\bibnamefont
  {Chen}}, \bibinfo {author} {\bibfnamefont {J.~G.}\ \bibnamefont {Bohnet}},
  \bibinfo {author} {\bibfnamefont {J.~M.}\ \bibnamefont {Weiner}}, \bibinfo
  {author} {\bibfnamefont {K.~C.}\ \bibnamefont {Cox}}, \ and\ \bibinfo
  {author} {\bibfnamefont {J.~K.}\ \bibnamefont {Thompson}},\ }\bibfield
  {title} {\enquote {\bibinfo {title} {Cavity-aided nondemolition measurements
  for atom counting and spin squeezing},}\ }\href {\doibase
  10.1103/PhysRevA.89.043837} {\bibfield  {journal} {\bibinfo  {journal} {Phys.
  Rev. A}\ }\textbf {\bibinfo {volume} {89}},\ \bibinfo {pages} {043837}
  (\bibinfo {year} {2014})}\BibitemShut {NoStop}%
\bibitem [{\citenamefont {Zhang}\ \emph {et~al.}(2012)\citenamefont {Zhang},
  \citenamefont {McConnell}, \citenamefont {\ifmmode~\acute{C}\else
  \'{C}\fi{}uk}, \citenamefont {Lin}, \citenamefont {Schleier-Smith},
  \citenamefont {Leroux},\ and\ \citenamefont {Vuleti\ifmmode~\acute{c}\else
  \'{c}\fi{}}}]{Zhang2012}%
  \BibitemOpen
  \bibfield  {author} {\bibinfo {author} {\bibfnamefont {H.}~\bibnamefont
  {Zhang}}, \bibinfo {author} {\bibfnamefont {R.}~\bibnamefont {McConnell}},
  \bibinfo {author} {\bibfnamefont {S.}~\bibnamefont {\ifmmode~\acute{C}\else
  \'{C}\fi{}uk}}, \bibinfo {author} {\bibfnamefont {Q.}~\bibnamefont {Lin}},
  \bibinfo {author} {\bibfnamefont {M.~H.}\ \bibnamefont {Schleier-Smith}},
  \bibinfo {author} {\bibfnamefont {I.~D.}\ \bibnamefont {Leroux}}, \ and\
  \bibinfo {author} {\bibfnamefont {V.}~\bibnamefont
  {Vuleti\ifmmode~\acute{c}\else \'{c}\fi{}}},\ }\bibfield  {title} {\enquote
  {\bibinfo {title} {Collective state measurement of mesoscopic ensembles with
  single-atom resolution},}\ }\href {\doibase 10.1103/PhysRevLett.109.133603}
  {\bibfield  {journal} {\bibinfo  {journal} {Phys. Rev. Lett.}\ }\textbf
  {\bibinfo {volume} {109}},\ \bibinfo {pages} {133603} (\bibinfo {year}
  {2012})}\BibitemShut {NoStop}%
\bibitem [{\citenamefont {Hobson}\ \emph {et~al.}(2019)\citenamefont {Hobson},
  \citenamefont {Bowden}, \citenamefont {Vianello}, \citenamefont {Hill},\ and\
  \citenamefont {Gill}}]{Hobson2019}%
  \BibitemOpen
  \bibfield  {author} {\bibinfo {author} {\bibfnamefont {R.}~\bibnamefont
  {Hobson}}, \bibinfo {author} {\bibfnamefont {W.}~\bibnamefont {Bowden}},
  \bibinfo {author} {\bibfnamefont {A.}~\bibnamefont {Vianello}}, \bibinfo
  {author} {\bibfnamefont {I.~R.}\ \bibnamefont {Hill}}, \ and\ \bibinfo
  {author} {\bibfnamefont {P.}~\bibnamefont {Gill}},\ }\bibfield  {title}
  {\enquote {\bibinfo {title} {Cavity-enhanced non-destructive detection of
  atoms for an optical lattice clock},}\ }\href {\doibase 10.1364/OE.27.037099}
  {\bibfield  {journal} {\bibinfo  {journal} {Opt. Express}\ }\textbf {\bibinfo
  {volume} {27}},\ \bibinfo {pages} {37099--37110} (\bibinfo {year}
  {2019})}\BibitemShut {NoStop}%
\bibitem [{\citenamefont {Kessler}\ \emph {et~al.}(2012)\citenamefont
  {Kessler}, \citenamefont {Hagemann}, \citenamefont {Grebing}, \citenamefont
  {Legero}, \citenamefont {Sterr}, \citenamefont {Riehle}, \citenamefont
  {Martin}, \citenamefont {Chen},\ and\ \citenamefont {Ye}}]{kessler2012sub}%
  \BibitemOpen
  \bibfield  {author} {\bibinfo {author} {\bibfnamefont {T.}~\bibnamefont
  {Kessler}}, \bibinfo {author} {\bibfnamefont {C.}~\bibnamefont {Hagemann}},
  \bibinfo {author} {\bibfnamefont {C.}~\bibnamefont {Grebing}}, \bibinfo
  {author} {\bibfnamefont {T.}~\bibnamefont {Legero}}, \bibinfo {author}
  {\bibfnamefont {U.}~\bibnamefont {Sterr}}, \bibinfo {author} {\bibfnamefont
  {F.}~\bibnamefont {Riehle}}, \bibinfo {author} {\bibfnamefont
  {M.}~\bibnamefont {Martin}}, \bibinfo {author} {\bibfnamefont
  {L.}~\bibnamefont {Chen}}, \ and\ \bibinfo {author} {\bibfnamefont
  {J.}~\bibnamefont {Ye}},\ }\bibfield  {title} {\enquote {\bibinfo {title} {A
  sub-40-mhz-linewidth laser based on a silicon single-crystal optical
  cavity},}\ }\href@noop {} {\bibfield  {journal} {\bibinfo  {journal} {Nature
  Photonics}\ }\textbf {\bibinfo {volume} {6}},\ \bibinfo {pages} {687--692}
  (\bibinfo {year} {2012})}\BibitemShut {NoStop}%
\bibitem [{\citenamefont {Davis}, \citenamefont {Bentsen},\ and\ \citenamefont
  {Schleier-Smith}(2016)}]{Davis2016}%
  \BibitemOpen
  \bibfield  {author} {\bibinfo {author} {\bibfnamefont {E.}~\bibnamefont
  {Davis}}, \bibinfo {author} {\bibfnamefont {G.}~\bibnamefont {Bentsen}}, \
  and\ \bibinfo {author} {\bibfnamefont {M.}~\bibnamefont {Schleier-Smith}},\
  }\bibfield  {title} {\enquote {\bibinfo {title} {Approaching the {Heisenberg}
  limit without single-particle detection},}\ }\href {\doibase
  10.1103/PhysRevLett.116.053601} {\bibfield  {journal} {\bibinfo  {journal}
  {Phys. Rev. Lett.}\ }\textbf {\bibinfo {volume} {116}},\ \bibinfo {pages}
  {053601} (\bibinfo {year} {2016})}\BibitemShut {NoStop}%
\bibitem [{\citenamefont {Kessler}\ \emph {et~al.}(2014)\citenamefont
  {Kessler}, \citenamefont {K\'om\'ar}, \citenamefont {Bishof}, \citenamefont
  {Jiang}, \citenamefont {S\o{}rensen}, \citenamefont {Ye},\ and\ \citenamefont
  {Lukin}}]{Kessler2014}%
  \BibitemOpen
  \bibfield  {author} {\bibinfo {author} {\bibfnamefont {E.~M.}\ \bibnamefont
  {Kessler}}, \bibinfo {author} {\bibfnamefont {P.}~\bibnamefont {K\'om\'ar}},
  \bibinfo {author} {\bibfnamefont {M.}~\bibnamefont {Bishof}}, \bibinfo
  {author} {\bibfnamefont {L.}~\bibnamefont {Jiang}}, \bibinfo {author}
  {\bibfnamefont {A.~S.}\ \bibnamefont {S\o{}rensen}}, \bibinfo {author}
  {\bibfnamefont {J.}~\bibnamefont {Ye}}, \ and\ \bibinfo {author}
  {\bibfnamefont {M.~D.}\ \bibnamefont {Lukin}},\ }\bibfield  {title} {\enquote
  {\bibinfo {title} {Heisenberg-limited atom clocks based on entangled
  qubits},}\ }\href {\doibase 10.1103/PhysRevLett.112.190403} {\bibfield
  {journal} {\bibinfo  {journal} {Phys. Rev. Lett.}\ }\textbf {\bibinfo
  {volume} {112}},\ \bibinfo {pages} {190403} (\bibinfo {year}
  {2014})}\BibitemShut {NoStop}%
\bibitem [{\citenamefont {Pezz\`e}\ and\ \citenamefont
  {Smerzi}(2020)}]{PezzeHybridClock2020}%
  \BibitemOpen
  \bibfield  {author} {\bibinfo {author} {\bibfnamefont {L.}~\bibnamefont
  {Pezz\`e}}\ and\ \bibinfo {author} {\bibfnamefont {A.}~\bibnamefont
  {Smerzi}},\ }\bibfield  {title} {\enquote {\bibinfo {title}
  {Heisenberg-limited noisy atomic clock using a hybrid coherent and squeezed
  state protocol},}\ }\href {\doibase 10.1103/PhysRevLett.125.210503}
  {\bibfield  {journal} {\bibinfo  {journal} {Phys. Rev. Lett.}\ }\textbf
  {\bibinfo {volume} {125}},\ \bibinfo {pages} {210503} (\bibinfo {year}
  {2020})}\BibitemShut {NoStop}%
\bibitem [{\citenamefont {Borregaard}\ and\ \citenamefont
  {S\o{}rensen}(2013)}]{Borregaard2013}%
  \BibitemOpen
  \bibfield  {author} {\bibinfo {author} {\bibfnamefont {J.}~\bibnamefont
  {Borregaard}}\ and\ \bibinfo {author} {\bibfnamefont {A.~S.}\ \bibnamefont
  {S\o{}rensen}},\ }\bibfield  {title} {\enquote {\bibinfo {title}
  {Near-heisenberg-limited atomic clocks in the presence of decoherence},}\
  }\href {\doibase 10.1103/PhysRevLett.111.090801} {\bibfield  {journal}
  {\bibinfo  {journal} {Phys. Rev. Lett.}\ }\textbf {\bibinfo {volume} {111}},\
  \bibinfo {pages} {090801} (\bibinfo {year} {2013})}\BibitemShut {NoStop}%
\bibitem [{\citenamefont {Schulte}\ \emph
  {et~al.}(2020{\natexlab{a}})\citenamefont {Schulte}, \citenamefont
  {Mart{\'\i}nez-Lahuerta}, \citenamefont {Scharnagl},\ and\ \citenamefont
  {Hammerer}}]{schulte2020ramsey}%
  \BibitemOpen
  \bibfield  {author} {\bibinfo {author} {\bibfnamefont {M.}~\bibnamefont
  {Schulte}}, \bibinfo {author} {\bibfnamefont {V.~J.}\ \bibnamefont
  {Mart{\'\i}nez-Lahuerta}}, \bibinfo {author} {\bibfnamefont {M.~S.}\
  \bibnamefont {Scharnagl}}, \ and\ \bibinfo {author} {\bibfnamefont
  {K.}~\bibnamefont {Hammerer}},\ }\bibfield  {title} {\enquote {\bibinfo
  {title} {Ramsey interferometry with generalized one-axis twisting echoes},}\
  }\href@noop {} {\bibfield  {journal} {\bibinfo  {journal} {Quantum}\ }\textbf
  {\bibinfo {volume} {4}},\ \bibinfo {pages} {268} (\bibinfo {year}
  {2020}{\natexlab{a}})}\BibitemShut {NoStop}%
\bibitem [{\citenamefont {Fr\"{o}wis}, \citenamefont {Sekatski},\ and\
  \citenamefont {D\"{u}r}(2016)}]{Frowis2016}%
  \BibitemOpen
  \bibfield  {author} {\bibinfo {author} {\bibfnamefont {F.}~\bibnamefont
  {Fr\"{o}wis}}, \bibinfo {author} {\bibfnamefont {P.}~\bibnamefont
  {Sekatski}}, \ and\ \bibinfo {author} {\bibfnamefont {W.}~\bibnamefont
  {D\"{u}r}},\ }\bibfield  {title} {\enquote {\bibinfo {title} {Detecting large
  quantum {Fisher} information with finite measurement precision},}\ }\href
  {\doibase 10.1103/PhysRevLett.116.090801} {\bibfield  {journal} {\bibinfo
  {journal} {Phys. Rev. Lett.}\ }\textbf {\bibinfo {volume} {116}},\ \bibinfo
  {pages} {090801} (\bibinfo {year} {2016})}\BibitemShut {NoStop}%
\bibitem [{\citenamefont {Leibfried}\ \emph {et~al.}(2004)\citenamefont
  {Leibfried}, \citenamefont {Barrett}, \citenamefont {Schaetz}, \citenamefont
  {Britton}, \citenamefont {Chiaverini}, \citenamefont {Itano}, \citenamefont
  {Jost}, \citenamefont {Langer},\ and\ \citenamefont
  {Wineland}}]{Leibfried2004}%
  \BibitemOpen
  \bibfield  {author} {\bibinfo {author} {\bibfnamefont {D.}~\bibnamefont
  {Leibfried}}, \bibinfo {author} {\bibfnamefont {M.~D.}\ \bibnamefont
  {Barrett}}, \bibinfo {author} {\bibfnamefont {T.}~\bibnamefont {Schaetz}},
  \bibinfo {author} {\bibfnamefont {J.}~\bibnamefont {Britton}}, \bibinfo
  {author} {\bibfnamefont {J.}~\bibnamefont {Chiaverini}}, \bibinfo {author}
  {\bibfnamefont {W.~M.}\ \bibnamefont {Itano}}, \bibinfo {author}
  {\bibfnamefont {J.~D.}\ \bibnamefont {Jost}}, \bibinfo {author}
  {\bibfnamefont {C.}~\bibnamefont {Langer}}, \ and\ \bibinfo {author}
  {\bibfnamefont {D.~J.}\ \bibnamefont {Wineland}},\ }\bibfield  {title}
  {\enquote {\bibinfo {title} {Toward {Heisenberg-limited} spectroscopy with
  multiparticle entangled states},}\ }\href@noop {} {\bibfield  {journal}
  {\bibinfo  {journal} {Science}\ }\textbf {\bibinfo {volume} {304}},\ \bibinfo
  {pages} {1476--1478} (\bibinfo {year} {2004})}\BibitemShut {NoStop}%
\bibitem [{\citenamefont {Toscano}\ \emph {et~al.}(2006)\citenamefont
  {Toscano}, \citenamefont {Dalvit}, \citenamefont {Davidovich},\ and\
  \citenamefont {Zurek}}]{Toscano2006}%
  \BibitemOpen
  \bibfield  {author} {\bibinfo {author} {\bibfnamefont {F.}~\bibnamefont
  {Toscano}}, \bibinfo {author} {\bibfnamefont {D.~A.~R.}\ \bibnamefont
  {Dalvit}}, \bibinfo {author} {\bibfnamefont {L.}~\bibnamefont {Davidovich}},
  \ and\ \bibinfo {author} {\bibfnamefont {W.~H.}\ \bibnamefont {Zurek}},\
  }\bibfield  {title} {\enquote {\bibinfo {title} {{Sub-Planck phase-space
  structures and Heisenberg-limited measurements}},}\ }\href {\doibase
  10.1103/PhysRevA.73.023803} {\bibfield  {journal} {\bibinfo  {journal} {Phys.
  Rev. A}\ }\textbf {\bibinfo {volume} {73}},\ \bibinfo {pages} {023803}
  (\bibinfo {year} {2006})}\BibitemShut {NoStop}%
\bibitem [{\citenamefont {Nolan}, \citenamefont {Szigeti},\ and\ \citenamefont
  {Haine}(2017)}]{Nolan2017}%
  \BibitemOpen
  \bibfield  {author} {\bibinfo {author} {\bibfnamefont {S.~P.}\ \bibnamefont
  {Nolan}}, \bibinfo {author} {\bibfnamefont {S.~S.}\ \bibnamefont {Szigeti}},
  \ and\ \bibinfo {author} {\bibfnamefont {S.~A.}\ \bibnamefont {Haine}},\
  }\bibfield  {title} {\enquote {\bibinfo {title} {Optimal and robust quantum
  metrology using interaction-based readouts},}\ }\href {\doibase
  10.1103/PhysRevLett.119.193601} {\bibfield  {journal} {\bibinfo  {journal}
  {Phys. Rev. Lett.}\ }\textbf {\bibinfo {volume} {119}},\ \bibinfo {pages}
  {193601} (\bibinfo {year} {2017})}\BibitemShut {NoStop}%
\bibitem [{\citenamefont {Macr\`{i}}, \citenamefont {Smerzi},\ and\
  \citenamefont {Pezz\`{e}}(2016)}]{Macri2016}%
  \BibitemOpen
  \bibfield  {author} {\bibinfo {author} {\bibfnamefont {T.}~\bibnamefont
  {Macr\`{i}}}, \bibinfo {author} {\bibfnamefont {A.}~\bibnamefont {Smerzi}}, \
  and\ \bibinfo {author} {\bibfnamefont {L.}~\bibnamefont {Pezz\`{e}}},\
  }\bibfield  {title} {\enquote {\bibinfo {title} {Loschmidt echo for quantum
  metrology},}\ }\href {\doibase 10.1103/PhysRevA.94.010102} {\bibfield
  {journal} {\bibinfo  {journal} {Phys. Rev. A}\ }\textbf {\bibinfo {volume}
  {94}},\ \bibinfo {pages} {010102} (\bibinfo {year} {2016})}\BibitemShut
  {NoStop}%
\bibitem [{\citenamefont {Volkoff}\ and\ \citenamefont
  {Martin}(2022)}]{Volkoff_twist-untwist_2022}%
  \BibitemOpen
  \bibfield  {author} {\bibinfo {author} {\bibfnamefont {T.~J.}\ \bibnamefont
  {Volkoff}}\ and\ \bibinfo {author} {\bibfnamefont {M.~J.}\ \bibnamefont
  {Martin}},\ }\bibfield  {title} {\enquote {\bibinfo {title} {Asymptotic
  optimality of twist-untwist protocols for heisenberg scaling in atom-based
  sensing},}\ }\href {\doibase 10.1103/PhysRevResearch.4.013236} {\bibfield
  {journal} {\bibinfo  {journal} {Phys. Rev. Research}\ }\textbf {\bibinfo
  {volume} {4}},\ \bibinfo {pages} {013236} (\bibinfo {year}
  {2022})}\BibitemShut {NoStop}%
\bibitem [{\citenamefont {Colombo}\ \emph {et~al.}(2022)\citenamefont
  {Colombo}, \citenamefont {Pedrozo-Pe\~{n}afiel}, \citenamefont {Adiyatullin},
  \citenamefont {Li}, \citenamefont {Mendez}, \citenamefont {Shu},\ and\
  \citenamefont {Vuleti\'{c}}}]{colombo2021time}%
  \BibitemOpen
  \bibfield  {author} {\bibinfo {author} {\bibfnamefont {S.}~\bibnamefont
  {Colombo}}, \bibinfo {author} {\bibfnamefont {E.}~\bibnamefont
  {Pedrozo-Pe\~{n}afiel}}, \bibinfo {author} {\bibfnamefont {A.~F.}\
  \bibnamefont {Adiyatullin}}, \bibinfo {author} {\bibfnamefont
  {Z.}~\bibnamefont {Li}}, \bibinfo {author} {\bibfnamefont {E.}~\bibnamefont
  {Mendez}}, \bibinfo {author} {\bibfnamefont {C.}~\bibnamefont {Shu}}, \ and\
  \bibinfo {author} {\bibfnamefont {V.}~\bibnamefont {Vuleti\'{c}}},\
  }\bibfield  {title} {\enquote {\bibinfo {title} {Time-reversal-based quantum
  metrology with many-body entangled states},}\ }\href {\doibase
  10.1038/s41567-022-01653-5} {\bibfield  {journal} {\bibinfo  {journal}
  {Nature Physics}\ } (\bibinfo {year} {2022}),\
  10.1038/s41567-022-01653-5}\BibitemShut {NoStop}%
\bibitem [{\citenamefont {Liu}\ \emph {et~al.}(2022)\citenamefont {Liu},
  \citenamefont {Wu}, \citenamefont {Cao}, \citenamefont {Mao}, \citenamefont
  {Li}, \citenamefont {Guo}, \citenamefont {Tey},\ and\ \citenamefont
  {You}}]{liu2022nonlinear}%
  \BibitemOpen
  \bibfield  {author} {\bibinfo {author} {\bibfnamefont {Q.}~\bibnamefont
  {Liu}}, \bibinfo {author} {\bibfnamefont {L.-N.}\ \bibnamefont {Wu}},
  \bibinfo {author} {\bibfnamefont {J.-H.}\ \bibnamefont {Cao}}, \bibinfo
  {author} {\bibfnamefont {T.-W.}\ \bibnamefont {Mao}}, \bibinfo {author}
  {\bibfnamefont {X.-W.}\ \bibnamefont {Li}}, \bibinfo {author} {\bibfnamefont
  {S.-F.}\ \bibnamefont {Guo}}, \bibinfo {author} {\bibfnamefont {M.~K.}\
  \bibnamefont {Tey}}, \ and\ \bibinfo {author} {\bibfnamefont
  {L.}~\bibnamefont {You}},\ }\bibfield  {title} {\enquote {\bibinfo {title}
  {Nonlinear interferometry beyond classical limit enabled by cyclic
  dynamics},}\ }\href@noop {} {\bibfield  {journal} {\bibinfo  {journal}
  {Nature Physics}\ }\textbf {\bibinfo {volume} {18}},\ \bibinfo {pages}
  {167--171} (\bibinfo {year} {2022})}\BibitemShut {NoStop}%
\bibitem [{\citenamefont {Li}\ \emph {et~al.}(2022{\natexlab{b}})\citenamefont
  {Li}, \citenamefont {da~Silva}, \citenamefont {Kain},\ and\ \citenamefont
  {Shahriar}}]{li2022generalized}%
  \BibitemOpen
  \bibfield  {author} {\bibinfo {author} {\bibfnamefont {J.}~\bibnamefont
  {Li}}, \bibinfo {author} {\bibfnamefont {G.~R.}\ \bibnamefont {da~Silva}},
  \bibinfo {author} {\bibfnamefont {S.}~\bibnamefont {Kain}}, \ and\ \bibinfo
  {author} {\bibfnamefont {S.~M.}\ \bibnamefont {Shahriar}},\ }\bibfield
  {title} {\enquote {\bibinfo {title} {A generalized echo squeezing protocol
  with near-heisenberg limit sensitivity and strong robustness against excess
  noise and variation in squeezing parameter},}\ }\href@noop {} {\bibfield
  {journal} {\bibinfo  {journal} {arXiv preprint arXiv:2204.08681}\ } (\bibinfo
  {year} {2022}{\natexlab{b}})}\BibitemShut {NoStop}%
\bibitem [{\citenamefont {Linnemann}\ \emph {et~al.}(2016)\citenamefont
  {Linnemann}, \citenamefont {Strobel}, \citenamefont {Muessel}, \citenamefont
  {Schulz}, \citenamefont {Lewis-Swan}, \citenamefont {Kheruntsyan},\ and\
  \citenamefont {Oberthaler}}]{Linnemann2016}%
  \BibitemOpen
  \bibfield  {author} {\bibinfo {author} {\bibfnamefont {D.}~\bibnamefont
  {Linnemann}}, \bibinfo {author} {\bibfnamefont {H.}~\bibnamefont {Strobel}},
  \bibinfo {author} {\bibfnamefont {W.}~\bibnamefont {Muessel}}, \bibinfo
  {author} {\bibfnamefont {J.}~\bibnamefont {Schulz}}, \bibinfo {author}
  {\bibfnamefont {R.~J.}\ \bibnamefont {Lewis-Swan}}, \bibinfo {author}
  {\bibfnamefont {K.~V.}\ \bibnamefont {Kheruntsyan}}, \ and\ \bibinfo {author}
  {\bibfnamefont {M.~K.}\ \bibnamefont {Oberthaler}},\ }\bibfield  {title}
  {\enquote {\bibinfo {title} {Quantum-enhanced sensing based on time reversal
  of nonlinear dynamics},}\ }\href {\doibase 10.1103/PhysRevLett.117.013001}
  {\bibfield  {journal} {\bibinfo  {journal} {Phys. Rev. Lett.}\ }\textbf
  {\bibinfo {volume} {117}},\ \bibinfo {pages} {013001} (\bibinfo {year}
  {2016})}\BibitemShut {NoStop}%
\bibitem [{\citenamefont {Holstein}\ and\ \citenamefont
  {Primakoff}(1940)}]{Holstein1940}%
  \BibitemOpen
  \bibfield  {author} {\bibinfo {author} {\bibfnamefont {T.}~\bibnamefont
  {Holstein}}\ and\ \bibinfo {author} {\bibfnamefont {H.}~\bibnamefont
  {Primakoff}},\ }\bibfield  {title} {\enquote {\bibinfo {title} {Field
  dependence of the intrinsic domain magnetization of a ferromagnet},}\ }\href
  {\doibase 10.1103/PhysRev.58.1098} {\bibfield  {journal} {\bibinfo  {journal}
  {Phys. Rev.}\ }\textbf {\bibinfo {volume} {58}},\ \bibinfo {pages}
  {1098--1113} (\bibinfo {year} {1940})}\BibitemShut {NoStop}%
\bibitem [{\citenamefont {Hosten}\ \emph
  {et~al.}(2016{\natexlab{b}})\citenamefont {Hosten}, \citenamefont
  {Krishnakumar}, \citenamefont {Engelsen},\ and\ \citenamefont
  {Kasevich}}]{Hosten2016a}%
  \BibitemOpen
  \bibfield  {author} {\bibinfo {author} {\bibfnamefont {O.}~\bibnamefont
  {Hosten}}, \bibinfo {author} {\bibfnamefont {R.}~\bibnamefont
  {Krishnakumar}}, \bibinfo {author} {\bibfnamefont {N.~J.}\ \bibnamefont
  {Engelsen}}, \ and\ \bibinfo {author} {\bibfnamefont {M.~A.}\ \bibnamefont
  {Kasevich}},\ }\bibfield  {title} {\enquote {\bibinfo {title} {Quantum phase
  magnification},}\ }\href {\doibase 10.1126/science.aaf3397} {\bibfield
  {journal} {\bibinfo  {journal} {Science}\ }\textbf {\bibinfo {volume}
  {352}},\ \bibinfo {pages} {1552--1555} (\bibinfo {year}
  {2016}{\natexlab{b}})}\BibitemShut {NoStop}%
\bibitem [{\citenamefont {Ramsey}(1950)}]{ramsey1950molecular}%
  \BibitemOpen
  \bibfield  {author} {\bibinfo {author} {\bibfnamefont {N.~F.}\ \bibnamefont
  {Ramsey}},\ }\bibfield  {title} {\enquote {\bibinfo {title} {A molecular beam
  resonance method with separated oscillating fields},}\ }\href@noop {}
  {\bibfield  {journal} {\bibinfo  {journal} {Physical Review}\ }\textbf
  {\bibinfo {volume} {78}},\ \bibinfo {pages} {695} (\bibinfo {year}
  {1950})}\BibitemShut {NoStop}%
\bibitem [{\citenamefont {Schulte}\ \emph
  {et~al.}(2020{\natexlab{b}})\citenamefont {Schulte}, \citenamefont {Lisdat},
  \citenamefont {Schmidt}, \citenamefont {Sterr},\ and\ \citenamefont
  {Hammerer}}]{schulte2020prospects}%
  \BibitemOpen
  \bibfield  {author} {\bibinfo {author} {\bibfnamefont {M.}~\bibnamefont
  {Schulte}}, \bibinfo {author} {\bibfnamefont {C.}~\bibnamefont {Lisdat}},
  \bibinfo {author} {\bibfnamefont {P.~O.}\ \bibnamefont {Schmidt}}, \bibinfo
  {author} {\bibfnamefont {U.}~\bibnamefont {Sterr}}, \ and\ \bibinfo {author}
  {\bibfnamefont {K.}~\bibnamefont {Hammerer}},\ }\bibfield  {title} {\enquote
  {\bibinfo {title} {Prospects and challenges for squeezing-enhanced optical
  atomic clocks},}\ }\href@noop {} {\bibfield  {journal} {\bibinfo  {journal}
  {Nature communications}\ }\textbf {\bibinfo {volume} {11}},\ \bibinfo {pages}
  {1--10} (\bibinfo {year} {2020}{\natexlab{b}})}\BibitemShut {NoStop}%
\bibitem [{\citenamefont {Braverman}, \citenamefont {Kawasaki},\ and\
  \citenamefont {Vuleti\'{c}}(2018)}]{Braverman2018}%
  \BibitemOpen
  \bibfield  {author} {\bibinfo {author} {\bibfnamefont {B.}~\bibnamefont
  {Braverman}}, \bibinfo {author} {\bibfnamefont {A.}~\bibnamefont {Kawasaki}},
  \ and\ \bibinfo {author} {\bibfnamefont {V.}~\bibnamefont {Vuleti\'{c}}},\
  }\bibfield  {title} {\enquote {\bibinfo {title} {Impact of non-unitary spin
  squeezing on atomic clock performance},}\ }\href
  {http://stacks.iop.org/1367-2630/20/i=10/a=103019} {\bibfield  {journal}
  {\bibinfo  {journal} {New J. Phys.}\ }\textbf {\bibinfo {volume} {20}},\
  \bibinfo {pages} {103019} (\bibinfo {year} {2018})}\BibitemShut {NoStop}%
\bibitem [{\citenamefont {Andr\'e}, \citenamefont {S\o{}rensen},\ and\
  \citenamefont {Lukin}(2004)}]{Andre2004}%
  \BibitemOpen
  \bibfield  {author} {\bibinfo {author} {\bibfnamefont {A.}~\bibnamefont
  {Andr\'e}}, \bibinfo {author} {\bibfnamefont {A.~S.}\ \bibnamefont
  {S\o{}rensen}}, \ and\ \bibinfo {author} {\bibfnamefont {M.~D.}\ \bibnamefont
  {Lukin}},\ }\bibfield  {title} {\enquote {\bibinfo {title} {Stability of
  atomic clocks based on entangled atoms},}\ }\href@noop {} {\bibfield
  {journal} {\bibinfo  {journal} {Phys.\ Rev.\ Lett.}\ }\textbf {\bibinfo
  {volume} {92}},\ \bibinfo {pages} {230801} (\bibinfo {year}
  {2004})}\BibitemShut {NoStop}%
\bibitem [{\citenamefont {Leroux}\ \emph {et~al.}(2017)\citenamefont {Leroux},
  \citenamefont {Scharnhorst}, \citenamefont {Hannig}, \citenamefont {Kramer},
  \citenamefont {Pelzer}, \citenamefont {Stepanova},\ and\ \citenamefont
  {Schmidt}}]{leroux2017line}%
  \BibitemOpen
  \bibfield  {author} {\bibinfo {author} {\bibfnamefont {I.~D.}\ \bibnamefont
  {Leroux}}, \bibinfo {author} {\bibfnamefont {N.}~\bibnamefont {Scharnhorst}},
  \bibinfo {author} {\bibfnamefont {S.}~\bibnamefont {Hannig}}, \bibinfo
  {author} {\bibfnamefont {J.}~\bibnamefont {Kramer}}, \bibinfo {author}
  {\bibfnamefont {L.}~\bibnamefont {Pelzer}}, \bibinfo {author} {\bibfnamefont
  {M.}~\bibnamefont {Stepanova}}, \ and\ \bibinfo {author} {\bibfnamefont
  {P.~O.}\ \bibnamefont {Schmidt}},\ }\bibfield  {title} {\enquote {\bibinfo
  {title} {On-line estimation of local oscillator noise and optimisation of
  servo parameters in atomic clocks},}\ }\href@noop {} {\bibfield  {journal}
  {\bibinfo  {journal} {Metrologia}\ }\textbf {\bibinfo {volume} {54}},\
  \bibinfo {pages} {307} (\bibinfo {year} {2017})}\BibitemShut {NoStop}%
\bibitem [{\citenamefont {Borregaard}\ and\ \citenamefont
  {S{\o}rensen}(2013)}]{borregaard2013efficient}%
  \BibitemOpen
  \bibfield  {author} {\bibinfo {author} {\bibfnamefont {J.}~\bibnamefont
  {Borregaard}}\ and\ \bibinfo {author} {\bibfnamefont {A.~S.}\ \bibnamefont
  {S{\o}rensen}},\ }\bibfield  {title} {\enquote {\bibinfo {title} {Efficient
  atomic clocks operated with several atomic ensembles},}\ }\href@noop {}
  {\bibfield  {journal} {\bibinfo  {journal} {Physical review letters}\
  }\textbf {\bibinfo {volume} {111}},\ \bibinfo {pages} {090802} (\bibinfo
  {year} {2013})}\BibitemShut {NoStop}%
\bibitem [{\citenamefont {Escher}, \citenamefont {de~Matos~Filho},\ and\
  \citenamefont {Davidovich}(2011)}]{escher2011general}%
  \BibitemOpen
  \bibfield  {author} {\bibinfo {author} {\bibfnamefont {B.}~\bibnamefont
  {Escher}}, \bibinfo {author} {\bibfnamefont {R.}~\bibnamefont
  {de~Matos~Filho}}, \ and\ \bibinfo {author} {\bibfnamefont {L.}~\bibnamefont
  {Davidovich}},\ }\bibfield  {title} {\enquote {\bibinfo {title} {General
  framework for estimating the ultimate precision limit in noisy
  quantum-enhanced metrology},}\ }\href@noop {} {\bibfield  {journal} {\bibinfo
   {journal} {Nature Physics}\ }\textbf {\bibinfo {volume} {7}},\ \bibinfo
  {pages} {406--411} (\bibinfo {year} {2011})}\BibitemShut {NoStop}%
\bibitem [{\citenamefont {Demkowicz-Dobrza{\'n}ski}, \citenamefont
  {Ko{\l}ody{\'n}ski},\ and\ \citenamefont
  {Gu{\c{t}}{\u{a}}}(2012)}]{demkowicz2012elusive}%
  \BibitemOpen
  \bibfield  {author} {\bibinfo {author} {\bibfnamefont {R.}~\bibnamefont
  {Demkowicz-Dobrza{\'n}ski}}, \bibinfo {author} {\bibfnamefont
  {J.}~\bibnamefont {Ko{\l}ody{\'n}ski}}, \ and\ \bibinfo {author}
  {\bibfnamefont {M.}~\bibnamefont {Gu{\c{t}}{\u{a}}}},\ }\bibfield  {title}
  {\enquote {\bibinfo {title} {The elusive {Heisenberg} limit in
  quantum-enhanced metrology},}\ }\href {\doibase
  https://doi.org/10.1038/ncomms2067} {\bibfield  {journal} {\bibinfo
  {journal} {Nat. Commun.}\ }\textbf {\bibinfo {volume} {3}},\ \bibinfo {pages}
  {1063} (\bibinfo {year} {2012})}\BibitemShut {NoStop}%
\bibitem [{\citenamefont {Komar}\ \emph {et~al.}(2014)\citenamefont {Komar},
  \citenamefont {Kessler}, \citenamefont {Bishof}, \citenamefont {Jiang},
  \citenamefont {S{\o}rensen}, \citenamefont {Ye},\ and\ \citenamefont
  {Lukin}}]{komar2014quantum}%
  \BibitemOpen
  \bibfield  {author} {\bibinfo {author} {\bibfnamefont {P.}~\bibnamefont
  {Komar}}, \bibinfo {author} {\bibfnamefont {E.~M.}\ \bibnamefont {Kessler}},
  \bibinfo {author} {\bibfnamefont {M.}~\bibnamefont {Bishof}}, \bibinfo
  {author} {\bibfnamefont {L.}~\bibnamefont {Jiang}}, \bibinfo {author}
  {\bibfnamefont {A.~S.}\ \bibnamefont {S{\o}rensen}}, \bibinfo {author}
  {\bibfnamefont {J.}~\bibnamefont {Ye}}, \ and\ \bibinfo {author}
  {\bibfnamefont {M.~D.}\ \bibnamefont {Lukin}},\ }\bibfield  {title} {\enquote
  {\bibinfo {title} {A quantum network of clocks},}\ }\href@noop {} {\bibfield
  {journal} {\bibinfo  {journal} {Nature Physics}\ }\textbf {\bibinfo {volume}
  {10}},\ \bibinfo {pages} {582--587} (\bibinfo {year} {2014})}\BibitemShut
  {NoStop}%
\bibitem [{\citenamefont {Nichol}\ \emph {et~al.}(2022)\citenamefont {Nichol},
  \citenamefont {Srinivas}, \citenamefont {Nadlinger}, \citenamefont {Drmota},
  \citenamefont {Main}, \citenamefont {Araneda}, \citenamefont {Ballance},\
  and\ \citenamefont {Lucas}}]{nichol2022quantum}%
  \BibitemOpen
  \bibfield  {author} {\bibinfo {author} {\bibfnamefont {B.}~\bibnamefont
  {Nichol}}, \bibinfo {author} {\bibfnamefont {R.}~\bibnamefont {Srinivas}},
  \bibinfo {author} {\bibfnamefont {D.}~\bibnamefont {Nadlinger}}, \bibinfo
  {author} {\bibfnamefont {P.}~\bibnamefont {Drmota}}, \bibinfo {author}
  {\bibfnamefont {D.}~\bibnamefont {Main}}, \bibinfo {author} {\bibfnamefont
  {G.}~\bibnamefont {Araneda}}, \bibinfo {author} {\bibfnamefont
  {C.}~\bibnamefont {Ballance}}, \ and\ \bibinfo {author} {\bibfnamefont
  {D.}~\bibnamefont {Lucas}},\ }\bibfield  {title} {\enquote {\bibinfo {title}
  {An elementary quantum network of entangled optical atomic clocks},}\
  }\href@noop {} {\bibfield  {journal} {\bibinfo  {journal} {Nature}\ ,\
  \bibinfo {pages} {1--6}} (\bibinfo {year} {2022})}\BibitemShut {NoStop}%
\bibitem [{\citenamefont {Monroe}\ \emph {et~al.}(2014)\citenamefont {Monroe},
  \citenamefont {Raussendorf}, \citenamefont {Ruthven}, \citenamefont {Brown},
  \citenamefont {Maunz}, \citenamefont {Duan},\ and\ \citenamefont
  {Kim}}]{monroe2014large}%
  \BibitemOpen
  \bibfield  {author} {\bibinfo {author} {\bibfnamefont {C.}~\bibnamefont
  {Monroe}}, \bibinfo {author} {\bibfnamefont {R.}~\bibnamefont {Raussendorf}},
  \bibinfo {author} {\bibfnamefont {A.}~\bibnamefont {Ruthven}}, \bibinfo
  {author} {\bibfnamefont {K.}~\bibnamefont {Brown}}, \bibinfo {author}
  {\bibfnamefont {P.}~\bibnamefont {Maunz}}, \bibinfo {author} {\bibfnamefont
  {L.-M.}\ \bibnamefont {Duan}}, \ and\ \bibinfo {author} {\bibfnamefont
  {J.}~\bibnamefont {Kim}},\ }\bibfield  {title} {\enquote {\bibinfo {title}
  {Large-scale modular quantum-computer architecture with atomic memory and
  photonic interconnects},}\ }\href@noop {} {\bibfield  {journal} {\bibinfo
  {journal} {Physical Review A}\ }\textbf {\bibinfo {volume} {89}},\ \bibinfo
  {pages} {022317} (\bibinfo {year} {2014})}\BibitemShut {NoStop}%
\bibitem [{\citenamefont {Budker}\ \emph {et~al.}(2014)\citenamefont {Budker},
  \citenamefont {Graham}, \citenamefont {Ledbetter}, \citenamefont
  {Rajendran},\ and\ \citenamefont {Sushkov}}]{budker2014proposal}%
  \BibitemOpen
  \bibfield  {author} {\bibinfo {author} {\bibfnamefont {D.}~\bibnamefont
  {Budker}}, \bibinfo {author} {\bibfnamefont {P.~W.}\ \bibnamefont {Graham}},
  \bibinfo {author} {\bibfnamefont {M.}~\bibnamefont {Ledbetter}}, \bibinfo
  {author} {\bibfnamefont {S.}~\bibnamefont {Rajendran}}, \ and\ \bibinfo
  {author} {\bibfnamefont {A.~O.}\ \bibnamefont {Sushkov}},\ }\bibfield
  {title} {\enquote {\bibinfo {title} {Proposal for a cosmic axion spin
  precession experiment (casper)},}\ }\href@noop {} {\bibfield  {journal}
  {\bibinfo  {journal} {Physical Review X}\ }\textbf {\bibinfo {volume} {4}},\
  \bibinfo {pages} {021030} (\bibinfo {year} {2014})}\BibitemShut {NoStop}%
\bibitem [{\citenamefont {Derevianko}\ and\ \citenamefont
  {Pospelov}(2014)}]{Derevianko2014DarkMatter}%
  \BibitemOpen
  \bibfield  {author} {\bibinfo {author} {\bibfnamefont {A.}~\bibnamefont
  {Derevianko}}\ and\ \bibinfo {author} {\bibfnamefont {M.}~\bibnamefont
  {Pospelov}},\ }\bibfield  {title} {\enquote {\bibinfo {title} {Hunting for
  topological dark matter with atomic clocks},}\ }\href
  {https://doi.org/10.1038/nphys3137} {\bibfield  {journal} {\bibinfo
  {journal} {Nature Physics}\ }\textbf {\bibinfo {volume} {10}},\ \bibinfo
  {pages} {933--936} (\bibinfo {year} {2014})}\BibitemShut {NoStop}%
\bibitem [{\citenamefont {Arvanitaki}, \citenamefont {Huang},\ and\
  \citenamefont {Van~Tilburg}(2015)}]{Arvanitaki2015DarkMatter}%
  \BibitemOpen
  \bibfield  {author} {\bibinfo {author} {\bibfnamefont {A.}~\bibnamefont
  {Arvanitaki}}, \bibinfo {author} {\bibfnamefont {J.}~\bibnamefont {Huang}}, \
  and\ \bibinfo {author} {\bibfnamefont {K.}~\bibnamefont {Van~Tilburg}},\
  }\bibfield  {title} {\enquote {\bibinfo {title} {Searching for dilaton dark
  matter with atomic clocks},}\ }\href {\doibase 10.1103/PhysRevD.91.015015}
  {\bibfield  {journal} {\bibinfo  {journal} {Phys. Rev. D}\ }\textbf {\bibinfo
  {volume} {91}},\ \bibinfo {pages} {015015} (\bibinfo {year}
  {2015})}\BibitemShut {NoStop}%
\bibitem [{\citenamefont {Wcis{\l}o}\ \emph {et~al.}(2018)\citenamefont
  {Wcis{\l}o}, \citenamefont {Ablewski}, \citenamefont {Beloy}, \citenamefont
  {Bilicki}, \citenamefont {Bober}, \citenamefont {Brown}, \citenamefont
  {Fasano}, \citenamefont {Ciury{\l}o}, \citenamefont {Hachisu}, \citenamefont
  {Ido}, \citenamefont {Lodewyck}, \citenamefont {Ludlow}, \citenamefont
  {McGrew}, \citenamefont {Morzy{\'n}ski}, \citenamefont {Nicolodi},
  \citenamefont {Schioppo}, \citenamefont {Sekido}, \citenamefont {Le~Targat},
  \citenamefont {Wolf}, \citenamefont {Zhang}, \citenamefont {Zjawin},\ and\
  \citenamefont {Zawada}}]{Wcislo2018}%
  \BibitemOpen
  \bibfield  {author} {\bibinfo {author} {\bibfnamefont {P.}~\bibnamefont
  {Wcis{\l}o}}, \bibinfo {author} {\bibfnamefont {P.}~\bibnamefont {Ablewski}},
  \bibinfo {author} {\bibfnamefont {K.}~\bibnamefont {Beloy}}, \bibinfo
  {author} {\bibfnamefont {S.}~\bibnamefont {Bilicki}}, \bibinfo {author}
  {\bibfnamefont {M.}~\bibnamefont {Bober}}, \bibinfo {author} {\bibfnamefont
  {R.}~\bibnamefont {Brown}}, \bibinfo {author} {\bibfnamefont
  {R.}~\bibnamefont {Fasano}}, \bibinfo {author} {\bibfnamefont
  {R.}~\bibnamefont {Ciury{\l}o}}, \bibinfo {author} {\bibfnamefont
  {H.}~\bibnamefont {Hachisu}}, \bibinfo {author} {\bibfnamefont
  {T.}~\bibnamefont {Ido}}, \bibinfo {author} {\bibfnamefont {J.}~\bibnamefont
  {Lodewyck}}, \bibinfo {author} {\bibfnamefont {A.}~\bibnamefont {Ludlow}},
  \bibinfo {author} {\bibfnamefont {W.}~\bibnamefont {McGrew}}, \bibinfo
  {author} {\bibfnamefont {P.}~\bibnamefont {Morzy{\'n}ski}}, \bibinfo {author}
  {\bibfnamefont {D.}~\bibnamefont {Nicolodi}}, \bibinfo {author}
  {\bibfnamefont {M.}~\bibnamefont {Schioppo}}, \bibinfo {author}
  {\bibfnamefont {M.}~\bibnamefont {Sekido}}, \bibinfo {author} {\bibfnamefont
  {R.}~\bibnamefont {Le~Targat}}, \bibinfo {author} {\bibfnamefont
  {P.}~\bibnamefont {Wolf}}, \bibinfo {author} {\bibfnamefont {X.}~\bibnamefont
  {Zhang}}, \bibinfo {author} {\bibfnamefont {B.}~\bibnamefont {Zjawin}}, \
  and\ \bibinfo {author} {\bibfnamefont {M.}~\bibnamefont {Zawada}},\
  }\bibfield  {title} {\enquote {\bibinfo {title} {New bounds on dark matter
  coupling from a global network of optical atomic clocks},}\ }\href {\doibase
  10.1126/sciadv.aau4869} {\bibfield  {journal} {\bibinfo  {journal} {Science
  Advances}\ }\textbf {\bibinfo {volume} {4}} (\bibinfo {year} {2018}),\
  10.1126/sciadv.aau4869},\ \Eprint
  {http://arxiv.org/abs/https://advances.sciencemag.org/content/4/12/eaau4869.full.pdf}
  {https://advances.sciencemag.org/content/4/12/eaau4869.full.pdf} \BibitemShut
  {NoStop}%
\bibitem [{\citenamefont {Safronova}\ \emph
  {et~al.}(2018{\natexlab{b}})\citenamefont {Safronova}, \citenamefont
  {Budker}, \citenamefont {DeMille}, \citenamefont {Kimball}, \citenamefont
  {Derevianko},\ and\ \citenamefont {Clark}}]{Safronova2018RevModPhys}%
  \BibitemOpen
  \bibfield  {author} {\bibinfo {author} {\bibfnamefont {M.~S.}\ \bibnamefont
  {Safronova}}, \bibinfo {author} {\bibfnamefont {D.}~\bibnamefont {Budker}},
  \bibinfo {author} {\bibfnamefont {D.}~\bibnamefont {DeMille}}, \bibinfo
  {author} {\bibfnamefont {D.~F.~J.}\ \bibnamefont {Kimball}}, \bibinfo
  {author} {\bibfnamefont {A.}~\bibnamefont {Derevianko}}, \ and\ \bibinfo
  {author} {\bibfnamefont {C.~W.}\ \bibnamefont {Clark}},\ }\bibfield  {title}
  {\enquote {\bibinfo {title} {Search for new physics with atoms and
  molecules},}\ }\href {\doibase 10.1103/RevModPhys.90.025008} {\bibfield
  {journal} {\bibinfo  {journal} {Rev. Mod. Phys.}\ }\textbf {\bibinfo {volume}
  {90}},\ \bibinfo {pages} {025008} (\bibinfo {year}
  {2018}{\natexlab{b}})}\BibitemShut {NoStop}%
\bibitem [{\citenamefont {Delva}\ \emph {et~al.}(2018)\citenamefont {Delva},
  \citenamefont {Puchades}, \citenamefont {Sch\"onemann}, \citenamefont
  {Dilssner}, \citenamefont {Courde}, \citenamefont {Bertone}, \citenamefont
  {Gonzalez}, \citenamefont {Hees}, \citenamefont {Le~Poncin-Lafitte},
  \citenamefont {Meynadier}, \citenamefont {Prieto-Cerdeira}, \citenamefont
  {Sohet}, \citenamefont {Ventura-Traveset},\ and\ \citenamefont
  {Wolf}}]{Delva2018_gravity}%
  \BibitemOpen
  \bibfield  {author} {\bibinfo {author} {\bibfnamefont {P.}~\bibnamefont
  {Delva}}, \bibinfo {author} {\bibfnamefont {N.}~\bibnamefont {Puchades}},
  \bibinfo {author} {\bibfnamefont {E.}~\bibnamefont {Sch\"onemann}}, \bibinfo
  {author} {\bibfnamefont {F.}~\bibnamefont {Dilssner}}, \bibinfo {author}
  {\bibfnamefont {C.}~\bibnamefont {Courde}}, \bibinfo {author} {\bibfnamefont
  {S.}~\bibnamefont {Bertone}}, \bibinfo {author} {\bibfnamefont
  {F.}~\bibnamefont {Gonzalez}}, \bibinfo {author} {\bibfnamefont
  {A.}~\bibnamefont {Hees}}, \bibinfo {author} {\bibfnamefont {C.}~\bibnamefont
  {Le~Poncin-Lafitte}}, \bibinfo {author} {\bibfnamefont {F.}~\bibnamefont
  {Meynadier}}, \bibinfo {author} {\bibfnamefont {R.}~\bibnamefont
  {Prieto-Cerdeira}}, \bibinfo {author} {\bibfnamefont {B.}~\bibnamefont
  {Sohet}}, \bibinfo {author} {\bibfnamefont {J.}~\bibnamefont
  {Ventura-Traveset}}, \ and\ \bibinfo {author} {\bibfnamefont
  {P.}~\bibnamefont {Wolf}},\ }\bibfield  {title} {\enquote {\bibinfo {title}
  {Gravitational redshift test using eccentric galileo satellites},}\ }\href
  {\doibase 10.1103/PhysRevLett.121.231101} {\bibfield  {journal} {\bibinfo
  {journal} {Phys. Rev. Lett.}\ }\textbf {\bibinfo {volume} {121}},\ \bibinfo
  {pages} {231101} (\bibinfo {year} {2018})}\BibitemShut {NoStop}%
\bibitem [{\citenamefont {Herrmann}\ \emph {et~al.}(2018)\citenamefont
  {Herrmann}, \citenamefont {Finke}, \citenamefont {L\"ulf}, \citenamefont
  {Kichakova}, \citenamefont {Puetzfeld}, \citenamefont {Knickmann},
  \citenamefont {List}, \citenamefont {Rievers}, \citenamefont {Giorgi},
  \citenamefont {G\"unther}, \citenamefont {Dittus}, \citenamefont
  {Prieto-Cerdeira}, \citenamefont {Dilssner}, \citenamefont {Gonzalez},
  \citenamefont {Sch\"onemann}, \citenamefont {Ventura-Traveset},\ and\
  \citenamefont {L\"ammerzahl}}]{Herrmann2018_gravity}%
  \BibitemOpen
  \bibfield  {author} {\bibinfo {author} {\bibfnamefont {S.}~\bibnamefont
  {Herrmann}}, \bibinfo {author} {\bibfnamefont {F.}~\bibnamefont {Finke}},
  \bibinfo {author} {\bibfnamefont {M.}~\bibnamefont {L\"ulf}}, \bibinfo
  {author} {\bibfnamefont {O.}~\bibnamefont {Kichakova}}, \bibinfo {author}
  {\bibfnamefont {D.}~\bibnamefont {Puetzfeld}}, \bibinfo {author}
  {\bibfnamefont {D.}~\bibnamefont {Knickmann}}, \bibinfo {author}
  {\bibfnamefont {M.}~\bibnamefont {List}}, \bibinfo {author} {\bibfnamefont
  {B.}~\bibnamefont {Rievers}}, \bibinfo {author} {\bibfnamefont
  {G.}~\bibnamefont {Giorgi}}, \bibinfo {author} {\bibfnamefont
  {C.}~\bibnamefont {G\"unther}}, \bibinfo {author} {\bibfnamefont
  {H.}~\bibnamefont {Dittus}}, \bibinfo {author} {\bibfnamefont
  {R.}~\bibnamefont {Prieto-Cerdeira}}, \bibinfo {author} {\bibfnamefont
  {F.}~\bibnamefont {Dilssner}}, \bibinfo {author} {\bibfnamefont
  {F.}~\bibnamefont {Gonzalez}}, \bibinfo {author} {\bibfnamefont
  {E.}~\bibnamefont {Sch\"onemann}}, \bibinfo {author} {\bibfnamefont
  {J.}~\bibnamefont {Ventura-Traveset}}, \ and\ \bibinfo {author}
  {\bibfnamefont {C.}~\bibnamefont {L\"ammerzahl}},\ }\bibfield  {title}
  {\enquote {\bibinfo {title} {Test of the gravitational redshift with galileo
  satellites in an eccentric orbit},}\ }\href {\doibase
  10.1103/PhysRevLett.121.231102} {\bibfield  {journal} {\bibinfo  {journal}
  {Phys. Rev. Lett.}\ }\textbf {\bibinfo {volume} {121}},\ \bibinfo {pages}
  {231102} (\bibinfo {year} {2018})}\BibitemShut {NoStop}%
\end{thebibliography}%

\end{document}